\documentclass[twocolumn,groupedaddress,amsmath,aps,longbibliography,nofootinbib]{revtex4-1}
\usepackage{hyperref}
\usepackage{graphicx}
\usepackage{amsmath,amsfonts}
\usepackage{dcolumn}
\usepackage{bm}
\usepackage{color}
\usepackage{multirow}
\usepackage{float}
\usepackage{url}
\usepackage{physics}
\usepackage{subfigure}
\usepackage[utf8]{inputenc}
\usepackage{siunitx}

\begin{document}

\title{Multiband superconductivity in the kagome-lattice superconductor Re$_2$Zr}

\author{Gabriel Kuderowicz}
\affiliation{Faculty of Physics and Applied Computer Science, AGH University of Krakow, Aleja Mickiewicza 30, 30-059 Krakow, Poland}

\author{Bartlomiej Wiendlocha}
\email{wiendlocha@fis.agh.edu.pl}
\affiliation{Faculty of Physics and Applied Computer Science, AGH University of Krakow, Aleja Mickiewicza 30, 30-059 Krakow, Poland}

\date{\today}

\begin{abstract}
We present a first-principles investigation of the electronic structure, lattice dynamics, electron–phonon coupling, and superconducting properties of the hexagonal kagome-lattice compound Re$_2$Zr, in which time-reversal symmetry breaking and unconventional pairing have recently been proposed. We examine whether superconductivity in Re$_2$Zr can be explained within a conventional phonon-mediated framework by performing fully ab initio calculations within the density functional theory for superconductors. 
The electronic structure is characterized by a mixture of seven two- and three-dimensional Fermi surface sheets, giving rise to multiband superconductivity with many {overlapping} superconducting gaps. 
A spin–orbit-induced van Hove singularity is identified in the density of states near $E_F$. The electron–phonon interaction is moderately strong, with a coupling constant $\lambda \simeq 0.88$.
Superconducting gap calculations reveal {significant} anisotropy and a {broad} distribution of gap values {across different Fermi surface sheets}. Spin fluctuations {introduce additional depairing effects and} reduce the critical temperature from 7.7 K to 6.4 K, in excellent agreement with the experimental value of 6.65 K. 
Taken together, our results show that the superconducting energy scale in Re$_2$Zr can be quantitatively reproduced within a conventional phonon-mediated framework, while the resulting state exhibits pronounced multiband and anisotropic gap structure. The origin of the experimentally suggested time-reversal symmetry breaking, however, remains an open question.
\end{abstract}

\maketitle

\section{Introduction}
Superconductivity in Re$_2$Zr compound with kagome lattice of Re atoms has recently been reported to reveal a time-reversal symmetry (TRS) breaking in the superconducting ground state, indicating the possibility of an unconventional nature of the order parameter \cite{Mandal2025}.
The transition temperature $T_c$ = 6.65 K and the electron-phonon coupling constant $\lambda_{Tc}=0.7$, estimated using the McMillan formula, classified it in the moderate coupling regime.
A relatively large value of the measured Sommerfeld coefficient $\gamma$=12.9 mJ mol$^{-1}$ K$^{-2}$ implies a high density of states at Fermi energy, $N(E_F)$.
The most unusual behavior was observed in the muon spin relaxation $\mu$SR experiment, where the minimum of the relaxation rate $\sigma$ at $T'$=3.1 K was explained by the presence of a spontaneous magnetic field that broke the time-reversal symmetry.
However, the other relaxation rate parameter $\Lambda$ exhibit a maximum at that temperature, thus the order parameter was suggested to be multicomponent and possibly with exotic $s+is$-wave symmetry,  with the superconducting gap changing sign.
Finally, $H_{c1}$ reported in another experimental work~\cite{re2zr-2} shows an upturn at low temperatures, which may be another hint of multigap superconductivity.

\begin{figure*}[t]
    \centering
    \includegraphics[width=0.99\linewidth]{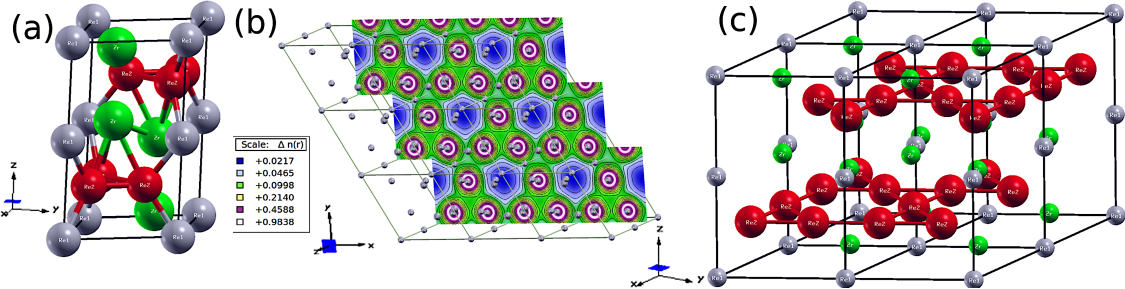}
    \caption{(a) Re$_2$Zr unit cell with gray Re-1, red Re-2 and green Zr atoms, (b) electron density map in a plane with Re-2 atoms forming the kagome lattice, (c) supercell 2x2x1 highlighting the kagome lattice. Visualized using XCrysDen \cite{Kokalj1999}.}
    \label{fig_crystal_struct}
\end{figure*}

Superconductivity of Re$_2$Zr has been known for a long time \cite{Giorgi1970,Roberts1976}, but recently reported TRS breaking in sister compound Re$_2$Hf~\cite{Re2Hf} initiated renewed interest in Re$_2$Zr~\cite{Mandal2025,re2zr-2}. 
These compounds belong to a family of C14 Laves phases, which is a centrosymmetric hexagonal system with space group P6$_3/mmc$ (no. 194).
Its unit cell contains 12 atoms. The characteristic feature of this crystal structure is the presence of the kagome lattice, made of Re-2 atoms arranged in the $ab$ plane in hexagons with triangles attached to their edges.
Figure \ref{fig_crystal_struct} shows the crystal structure and an electron density map in the plane with the kagome lattice.
The Re-2 atomic kagome planes are connected along the $c$ axis by Re-1 atoms: in the unit cell, we can distinguish that two Re-2 atoms from the triangle in the c=1/4 plane are bounded by two Re-1 atoms, located at c=1/2 to the two Re-2 atoms from the kagome triangle in the c=3/4 plane. These six Re atoms form a slightly distorted hexagon, slightly stretched along $c$ and perpendicular to the Zr-Zr atomic pair. 
This structure is geometrically frustrated, which is one of the reasons why kagome lattice materials often exhibit exotic properties, including flat bands, Dirac cones, topological surface states, strong electronic correlations, and time-reversal symmetry breaking \cite{Jiang2021,Ortiz2021,Yu2021,Neupert2022,Bolens2019,Peng2021,Yin2022,Jiang2022,Zhao2021,Guguchia2023,Chaundhary2023,Gui2022,Mielke2021,Kang2023,Gornicka2023,Garmroudi_ni3in}.
In particular many series were synthesized to systematically study unconventional superconductivity.

{The existing literature on kagome superconductors is very rich and presents different phenomena including those competing with superconductivity.
For example in AV$_3$Sb$_5$ \cite{Tan2021} the electron-phonon coupling constant is too small to explain the $T_c$ on the order of 1K, especially in CsV$_3$Sb$_5$.
Those compounds exhibit topological surface states and charge density wave order below about 100 K which suggest unconventional pairing mechanism.
Moreover, CsV$_3$Sb$_5$ shows coupling between kagome layers since very small difference in the c/a ratio resulted in different symmetries in the band structure \cite{Watkins2024}, where the GW calculations reveal dynamical correlations and nontrivial topological character.
On the other hand, high throughput ab initio calculation of kagome materials \cite{Yi2022} found 14 candidates for superconductors with conventional electron-phonon coupling so it should be fairly common in this family.}

To our knowledge, first-principle studies on the lattice dynamics, electron-phonon interaction, and superconductivity of Re$_2$Zr have not yet been reported, and we aim to fill this gap in the current work.
{
The primary goal of the present work is to establish a fully ab initio baseline description of superconductivity in Re$_2$Zr within a conventional phonon-mediated framework, including spin–orbit coupling, multiband effect, and Coulomb repulsion. Such a quantitative reference is essential for determining whether the experimentally observed superconducting properties can be explained within conventional theory or require invoking unconventional pairing mechanisms.
The anisotropic density functional theory for superconductors (SCDFT) within the decoupling approximation and assuming singlet pairing is used \cite{Oliveira1988,Luders2005,Kawamura2017,Kawamura2020}.}
Since electronic Coulomb interactions are expected to significantly affect the superconducting state, ab initio computation of the electron-electron interaction is needed and a typical simplification of the Coulomb interaction with a single parameter $\mu^*$ may be insufficient.
{Longitudinal} spin fluctuations are also considered {as an additional de-pairing mechanism}.
Increased muon relaxation rate below $T_c$ is also correlated with nuclear spin fluctuations in La$_7$Ir$_3$, La$_7$Pd$_3$, La$_7$Rh$_3$, Pr$_{1-x}$La$_x$Pt$_4$Ge$_{12}$, Pr(Os$_{1-x}$Ru$_x$)$_4$Sb$_{12}$, Pr$_{1-y}$La$_y$Os$_4$Sb$_{12}$, Lu$_3$Os$_4$Ge$_{13}$ \cite{Barker2015,Mayoh2021,Singh2020,Shu2021,Zhang2019,Kataria2023} thus electron spin fluctuations may follow and influence superconductivity, especially for the frustrated lattice geometry.

\section{Computation details}

The density functional theory calculations were performed in the {\sc quantum espresso} (QE) package \cite{Giannozzi2009,Giannozzi2017}.
The generalized gradient approximation \cite{Perdew1996} was used for the exchange-correlation potential with projector-augmented wave pseudopotentials from pslibrabry.1.0.0 \cite{DalCorso2014,psnames}.
The wave function and charge density expansion cut-offs were set to 55 Ry and 400 Ry, respectively.
Self-consistent field calculations were performed on a 14x14x6 \textbf{k}-point grid. Density of states, Fermi surface and electron-phonon matrix elements were calculated on a 28x28x12 grid.
{We used the Marzari-Vanderbilt smearing for the selfconsistent field calculations with the default width $\sigma$=0.02 Ry. Electronic DOS was calculated with the optimized tetrahedron method.}
The phonons and Eliashberg functions were calculated using the Density Functional Perturbation Theory (DFPT) \cite{dfpt} on a 7x7x3 \textbf{q}-point grid that gives 16 independent dynamical matrices.
{Electron-phonon matrix elements for the Eliashberg functions were obtained with a smearing $\sigma$=0.005 Ry. }

Ab initio calculations of the superconducting gap were done using the {\sc sctk} Superconducting-Toolkit \cite{Kawamura2020}.
Since the code uses shifted phonon grids, the shifted 7x7x3 q-grid was used, which increased the number of inequivalent \textbf{q}-points to be included in the calculations to 32.
Due to the larger computational requirements, the Fermi surface in the SCDFT calculations was sampled using the same 14x14x6 \textbf{k}-point grid as used for the self-consistent cycle.
{Here, the optimized tetrahedron method is used for all of electronic structure, phonons and electron-phonon matrix elements calculations.}
As the electron-phonon coupling parameter $\lambda$ from the {\sc sctk} calculations matched well the value obtained from QE with the denser Fermi surface sampling, and also the obtained Fermi surface used to visualize the superconducting gap was smooth, the less dense k-grid in SCDFT calculations { was considered as sufficient. Convergence of the calculations is discussed in Suppelemental Material~\cite{suppl}}.

The crystal structure was relaxed separately with and without the spin-orbit coupling (SOC).
The relaxed unit cell parameters (with SOC) are $a$=5.2888~\AA\ and $c/a$=1.64027,
and the atomic positions are displayed in the Supplemental Material~\cite{suppl}.
SOC slightly expands the cell because relaxation without it results in 5.2861~\AA\ and $c/a$=1.63943.
Note that the details of the kagome lattice geometry are crucial for the lattice dynamics and thus superconductivity because our phonon calculations could not converge using the $c/a$ lattice parameter increased only by 0.03\% from the relaxed one.

{In addition, spin-polarized calculations with finite starting magnetization were done, and Re$_2$Zr quickly converged to a nonmagnetic ground state, in agreement with experimental reports. 
However, presence of ferromagnetic spin fluctuations is noticed. They were included in SCDFT calculations using the spin susceptibilities methodology discussed in details in Ref. \cite{Kawamura2020}.}

\section{Electronic structure}

\begin{figure}
    \centering
    \includegraphics[width=0.99\linewidth]{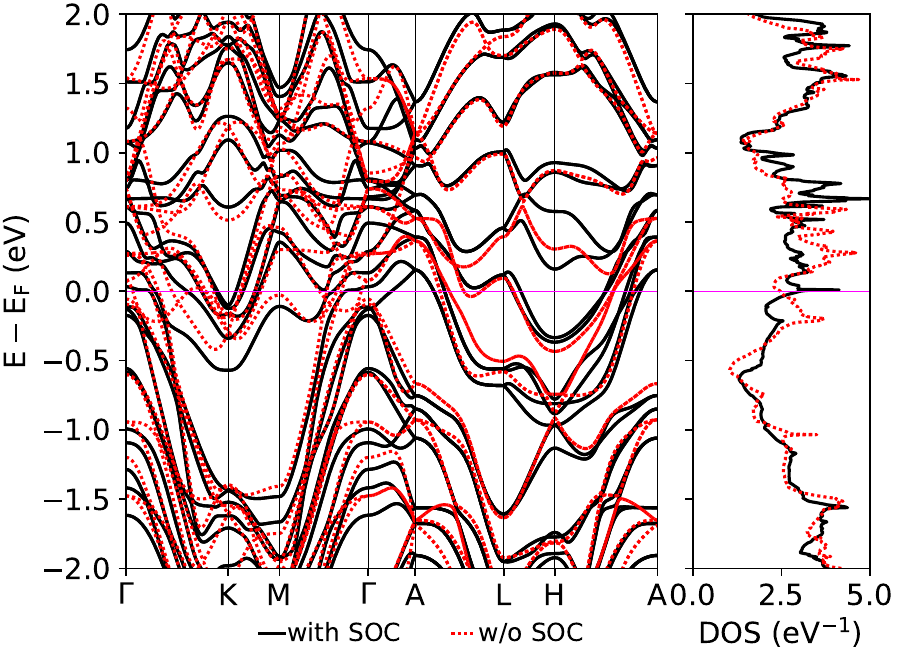}
    \caption{Re$_2$Zr electronic bands and total density of states calculated with and without spin-orbit coupling.}
    \label{fig_elbands}
\end{figure}

Firstly, the electronic band structure was calculated to look for flat bands and van Hove singularities enhancing $N(E_F)$.
Figure \ref{fig_elbands} shows the dispersion relations and the total density of states around the Fermi level, calculated with and without spin-orbit coupling (SOC). 
As we can notice, the impact of SOC is huge, since the band structure without this interaction is much different.
Most importantly, the flat bands appear only with SOC included, very close to $E_F$ (see also Supplemental Material~\cite{suppl} for additional figures).
Due to SOC the Fermi level moves from a local minimum towards the peak, so $N(E_F)$ increases from 2.47 eV$^{-1}$ (scalar-relativistic, per Re$_2$Zr formula unit) to 3.03 eV$^{-1}$ (with SOC). 
Therefore, the following calculations were performed with included spin-orbit coupling.

\begin{figure}
    \centering
    \includegraphics[width=0.99\linewidth]{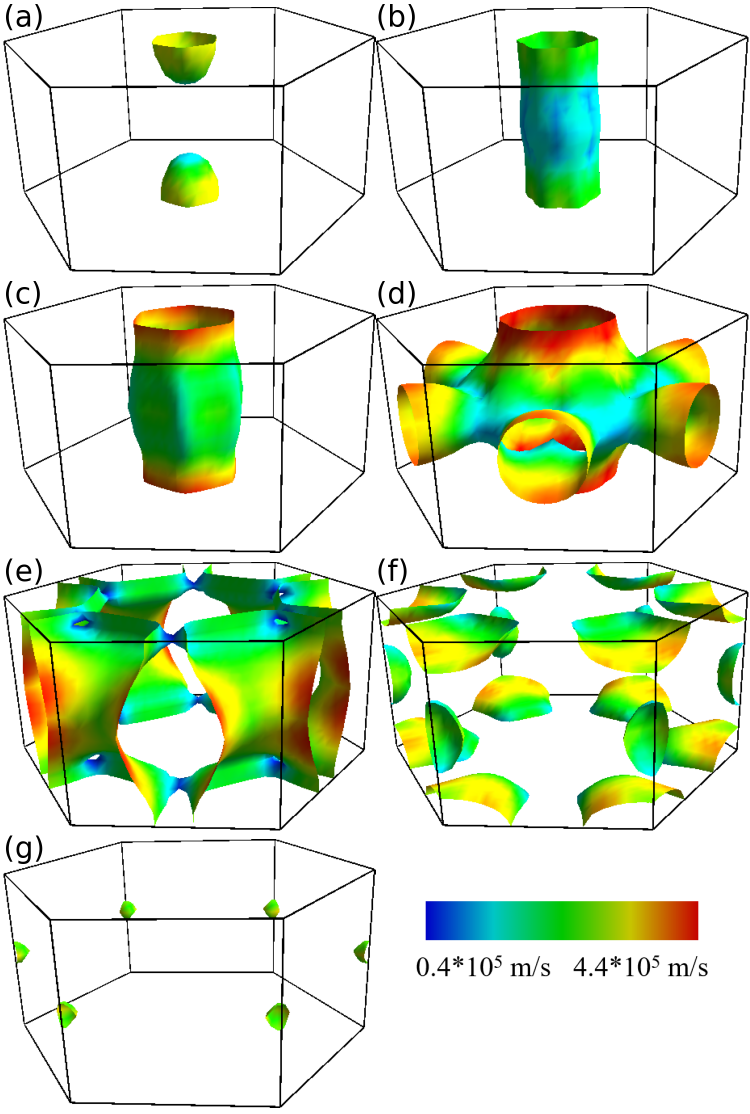}
    \caption{Re$_2$Zr Fermi surface colored with Fermi velocity. Visualized using Fermisurfer \cite{Kawamura2019}.}
    \label{fig_fermivel}
\end{figure}

A large number of bands are expected from 12 atoms in the unit cell of Re$_2$Zr; however, as many as 7 bands crossing the Fermi level are worth noting. Fermi surface sheets are shown in Fig. \ref{fig_fermivel} and five of the total number of seven FS pieces are well developed. 
Fermi velocities are not large, reaching up to $4.4\times 10^5$ m/s, with the average $v_F = 2.32\times 10^5$ m/s.
Generally, variations in shapes of the FS and Fermi velocities are expected to influence the superconducting gap.

Importantly, the Fermi surface of Re$_2$Zr shows that this compound exhibits a mixed 2D-3D character, reflecting the mixed 2D-3D character of the crystal structure. 
Tube-like sheets with the $c$ symmetry axis, shown in Figs.~\ref{fig_fermivel}(b,c), are characteristic for layered two-dimensional electronic systems, where the electronic bands have no dispersion along $k_z$. These tubes are similar to those found in iron-based superconductors~\cite{iron_sc,Carrington_2011}.
In Re$_2$Zr the 2D planes are formed by the Re-2 atoms arranged in the kagome lattice. 
They are perpendicular to the $c$ axis and are responsible for the formation of the FS tubes as Zr atoms contribution to those Fermi surface sheets is smaller and Re-1 is negligible (see the Supplemental Material {with the electronic structure} orbital character).

{On top of the 2D features of the system,} the kagome Re-2 planes are coupled along the $c$ axis via Re-1 atoms, which together with Re-2 atoms form distorted hexagons, perpendicular to the Zr-Zr atomic pair (see Fig.~\ref{fig_crystal_struct}). This bonding contributes to the three-dimensional character of the structure, and hence to the rest of the Fermi surface.

\begin{figure*}[htb]
    \centering
    \includegraphics[width=0.99\linewidth]{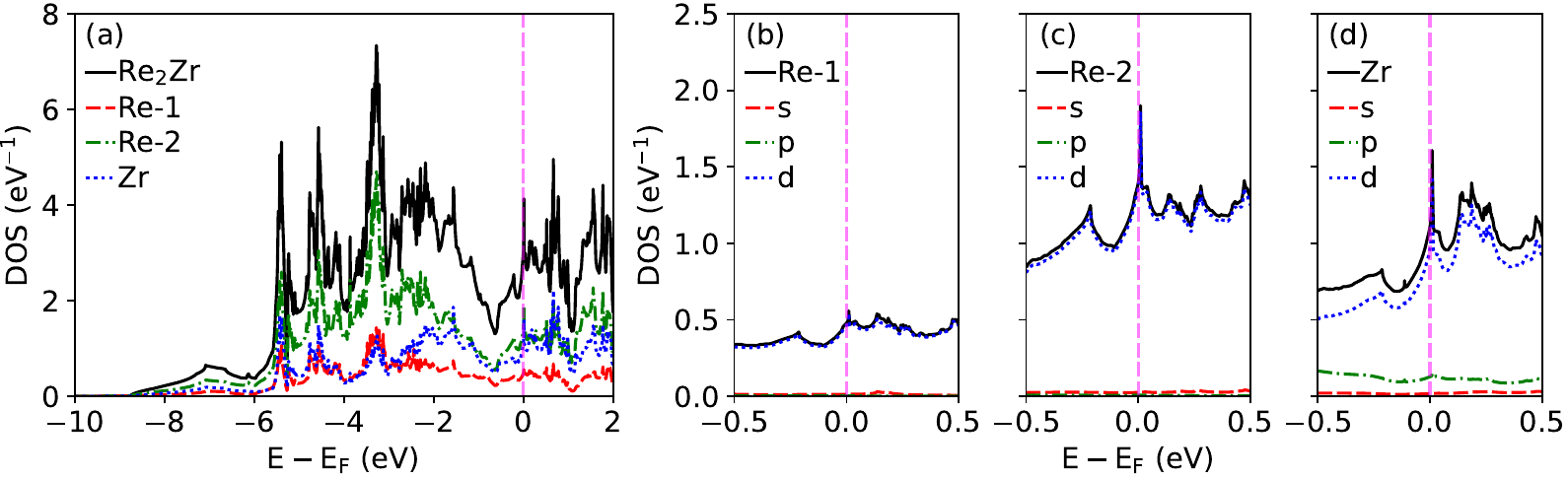}
    \caption{Re$_2$Zr electronic density of states projected onto atomic orbitals.}
    \label{fig_eldos}
\end{figure*}

The total density of states with projection on atomic orbitals is plotted in Fig. \ref{fig_eldos}.
We observe many peaks in the DOS that result from many locally flat bands in the electronic band structure.
For both Re and Zr atoms, d orbitals dominate the contribution to the Fermi surface, with only a small addition of Zr-p states.
The contribution of each atom is of the same magnitude because there are 2 Re-1 positions, 6 Re-2 and 4 Zr, so on average each atom contributes to $N(E_F)$ in approximately 0.25 eV$^{-1}$.
The $E_F$ is located on an ascending slope only 10 meV below the van Hove singularity (vHs) peak that is built mainly from the Re-2 {5d} and Zr {4d} states.
The singularity is related to flat bands seen just above the Fermi energy e.g. near the $\Gamma$ point (see also Supplemental Material ~\cite{suppl}). 
This enhancement of the density of states near $E_F$ increases the number of electronic states available for pairing and contributes to strengthening the electron–phonon interaction, although its effect is moderated by band-dependent Coulomb repulsion, discussed below.
As only Re-2 positions reside on the kagome lattice, the singularity is caused by the interplay of Re with Zr atoms that lay above the center of hexagons on a line perpendicular to them. 

Note that the van Hove singularity disappears when the SOC is not included in calculations.
Importance of spin-orbit splitting of $d$ states in formation of this vHs is additionally seen as the states in flat bands with lower $J=3/2$ dominate ones with $J=5/2$, as shown in Fig. S5~\cite{suppl}. The Fermi level could be moved even closer to the vHs with slight electron doping (assuming the rigid band approximation), which appears as interesting path for future experimental and theoretical works. Also, external hydrostatic pressure could shift the Fermi level and help studying the role of van Hove singularity on superconductivity in Re$_2$Zr. 
Comprehensive electron-phonon calculations of Re$_2$Zr under pressure or upon doping are beyond the scope of this work, however we tested whether formation of vHs is robust against small changes of the structural parameters by performing structure relaxation and total density of states calculations under a hydrostatic pressures of 2, 5 and 10 GPa. Results are shown in Supplemental Material~\cite{suppl}, and we see that the DOS peak survives and is just slightly pushed towards higher energy. 
We also note that this singularity was not identified in the calculations in Ref.~\cite{re2zr-2}.

Based on the calculated $N(E_F)$, the electron-phonon coupling constant $\lambda$ can be calculated from the renormalization of the experimentally determined Sommerfeld coefficient:
\begin{equation}
\gamma_{\rm expt} = \gamma_{\rm band} (1+\lambda_{\gamma}),
\end{equation}
where $\gamma_{\rm band}$ is calculated from the band structure
\begin{equation}
\gamma_{\rm band} = \frac{\pi^2}{3} k_B^2 N(E_F).
\end{equation}
In case of Re$_2$Zr $\gamma_{\rm band}$ = 7.15 mJ mol$^{-1}$ K$^{-2}$ so $\lambda_{\gamma}$ = 0.80.
This value is slightly higher than the electron-phonon coupling constant estimated in~\cite{Mandal2025} from the McMillan formula and $T_c$ (using $\mu^* = 0.13)$, $\lambda_{Tc}=0.7$. 
This suggests that
the superconductivity is not strongly coupled.
Electronic density of states calculations are summarized in Table \ref{tab_eldos}.

\begin{table}
\caption{Re$_2$Zr total and projected $N(E_F)$ (eV$^{-1}$) per f.u..\label{tab_eldos}}
\begin{center}
\begin{ruledtabular}
\begin{tabular}{ccccc}
     & total & s & p & d \\
    \hline
    Re$_2$Zr & 3.028 &  &  & \\
    Re-1 & 0.477 & 0.013 & 0.002 & 0.462 \\
    Re-2 & 1.393 & 0.026 & 0.008 & 1.359 \\
    Zr & 1.103 & 0.018 & 0.127 & 0.958 \\
	\end{tabular}
\end{ruledtabular}
\end{center}
\end{table}

\section{Phonons and isotropic Eliashberg functions}

\begin{figure}[b]
    \centering
    \includegraphics[width=0.99\linewidth]{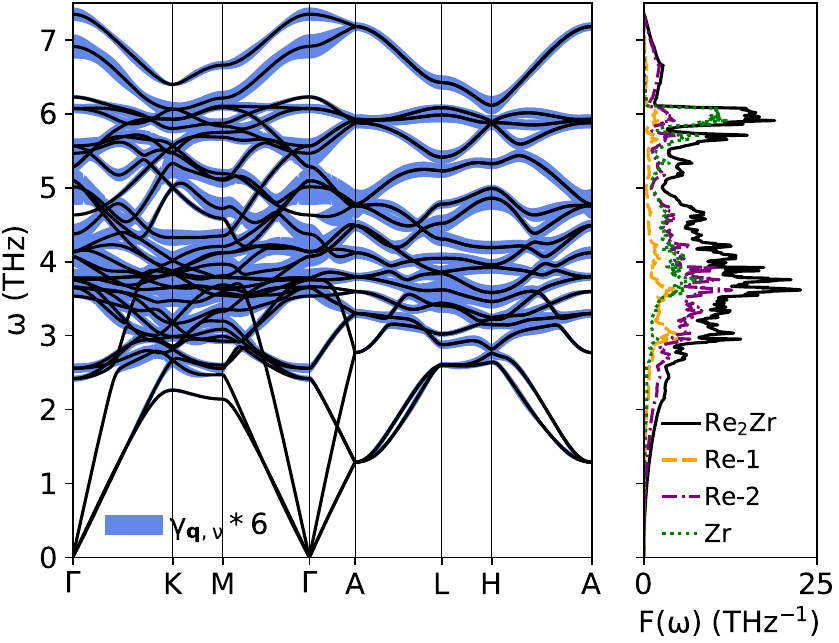}
    \caption{Re$_2$Zr phonon dispersion and density of states with atomic contributions. Blue fatbands represent phonon linewidth multiplied by a factor 6.}
    \label{fig_phgamlines}
\end{figure}

The phonon dispersion relations and the phonon density of states $F(\omega)$ are shown in Fig. \ref{fig_phgamlines}.
Several interesting observations can be made. 
The phonon structure exhibits many flat phonon bands, especially near 4 THz.
Although Re atoms are two times heavier than Zr atoms, there are no gaps in the phonon spectrum.
Zr vibrations are visible in the partial phonon density of states even for the low-frequency acoustic modes. 
The acoustic modes are softened in the A point, with frequencies below 1.5 THz.
The highest optical modes, on the other hand, are contributed by the heavy Re-2 atom vibrations. These are high-energy breathing modes (see Supplemental Material~\cite{suppl} for animations of the phonon modes) analogous to those observed in cubic Laves phases like SrIr$_2$~\cite{gutowska2021}, where the heavier atom also vibrates with the highest frequency. At the $\Gamma$ point, the highest energy mode consists of in-plane Re-2 atoms' vibration toward the center of the kagome triangles, whereas in the cubic Laves case, the analogical mode consisted of transition metals' vibrations towards the centers of tetrahedrons.

The blue shading of the bands in Fig.~\ref{fig_phgamlines} represents the phonon linewidths, which are calculated using the equation:
\begin{align}
	\gamma_{\mathbf{q}}^{\nu} = 2\pi \omega_{\mathbf{q}\nu} \sum_{ij} \int& \frac{d^3k}{\Omega_{BZ}}  |g_{\mathbf{q}}^{\nu}(\mathbf{k},i,j)|^2 \nonumber \\
\times& \delta(E_{\mathbf{q}}^{i} - E_F) \delta(E_{\mathbf{k}+\mathbf{q}}^{j} - E_F). \label{eq_gammaq}
\end{align} 
The electron-phonon interaction matrix element $g$ is calculated as
\cite{Grimvall1981,Wierzbowska2005,heid-pb}
\begin{equation}
 g_{\mathbf{q}}^\nu(\mathbf{k},i,j)=\sum_s \left ( \frac{\hbar}{2M_s\omega_{\mathbf{q}\nu}} \right ) ^{1/2} \langle \psi_{i,\mathbf{k}}| \frac{dV_{scf}}{d\hat{u}_{\nu,s}} \cdot \hat{\epsilon}_\nu | \psi _{j,\mathbf{k+q}} \rangle,
\end{equation}
for phonon vector $\mathbf{q}$, phonon mode $\nu$, atom with index $s$ and a self-consistently calculated deviation of potential $\frac{dV_{scf}}{d\hat{u}_{\nu,s}}$ by a phonon with a polarization vector $\hat{\epsilon}_\nu$.
The lifetime of phonons is inversely proportional to $\gamma_{\mathbf{q}}^{\nu}$.
Dirac deltas select only electrons on the Fermi surface to participate in the interaction. 

\begin{table*}[t]
\caption{Parameters of the electron-phonon interaction in Re$_2$Zr: electron phonon coupling constant $\lambda$ with contributions from each phonon mode, logarithmic average phonon frequency $\omega_{\rm ln}$, and transition temperature $T_c$ for two values of Coulomb pseudopotential parameter $\mu^*$.\label{tab_elphparams}}
\begin{center}
\begin{ruledtabular}
\begin{tabular}{c|cccccccccc}
$\lambda_{tot}$ = 0.868 & $\lambda_{1-9}$ & 0.050 & 0.051 & 0.041 & 0.036 & 0.033 & 0.032 & 0.032 & 0.035 & 0.032 \\
$\omega_{\rm ln}$ = 3.389 THz & $\lambda_{10-18}$ & 0.030 & 0.031 & 0.030 & 0.030 & 0.030 & 0.027 & 0.026 & 0.026 & 0.027 \\
$T_c$ = 8.9 K ($\mu^*$ = 0.1) & $\lambda_{19-27}$ & 0.024 & 0.021 & 0.022 & 0.021 & 0.021 & 0.018 & 0.019 & 0.015 & 0.015 \\
$T_c$ = 6.7 K ($\mu^*$ = 0.145) & $\lambda_{28-36}$ & 0.015 & 0.011 & 0.012 & 0.010 & 0.009 & 0.008 & 0.008 & 0.011 & 0.008
\end{tabular}
\end{ruledtabular}
\end{center}
\end{table*}

The magnitude of the phonon linewidths, which describe the local strength of the electron-phonon coupling, are distributed in the \textbf{q}-space in Fig.~\ref{fig_phgamlines} with moderate anisotropy and are comparable for each optical mode.

The Eliashberg function $\alpha^2F(\omega)$ is computed as
\begin{equation}
	\alpha^2F(\omega) = \frac{1}{2\pi N(E_F)} \sum_{\mathbf{q}\nu} \delta(\omega - \omega_{\mathbf{q}\nu}) \frac{\gamma_{\mathbf{q}\nu}}{\hbar \omega_{\mathbf{q}\nu}}, \label{eq_a2F}
\end{equation}
and describes the overall electron-phonon interaction strength as a function of the phonon frequency. Its integral gives the electron-phonon coupling constant $\lambda$:
\begin{equation}
	\lambda = 2 \int_0^{\omega_{\rm max}} \frac{\alpha^2F(\omega)}{\omega} d\omega, \label{eq_lambda_a2F}
\end{equation}
which is a single parameter conveniently characterizing the average coupling strength in a superconductor. 
The cumulative electron-phonon coupling constant $\lambda(\omega)$ is also calculated as a function of the upper integration limit $\omega_{max}$ to visualize the gradual contribution of the phonon spectrum.

\begin{figure}[b]
    \centering
    \includegraphics[width=0.99\linewidth]{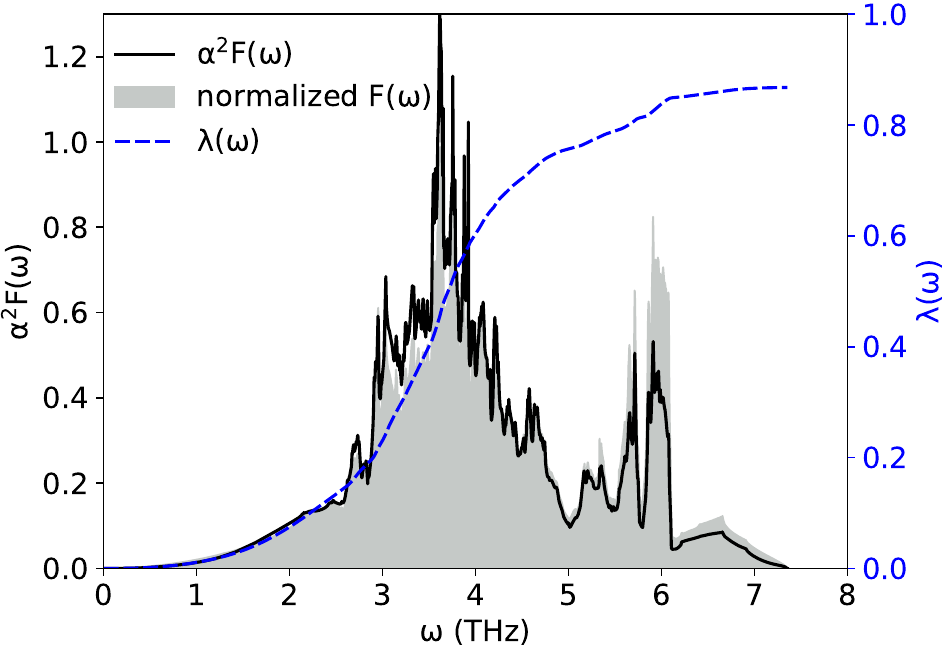}
    \caption{(left axis) Re$_2$Zr Eliashberg function with normalized phonon density of states ($\int_0^{\omega_{\rm max}}F(\omega) d\omega = \int_0^{\omega_{\rm max}}\alpha^2F(\omega) d\omega$). (right axis) Cumulative electron-phonon coupling constant $\lambda(\omega) = 2\int_0^{\omega}\frac{\alpha^2F(\omega')}{\omega'} d\omega'$.}
    \label{fig_a2flamb}
\end{figure}

Figure \ref{fig_a2flamb} presents the Eliashberg function, the phonon density of states (renormalized to have the same area under the curve) and the cumulative electron-phonon coupling constant.
In general, we can notice a similar shape of $\alpha^2F(\omega)$ and $F(\omega)$ with the exception near 6 THz, where the Eliashberg function is greatly enhanced above $F(\omega)$. Here, flat optical phonon bands are found, contributed mainly by Zr. In general, $\lambda(\omega)$ increases smoothly, so most phonon modes contribute equally to $\lambda$, with the above-mentioned exception near 6 THz where a steep increase in $\lambda(\omega)$ is found.
The total calculated value of $\lambda$ for Re$_2$ Zr is $\lambda=0.87$, placing this compound in moderate coupling strength, in agreement with the electronic specific heat analysis ($\lambda_{\gamma} = 0.80$).

For compounds with elements as heavy as Re ($Z=75$, $M = 186$ u), one could expect the low-lying acoustic modes to dominate the contribution to $\lambda$, as e.g. in the cubic Laves phase superconductor SrIr$_2$ ($\lambda = 1.1, T_c = 6.1$~K) which has a similar mass asymmetry. 
In contrast, a relatively strong interaction that leads to large phonon linewidths for optical modes indicates that they contribute significantly to $\lambda$ despite their high frequencies (see also Table \ref{tab_elphparams} for the contribution of specific modes, lower frequency ones have larger $\lambda_i$ but these differences are not that large).
Therefore, the characteristic logarithmic average phonon frequency weighted by the electron-phonon interaction:
\begin{equation}
	\omega_{\rm ln} = \exp \left( \frac{2}{\lambda}\int_0^{\omega_{\rm max}} \alpha^2F(\omega) \frac{\ln \omega}{\omega} d\omega \right)
\label{eq_omloga2F}
\end{equation}
is equal to 3.39 THz (163 K). This is a relatively high value, much higher than the 2 THz (97 K) found in SrIr$_2$~\cite{gutowska2021} and enhances $T_c$, which depends linearly on $\omega_{\rm ln}$ in the Allen-Dynes formula: \cite{Allen1975}
\begin{equation}
\label{eq_Allen_Dynes}
k_{B}T_{c}=\frac{f_1 f_2 \hbar\omega_{\rm ln}}{1.20}\,
\exp\left\{-\frac{1.04(1+\lambda)}{\lambda-\mu^{*}(1+0.62\lambda)}\right\}.
\end{equation}
Due to the moderate-coupling value of $\lambda=0.87$ the correction factors $f_1 f_2$ in the Allen-Dynes formula are not necessary ($f_1 f_2 \simeq 1.0$).
The Coulomb pseudopotential parameter $\mu^*$ is typically assumed for metals in the range 0.10-0.13.
The large $\omega_{\rm ln}$ results in a relatively high critical temperature as for the compound with $\lambda < 1.0$. With the lower bound of $\mu^*=0.10$ the $T_c=8.9$ K is obtained, which is much higher than the experimental value of 6.65 K. The experimental $T_c$ can be reproduced for $\mu^*$=0.145. This suggests that either the isotropic Eliashberg formalism, which is behind the Allen-Dynes formula, is not adequate for Re$_2$Zr, or the electron-electron interactions are slightly enhanced compared to a typical metal. 
These issues are discussed in the next part of the paper. 

The most important parameters calculated in this Section are collected in Table \ref{tab_elphparams}.

\section{Structure of the superconducting gaps}

\begin{figure*}
    \centering
    \includegraphics[width=0.99\linewidth]{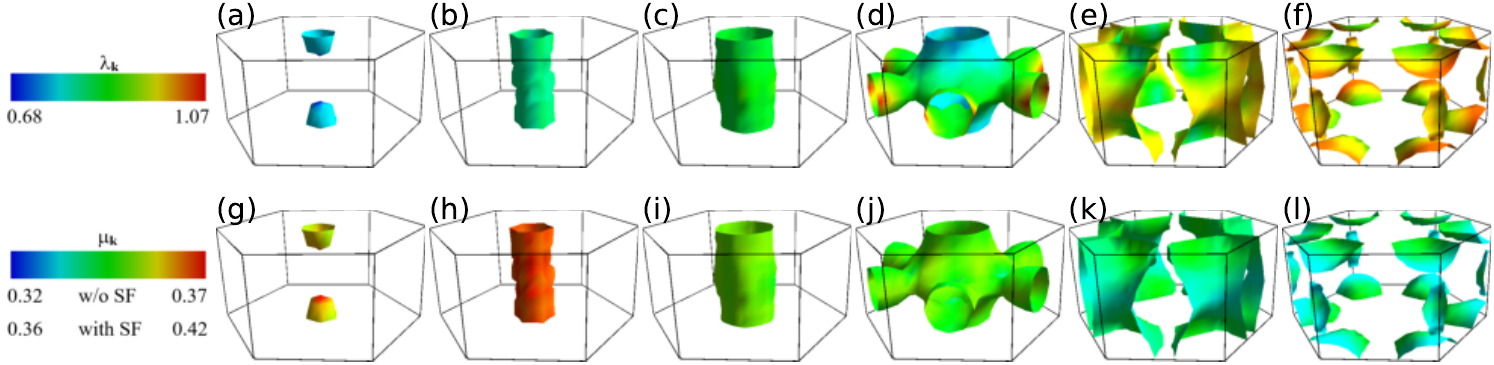}
    \caption{Re$_2$Zr $\mathbf{k}$-dependent electron-phonon coupling constant (top row) and Coulomb (or Coulomb with SF) potential (bottom row).}
    \label{fig_scdist}
\end{figure*}

The superconducting gap structure and the corresponding fully ab initio estimate of $T_c$ are calculated within the density functional theory for superconductors as implemented in the {\sc superconducting toolkit}. The decoupling approximation is employed and calculations are restricted to singlet pairing \cite{Oliveira1988,Luders2005,Kawamura2017,Kawamura2020}. Since Re$_2$Zr converges to a nonmagnetic ground state, the system preserves time-reversal symmetry within this framework, and TRS-breaking states are not considered. 
Our objective is to assess whether the experimentally observed superconducting properties of Re$_2$Zr can be quantitatively reproduced within a conventional phonon-mediated description, thereby establishing a reference point for evaluating possible unconventional scenarios.

Calculations begin with obtaining the electronic eigenvalues $\xi_{n\mathbf{k}}$ (relative to $E_F$), the phonon frequencies and elements of the electron-phonon interaction matrix from the DFPT~\cite{Kawamura2017,Kawamura2020}.
They enter the self-consistent equation for the superconducting gap $\Delta_{n\mathbf{k}}$
\begin{align}\label{eq:delta}
    \Delta_{n\mathbf{k}} = & -\frac{1}{2}\sum_{n'\mathbf{k}'} \frac{K_{n\mathbf{k}n'\mathbf{k}'}(\xi_{n\mathbf{k}},\xi_{n'\mathbf{k}'})}{1+Z_{n\mathbf{k}}(\xi_{n\mathbf{k}})} \nonumber \\
    & \times \frac{\Delta_{n'\mathbf{k}'}}{\sqrt{\xi_{n'\mathbf{k}'}^2+\Delta_{n'\mathbf{k}'}^2}} \tanh \frac{\sqrt{\xi_{n'\mathbf{k}'}^2+\Delta_{n'\mathbf{k}'}^2}}{2T},
\end{align}
where the integration kernel $K$
\begin{align}
    K_{n\mathbf{k}n'\mathbf{k}'}(\xi,\xi') & \equiv K^{ep}_{n\mathbf{k}n'\mathbf{k}'}(\xi,\xi') \nonumber \\
    & + K^{ee}_{n\mathbf{k}n'\mathbf{k}'}(\xi,\xi') + K^{sf}_{n\mathbf{k}n'\mathbf{k}'}(\xi,\xi')\label{eq_K} 
\end{align}
consist of electron-phonon, electron-electron and spin-fluctuation terms.
The dynamically screened Coulomb interaction is calculated by using the random phase approximation.
The quantity $Z_{n\mathbf{k}}$ is a renormalization factor that accounts only for electron-phonon and spin-fluctuation terms because Coulomb repulsion is already present in the calculation of $\xi_{n\mathbf{k}}$.
Longitudinal spin fluctuations, considered here, 
are calculated from Kohn-Sham orbitals by constructing spin susceptibilities, and
have a purely de-pairing role, competing with Cooper pairs formation \cite{Tsutsumi2020}.
More complex pairing-enhancing spin-fluctuation mechanisms, which may occur in strongly correlated systems, are beyond the scope of the present phonon-mediated SCDFT framework, and are not considered.
The gap equation is solved as a function of temperature and $T_c$ is found when the averaged gap vanishes.
A more detailed discussion of this computational framework can be found in \cite{Kawamura2017,Kawamura2020}.

\begin{figure*}[ht]
    \centering
    \includegraphics[width=0.99\linewidth]{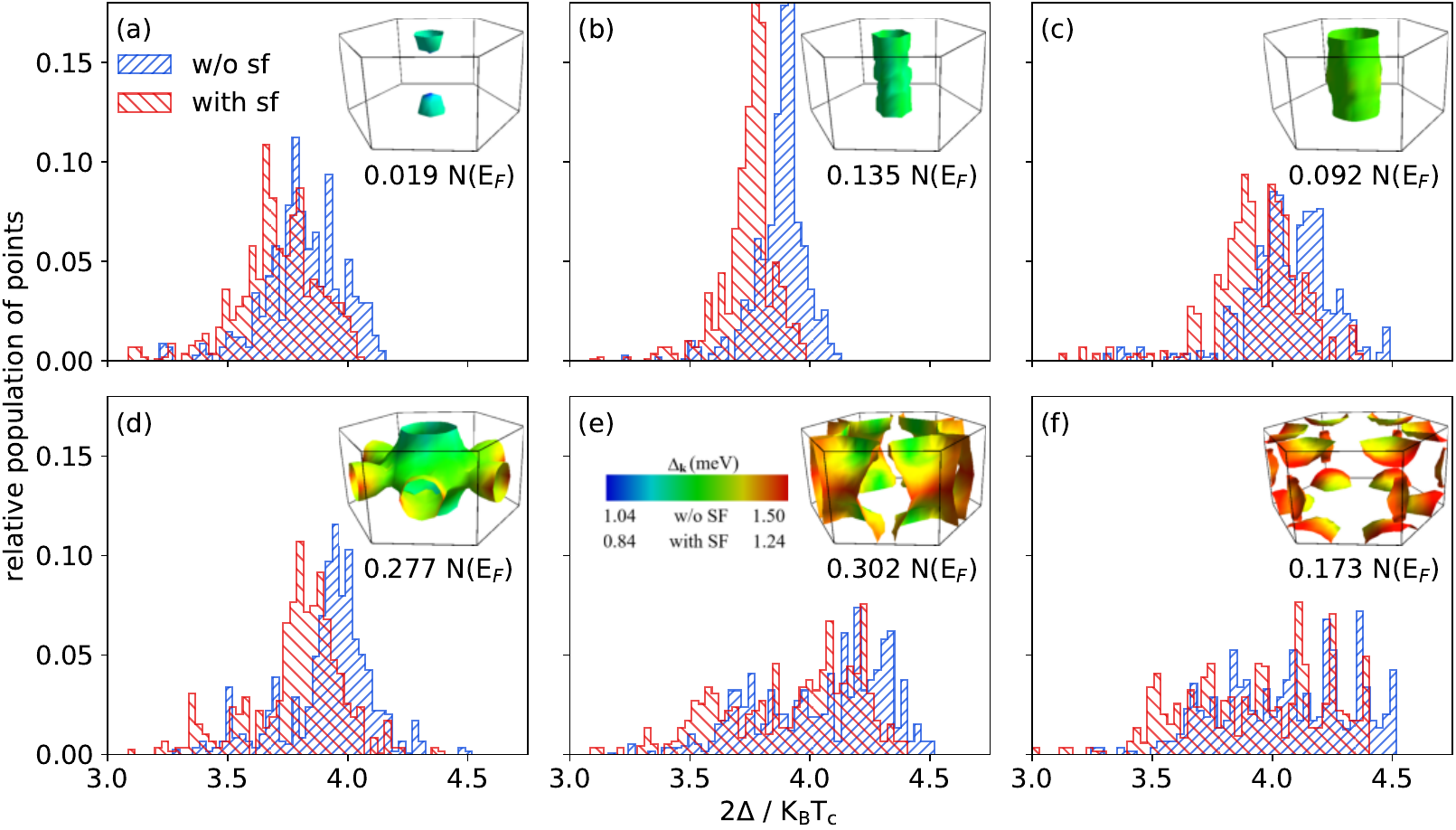}
    \caption{Re$_2$Zr distribution of superconducting gap values (at 0 K) in the reciprocal space. Insets present the superconducting gap plotted on the Fermi surface with their contributions to $N(E_F)$ written below. Band 7 with small pockets is not shown for brevity because it contributes less than 0.5\% to $N(E_F)$.}
    \label{fig_gap}
\end{figure*}

Figure~\ref{fig_scdist} presents the $\mathbf{k}$-dependent electron-phonon coupling parameter $\lambda_{n\mathbf{k}}$ and the repulsion parameter $\mu_{n\mathbf{k}}$ (either Coulomb only or with SF included) plotted on the Fermi surface.
The last FS sheet with the smallest pockets (from Fig.~\ref{fig_fermivel}(g)) is omitted for brevity because it contributes less than 0.5\% to $N(E_F)$.
The distribution of the superconducting gaps, along with their histograms, are shown in Fig. \ref{fig_gap}. 
To better quantify $\Delta$ and other properties for each Fermi surface sheet, Fermi surface averages are calculated using the equation:
\begin{equation}
   \overline{\Delta}_n = \frac{\sum_{\mathbf{k}} \Delta_{n\mathbf{k}}\delta(E_{\mathbf{k}} - E_F)}{\sum_{\mathbf{k}}{\delta(E_{\mathbf{k}} - E_F})},
\end{equation}
where Dirac deltas are approximated with the cold smearing method \cite{Marzari1999,Marzari2023}
\begin{equation}
    \tilde{\delta}(x) = \frac{1}{\sqrt{\pi}} e^{- \left[ x -1/\sqrt{2} \right]^2}(2-\sqrt{2}x),
\end{equation}
$x\equiv x_{n\mathbf{k}} = (E_F-E_{n\mathbf{k}})/\sigma$, $E_{n\mathbf{k}}$ is the energy eigenvalue for given $\mathbf{k}$ and band $n$, and a broadening $\sigma=0.005$ Ry was chosen.
$\lambda$ and $\mu$ were analyzed in the same way.
Global averages are obtained as a weighted sum with contributions to $N(E_F)$ of each Fermi surface sheet as weights. 
The average values are presented in Table \ref{tab_sctk_res}.

Looking now at the distribution of the electron-phonon coupling parameter in Fig.~\ref{fig_scdist}(a-f) one can see that the interaction strength is different for different FS sheets, being the strongest on the well-developed sheets in panels (d-f), with intermediate strength on cylindrical ones [panels (b,c)].
The $\lambda_{\mathbf k}$ values span the range from 0.68 to 1.07 and exhibit moderate anisotropy in sheets (d-f).
{The anisotropy of $\lambda_{\mathbf k}$ to certain degree follows the anisotropy of the Fermi velocity $d E/d\mathbf{k}$ (see Fig.~\ref{fig_fermivel}) as in general $\lambda_{\mathbf k}$ is inversely proportional to $v_F$ and there are no strongly nested parts of FS. However, the regions of FS with the largest $\lambda_{\mathbf k}$ in  Fig.~\ref{fig_scdist}(e-f) appear in spite of having also a large $v_F$. Here, the  electron-phonon interaction matrix elements are enhanced.}

The averaged $\overline{\lambda_i}$ for the seven bands that form the Fermi surface (Table~\ref{tab_sctk_res}) vary between 0.76 and 0.98 and the total average $\lambda = 0.89$, confirming the previous result and the moderate coupling regime. 
The difference in $\lambda$ between calculations done as an input to SCDFT and the earlier obtained value from the integrated Eliashberg function ($\lambda = 0.87$) is small, which confirms the convergence of our results.
This difference, approximately equal to 3\%, stems from the different grids and integration methods, and 
we can consider it as a measure of the numerical accuracy of our results.
Based on convergence tests of electron–phonon coupling strength and phonon frequencies, we estimate an overall uncertainty of approximately 5–7\% in the calculated $T_c$, which does not affect the main conclusions of this work.

The $\mathbf{k}$-distribution of the Coulomb repulsion parameter $\mu$, shown in Fig.~\ref{fig_scdist}(g-h) shows that it is the largest in the cylindrical sheet in panel (h). This FS sheet is {formed by the flat band which builds the van Hove singularity and is} contributed mainly by the 5d states of Re-2 atoms, which form the kagome lattice (see Supplemental Material~\cite{suppl} for the atomic orbital decomposition of FS).
Thus, electronic repulsion is substantially stronger in the kagome lattice, competing with superconductivity. 
Spin fluctuations additionally increase the electron repulsion parameter $\mu$, and the strongest increase is observed also for the cylindrical FS sheets, which is expected due to the frustrated nature of the kagome lattice.
{Interestingly, the anisotropy of the Coulomb repulsion parameter $\mu$ does not follow the same trend as the electron–phonon coupling constant $\lambda$, but instead shows a partially opposite behavior (see average values in Table~\ref{tab_sctk_res}). 
In conventional multiband superconductors, both $\lambda$ and $\mu$ tend to be positively correlated, being enhanced on the same Fermi surface sheets with high density of states, see e.g.~\cite{Kawamura2017,Gutowska2024}. 
In contrast, in Re$_2$Zr we observe opposite behavior due to enhanced Coulomb repulsion on the kagome lattice.
This partial anticorrelation enhances superconducting gap anisotropy discussed below, since bands with stronger depairing interaction experience a greater suppression of the superconducting gap
and represents an important mechanism shaping the multigap superconducting state in Re$_2$Zr.}

The global average values of the repulsion parameter $\mu$, which is the
dimensionless product of the average interaction kernel times the density of states at $E_F$, are 0.346 (sole Coulomb interaction) and 0.393 (with SF included). 
These are quite enhanced values compared to elemental Re (0.276 including SF)~\cite{Kawamura2020} or to a strongly-coupled ScAu$_2$Al compound (0.297)~\cite{Kuderowicz2023} also having 5d orbitals at $E_F$.
Therefore, the previous observation of a higher value $\mu^*=0.145$ that restores experimental $T_c$ in the Allen-Dynes formula is supported by strong electron-electron interactions enhanced by spin fluctuations. 
However, for $\mu$ to be directly compared with the commonly used $\mu^*$ Coulomb pseudopotential parameter, which enters the Allen-Dynes formula~\ref{eq_Allen_Dynes}, retardation effects must be included. 
In the superconducting state, due to the retarded nature of electron-phonon coupling, repulsive interactions are effectively weakened, and this effect is intrinsically included in the SCDFT~\cite{scdft-metals}.
Retardation effects may be approximately included to obtain the effective $\mu^*$ parameter~\cite{Bogoljubov1958,Morel1962,mcmillan,Allen1975,Carbotte1990,chang_nb,gunnarsson2013} through the equation:
\begin{equation}
    \label{eq_mu*}
    \mu^* = \frac{\mu}{1+\mu\ \ln(\omega_{\rm el}/\omega_{\rm ph})},
\end{equation}
where $\omega_{el}$ and $\omega_{ph}$ set the electron and phonon energy scales, respectively.
These are usually set as \cite{Allen1975} the plasmon frequency or the  bandwidth for electrons, and the maximum frequency $\omega_{\rm max}$ as $\omega_{\rm ph}$ for phonons.
Taking the electron bandwidth of 9 eV as $\omega_{el}$ and the maximum phonon frequency of 7.5 THz (31 meV) we obtain $\mu^* = 0.122$ (with SF) and $\mu^* = 0.117$ (w/o SF).
Thus, partially the increase in $\mu^*$ required in the Allen-Dynes formula to recover the experimental $T_c$ can be attributed to the increase in depairing interactions above the $\mu^* = 0.10$ value recommended by Allen and Dynes~\cite{Allen1975} and partially to anisotropic features, not covered by this formula.

\begin{table}[b]
    \centering
    \caption{Results of SCDFT calculations ($T =  0$ K) in a form of Fermi surface averaged electron-phonon coupling constant $\overline{\lambda_i}$, Coulomb parameter $\overline{\mu_i}$ and superconducting gap $\overline{\Delta_i}$ for each Fermi surface sheet b$_i$, and the global average (weighted with each sheet contribution to the $N(E_F)$).}
    \begin{ruledtabular}
    \begin{tabular}{cccccc}
    band & $\overline{\lambda_i}$ & \multicolumn{2}{c}{$\overline{\mu_i}$} & \multicolumn{2}{c}{$\overline{\Delta_i}$ (meV)} \\
     & & w/o SF & with SF & w/o SF & with SF \\
     \hline 
    b1 & 0.762 & 0.363 & 0.414 & 1.162 & 0.940 \\
    b2 & 0.827 & 0.366 & 0.419 & 1.232 & 0.998 \\
    b3 & 0.875 & 0.349 & 0.399 & 1.323 & 1.078 \\
    b4 & 0.862 & 0.349 & 0.395 & 1.310 & 1.071 \\
    b5 & 0.932 & 0.340 & 0.387 & 1.393 & 1.142 \\
    b6 & 0.976 & 0.338 & 0.384 & 1.435 & 1.179 \\
    b7 & 0.954 & 0.323 & 0.367 & 1.443 & 1.186 \\
    tot. avg. & 0.895 & 0.346 & 0.393 & 1.340 & 1.096\\
    \hline \\
   \multicolumn{6}{c}{$T_c = 7.7$ K (w/o SF),  6.4 K (with SF), 6.65 K (expt.)}
    \end{tabular}
    \end{ruledtabular}
    \label{tab_sctk_res}
\end{table}

Conversely, the explicit and $\mathbf{k}$-resolved treatment of the electron–phonon, electron–electron, and spin-fluctuation contributions in Re$_2$Zr within the SCDFT formalism enables the superconducting critical temperature to be predicted with high accuracy and without the use of empirical parameters. The computed critical temperatures are $T_c = 7.7$ K (without spin fluctuations) and $T_c = 6.4$ K (including spin fluctuations). This comparison indicates that spin fluctuations play a significant role in this compound, leading to a reduction of the critical temperature and yielding quantitative agreement with the experimental value $T_c = 6.65$ K.

Histograms of superconducting gaps are plotted for each FS sheet as $2\Delta_i/k_BT_c$ in Fig. \ref{fig_gap}, with insets showing their anisotropic distribution. In each case, the $x$ axis is normalized by the calculated corresponding $T_c$ (with or without SF).
As the orbital character of Re-2 is clearly dominant for large Fermi surface sheets (see Fig.S4 l-m in the Supplemental Material~\cite{suppl}) for which the superconducting gap is also the largest, therefore, Re-2 atoms in the kagome lattice indeed have noticeably higher contribution to superconductivity than Re-1 atoms.
The multiband - multigap superconductivity of Re$_2$Zr is evident because $\Delta$ distributions are centered at different energies and their spread is wide, especially for more developed sheets in panels (d-f). 
Nevertheless, superconducting gaps overlap and do not form disjointed parts, such as, for example, in MgB$_2$ \cite{mgb2} or Pb~\cite{pb-gap1,Gutowska2024}.
There are no nodal lines or points with $\Delta=0$ so Re$_2$Zr is fully gapped, while spin fluctuations uniformly decrease $\Delta$ values, even if normalized by $k_BT_c$.
The gap values spread the range between 3.0 and 4.5 $k_BT_c$, thus the distribution is very broad.
The averaged gaps (Table~\ref{tab_sctk_res}) show enhanced $2\Delta/k_BT_c$ ratios above the weak-coupling limit of 3.53: 4.02 (w/o SF) and 3.96 (with SF), at the border of the strong-coupling range.
{This is achieved even though $\lambda < 1$ as an anisotropic electron–phonon coupling and band-dependent Coulomb repulsion can increase $2\Delta/k_BT_c$ while keeping $\lambda$ in the moderate-coupling regime.}
The distribution of the overall gap (sum on all FS sheets) is shown in Supplemental Material~\cite{suppl}.

\begin{figure}
    \centering
    \includegraphics[width=0.9\linewidth]{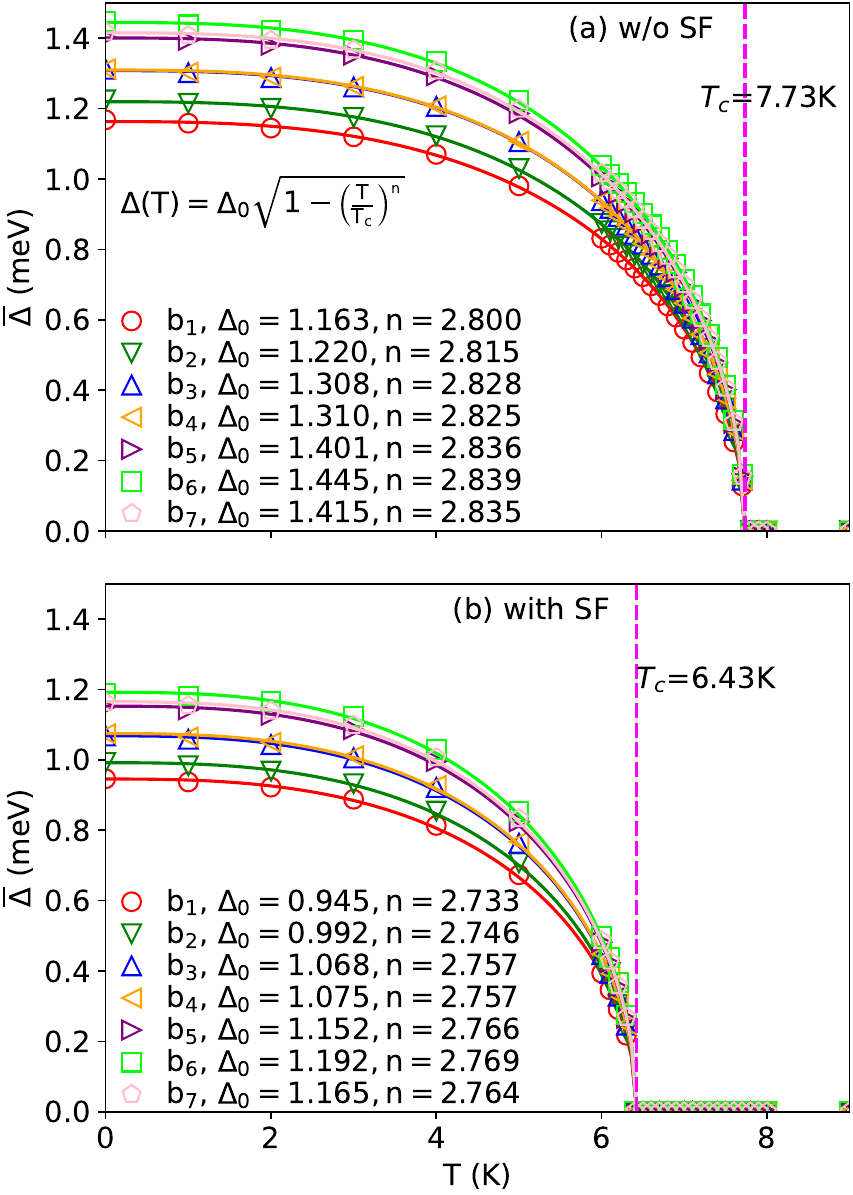}
    \caption{Re$_2$Zr Fermi surface averaged superconducting gap, for each Fermi surface sheet $b_i$, at selected temperatures.}
    \label{fig_deltavsT}
\end{figure}

\begin{figure}
    \centering
    \includegraphics[width=0.99\linewidth]{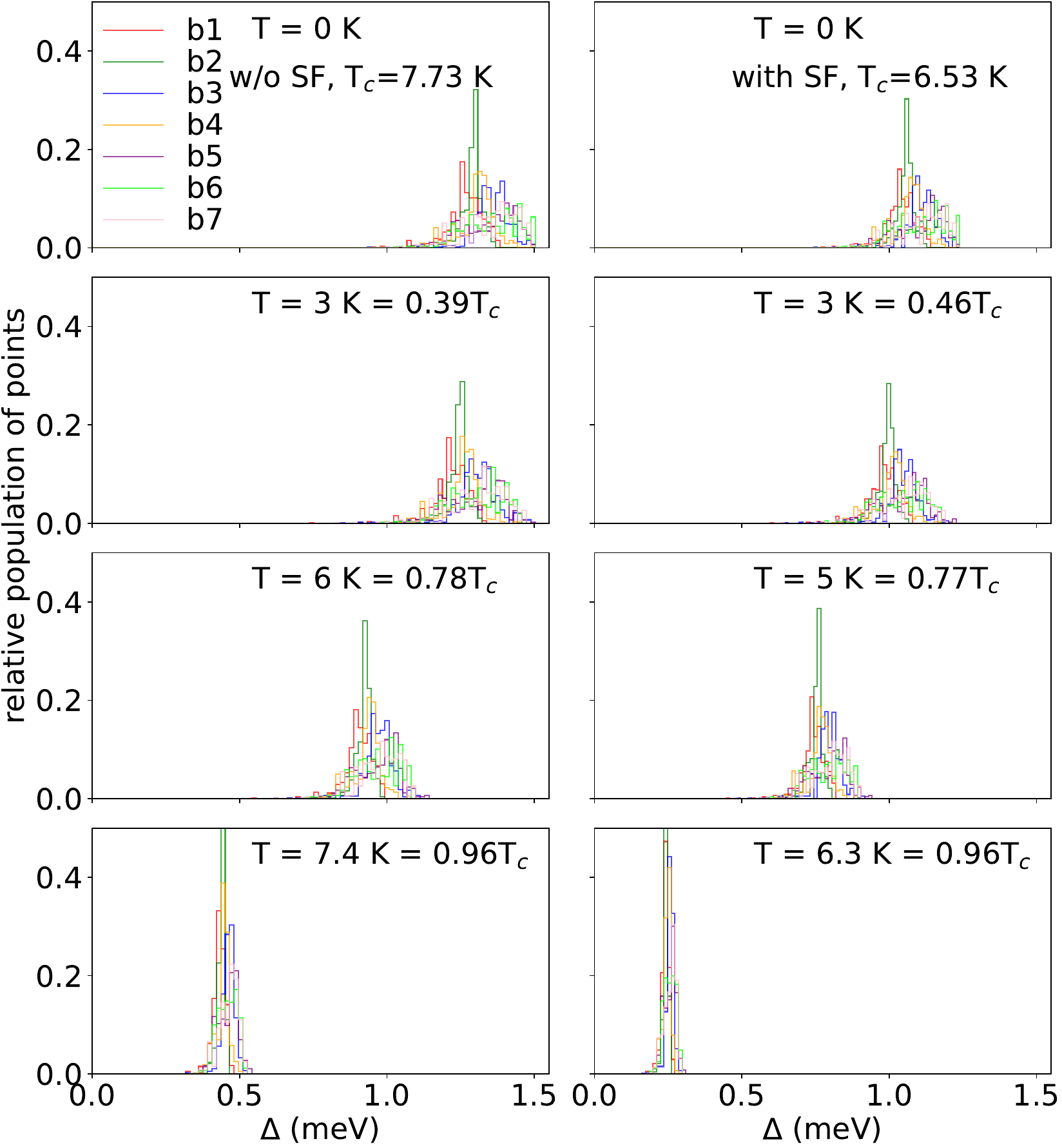}
    \caption{Re$_2$Zr superconducting gap distributions, for each Fermi surface sheet $b_i$, at selected temperatures.}
    \label{fig_deltahistvsT}
\end{figure}

{Temperature dependence of the superconducting gaps is presented in Figs. \ref{fig_deltavsT} and \ref{fig_deltahistvsT}.
Temperature-dependent Fermi surface averaged values of gaps for each band can be fitted with an equation \cite{Kuderowicz2021}:
\begin{equation}
    \Delta(T) = \Delta(0)\sqrt{1-\left(\frac{T}{T_c}\right)^n}.
\end{equation}
The characteristic dome shape of $\Delta(T)$ in the BCS theory is obtained for the exponent $n \approx 3$ which is slightly higher than $n=2.73-2.84$ obtained here in the case of Re$_2$Zr.
The exponent is smaller when the spin fluctuations are present.
The anisotropic multiple gap structure is well preserved to temperatures up to $0.8T_c$, for larger temperature gaps distribution naturally shrink to a very narrow energy range, especially when SFs are included.}

\begin{figure}
    \centering
    \includegraphics[width=0.9\linewidth]{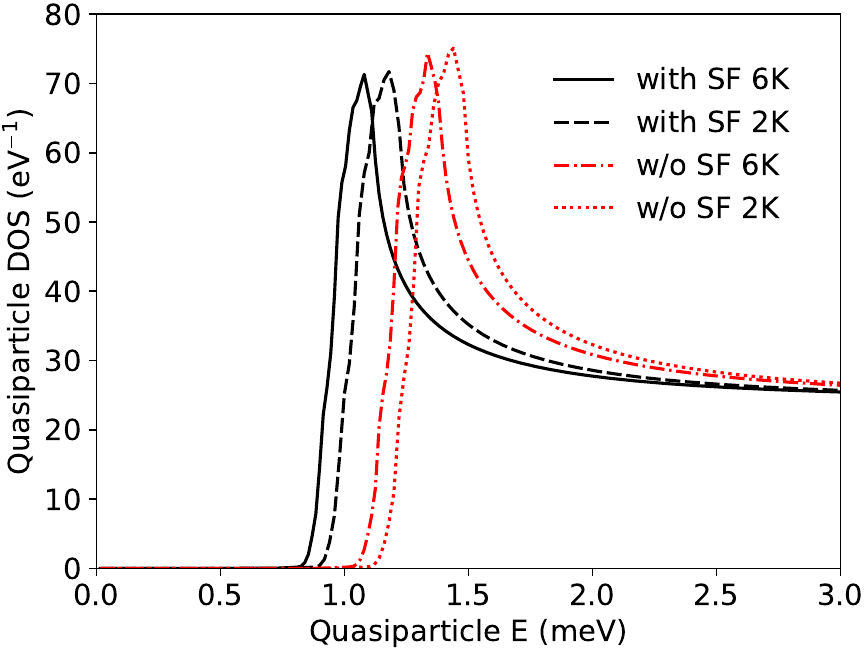}
    \caption{Re$_2$Zr quasiparticle density of states for two temperatures $T = 2$ K and $T = 6$ K.}
    \label{fig_qpdos}
\end{figure}

Finally, the quasiparticle density of states is calculated and presented in  Fig. \ref{fig_qpdos}.
It has a broadened structure over the usual BCS-like peak reflecting the broad energy spectrum of the gaps, however, due to the overlap of the gaps and the triangular shape of the global gap distribution~\cite{suppl}, there is only one maximum in the spectrum. 
Experimental techniques such as scanning tunneling microscopy (STM), point-contact spectroscopy, or tunneling measurements could directly probe this quasiparticle density of states and test the predicted multigap structure.

\section{Summary and Conclusions}

In summary, {we have investigated the electronic structure, lattice dynamics, electron–phonon interaction, and superconducting properties of the kagome-lattice compound Re$_2$Zr using first-principles methods. Flat electronic bands are identified near the Fermi level, and a van Hove singularity in the density of states is found to be induced by spin–orbit coupling.}
The Fermi surface of Re$_2$Zr consists of seven sheets built mainly from Re-5d and Zr-4d orbitals, five of which provide substantial contributions to the density of states.
The kagome lattice formed by Re-2 atoms in the $ab$ plane gives rise to quasi-two-dimensional Fermi surface sheets, although hybridization with Zr-4d states leads to a mixed orbital character.

The phonon spectrum extends up to 7 THz, and heavy Re-atom vibrations contribute across almost the entire frequency range except for a narrow band near 6 THz.
Despite the significant mass difference between Re and Zr, the phonon spectrum does not separate into distinct low- and high-frequency parts.
The highest-frequency modes correspond to characteristic breathing vibrations of Re atoms within the kagome lattice.
Contributions to the electron–phonon coupling constant $\lambda$ are distributed relatively uniformly across the phonon spectrum, resulting in a high logarithmic average phonon frequency $\omega_{\rm ln} = 163$~K that enhances $T_c$.
The calculated electron–phonon coupling constant $\lambda \simeq 0.88$ (0.87 from the integrated Eliashberg function and 0.89 from Fermi-surface averaging) places Re$_2$Zr in the moderate-coupling regime.
Within the isotropic Allen–Dynes formalism, $T_c = 8.9$ K is obtained for $\mu^* = 0.10$, while an enhanced value $\mu^* = 0.145$ is required to reproduce the experimental $T_c = 6.65$~K.

Superconducting gap calculations within density functional theory for superconductors, including explicit electron–electron interaction, yield $T_c = 7.7$ K and predict an anisotropic, multiband, and fully gapped superconducting state. The distribution of gap values corresponds to $2\Delta/k_BT_c$ ratios spanning approximately 3.0–4.5 across different Fermi surface sheets.
When spin fluctuations are included as additional depairing interactions, $T_c$ is reduced to 6.4 K, in excellent agreement with experiment. These results demonstrate that the superconducting critical temperature and gap structure of Re$_2$Zr can be quantitatively described within a conventional phonon-mediated framework, despite its kagome lattice and proposed unconventional features.

While the superconducting energy scale in Re$_2$Zr, reflected in both the agreement of the calculated $T_c$ and the electronic specific heat coefficient $\gamma$ with experiment, is well reproduced within a phonon-mediated singlet framework, the experimentally suggested time-reversal symmetry breaking \cite{Mandal2025} is not captured within the present singlet, time-reversal-symmetric approach and remains an open issue requiring further theoretical and experimental investigation.

\section*{Acknowledgments}
This work was supported by the program "Excellence Initiative - Research University" for the AGH University of Krakow.
The authors gratefully acknowledge the Polish high-performance computing infrastructure PLGrid (HPC Center: ACK Cyfronet AGH) for providing computer facilities and support within computational grant no. PLG/2025/018497.

\bibliography{refs-arxiv}

\begin{thebibliography}{65}%
\makeatletter
\providecommand \@ifxundefined [1]{%
 \@ifx{#1\undefined}
}%
\providecommand \@ifnum [1]{%
 \ifnum #1\expandafter \@firstoftwo
 \else \expandafter \@secondoftwo
 \fi
}%
\providecommand \@ifx [1]{%
 \ifx #1\expandafter \@firstoftwo
 \else \expandafter \@secondoftwo
 \fi
}%
\providecommand \natexlab [1]{#1}%
\providecommand \enquote  [1]{``#1''}%
\providecommand \bibnamefont  [1]{#1}%
\providecommand \bibfnamefont [1]{#1}%
\providecommand \citenamefont [1]{#1}%
\providecommand \href@noop [0]{\@secondoftwo}%
\providecommand \href [0]{\begingroup \@sanitize@url \@href}%
\providecommand \@href[1]{\@@startlink{#1}\@@href}%
\providecommand \@@href[1]{\endgroup#1\@@endlink}%
\providecommand \@sanitize@url [0]{\catcode `\\12\catcode `\$12\catcode `\&12\catcode `\#12\catcode `\^12\catcode `\_12\catcode `\%12\relax}%
\providecommand \@@startlink[1]{}%
\providecommand \@@endlink[0]{}%
\providecommand \url  [0]{\begingroup\@sanitize@url \@url }%
\providecommand \@url [1]{\endgroup\@href {#1}{\urlprefix }}%
\providecommand \urlprefix  [0]{URL }%
\providecommand \Eprint [0]{\href }%
\providecommand \doibase [0]{http://dx.doi.org/}%
\providecommand \selectlanguage [0]{\@gobble}%
\providecommand \bibinfo  [0]{\@secondoftwo}%
\providecommand \bibfield  [0]{\@secondoftwo}%
\providecommand \translation [1]{[#1]}%
\providecommand \BibitemOpen [0]{}%
\providecommand \bibitemStop [0]{}%
\providecommand \bibitemNoStop [0]{.\EOS\space}%
\providecommand \EOS [0]{\spacefactor3000\relax}%
\providecommand \BibitemShut  [1]{\csname bibitem#1\endcsname}%
\let\auto@bib@innerbib\@empty
\bibitem [{\citenamefont {Mandal}\ \emph {et~al.}(2025)\citenamefont {Mandal}, \citenamefont {Kataria}, \citenamefont {Meena}, \citenamefont {Kushwaha}, \citenamefont {Singh}, \citenamefont {Biswas}, \citenamefont {Stewart}, \citenamefont {Hillier},\ and\ \citenamefont {Singh}}]{Mandal2025}%
  \BibitemOpen
  \bibfield  {author} {\bibinfo {author} {\bibfnamefont {Manasi}\ \bibnamefont {Mandal}}, \bibinfo {author} {\bibfnamefont {A.}~\bibnamefont {Kataria}}, \bibinfo {author} {\bibfnamefont {P.~K.}\ \bibnamefont {Meena}}, \bibinfo {author} {\bibfnamefont {R.~K.}\ \bibnamefont {Kushwaha}}, \bibinfo {author} {\bibfnamefont {D.}~\bibnamefont {Singh}}, \bibinfo {author} {\bibfnamefont {P.~K.}\ \bibnamefont {Biswas}}, \bibinfo {author} {\bibfnamefont {R.}~\bibnamefont {Stewart}}, \bibinfo {author} {\bibfnamefont {A.~D.}\ \bibnamefont {Hillier}}, \ and\ \bibinfo {author} {\bibfnamefont {R.~P.}\ \bibnamefont {Singh}},\ }\bibfield  {title} {\enquote {\bibinfo {title} {{Time-reversal symmetry breaking in a Re-based kagome lattice superconductor}},}\ }\href {\doibase 10.1103/PhysRevB.111.054511} {\bibfield  {journal} {\bibinfo  {journal} {Phys. Rev. B}\ }\textbf {\bibinfo {volume} {111}},\ \bibinfo {pages} {054511} (\bibinfo {year} {2025})}\BibitemShut {NoStop}%
\bibitem [{\citenamefont {Yu}\ \emph {et~al.}(2024)\citenamefont {Yu}, \citenamefont {Liu}, \citenamefont {Li}, \citenamefont {Chen}, \citenamefont {Wang}, \citenamefont {Hao}, \citenamefont {Yang}, \citenamefont {Wang}, \citenamefont {Gong}, \citenamefont {Chen}, \citenamefont {Xie}, \citenamefont {Zhou}, \citenamefont {Ren}, \citenamefont {Chen},\ and\ \citenamefont {Jin}}]{re2zr-2}%
  \BibitemOpen
  \bibfield  {author} {\bibinfo {author} {\bibfnamefont {Yingpeng}\ \bibnamefont {Yu}}, \bibinfo {author} {\bibfnamefont {Zhaolong}\ \bibnamefont {Liu}}, \bibinfo {author} {\bibfnamefont {Qi}~\bibnamefont {Li}}, \bibinfo {author} {\bibfnamefont {Zhaoxu}\ \bibnamefont {Chen}}, \bibinfo {author} {\bibfnamefont {Yulong}\ \bibnamefont {Wang}}, \bibinfo {author} {\bibfnamefont {Munan}\ \bibnamefont {Hao}}, \bibinfo {author} {\bibfnamefont {Yaling}\ \bibnamefont {Yang}}, \bibinfo {author} {\bibfnamefont {Xuhui}\ \bibnamefont {Wang}}, \bibinfo {author} {\bibfnamefont {Chunsheng}\ \bibnamefont {Gong}}, \bibinfo {author} {\bibfnamefont {Long}\ \bibnamefont {Chen}}, \bibinfo {author} {\bibfnamefont {Zhenkai}\ \bibnamefont {Xie}}, \bibinfo {author} {\bibfnamefont {Kaiyao}\ \bibnamefont {Zhou}}, \bibinfo {author} {\bibfnamefont {Huifen}\ \bibnamefont {Ren}}, \bibinfo {author} {\bibfnamefont {Xu}~\bibnamefont {Chen}}, \ and\ \bibinfo {author} {\bibfnamefont {Shifeng}\ \bibnamefont {Jin}},\ }\bibfield  {title} {\enquote
  {\bibinfo {title} {{Weakly coupled Type-II superconductivity in a breathing Kagome metal ZrRe2}},}\ }\href {\doibase https://doi.org/10.1016/j.physc.2024.1354604} {\bibfield  {journal} {\bibinfo  {journal} {Physica C: Superconductivity and its Applications}\ }\textbf {\bibinfo {volume} {626}},\ \bibinfo {pages} {1354604} (\bibinfo {year} {2024})}\BibitemShut {NoStop}%
\bibitem [{\citenamefont {Giorgi}\ and\ \citenamefont {Szklarz}(1970)}]{Giorgi1970}%
  \BibitemOpen
  \bibfield  {author} {\bibinfo {author} {\bibfnamefont {A.L.}\ \bibnamefont {Giorgi}}\ and\ \bibinfo {author} {\bibfnamefont {E.G.}\ \bibnamefont {Szklarz}},\ }\bibfield  {title} {\enquote {\bibinfo {title} {{Superconductivity and lattice parameters of the dirhenides and ditechnides of thorium, hafnium and zirconium}},}\ }\href {\doibase https://doi.org/10.1016/0022-5088(70)90027-5} {\bibfield  {journal} {\bibinfo  {journal} {Journal of the Less Common Metals}\ }\textbf {\bibinfo {volume} {22}},\ \bibinfo {pages} {246--248} (\bibinfo {year} {1970})}\BibitemShut {NoStop}%
\bibitem [{\citenamefont {Roberts}(1976)}]{Roberts1976}%
  \BibitemOpen
  \bibfield  {author} {\bibinfo {author} {\bibfnamefont {B.~W.}\ \bibnamefont {Roberts}},\ }\bibfield  {title} {\enquote {\bibinfo {title} {{Survey of superconductive materials and critical evaluation of selected properties}},}\ }\href {\doibase 10.1063/1.555540} {\bibfield  {journal} {\bibinfo  {journal} {Journal of Physical and Chemical Reference Data}\ }\textbf {\bibinfo {volume} {5}},\ \bibinfo {pages} {581--822} (\bibinfo {year} {1976})},\ \Eprint {http://arxiv.org/abs/https://pubs.aip.org/aip/jpr/article-pdf/5/3/581/12172727/581\_1\_online.pdf} {https://pubs.aip.org/aip/jpr/article-pdf/5/3/581/12172727/581\_1\_online.pdf} \BibitemShut {NoStop}%
\bibitem [{\citenamefont {Mandal}\ \emph {et~al.}(2022)\citenamefont {Mandal}, \citenamefont {Kataria}, \citenamefont {Patra}, \citenamefont {Singh}, \citenamefont {Biswas}, \citenamefont {Hillier}, \citenamefont {Das},\ and\ \citenamefont {Singh}}]{Re2Hf}%
  \BibitemOpen
  \bibfield  {author} {\bibinfo {author} {\bibfnamefont {Manasi}\ \bibnamefont {Mandal}}, \bibinfo {author} {\bibfnamefont {Anshu}\ \bibnamefont {Kataria}}, \bibinfo {author} {\bibfnamefont {Chandan}\ \bibnamefont {Patra}}, \bibinfo {author} {\bibfnamefont {D.}~\bibnamefont {Singh}}, \bibinfo {author} {\bibfnamefont {P.~K.}\ \bibnamefont {Biswas}}, \bibinfo {author} {\bibfnamefont {A.~D.}\ \bibnamefont {Hillier}}, \bibinfo {author} {\bibfnamefont {Tanmoy}\ \bibnamefont {Das}}, \ and\ \bibinfo {author} {\bibfnamefont {R.~P.}\ \bibnamefont {Singh}},\ }\bibfield  {title} {\enquote {\bibinfo {title} {{Time-reversal symmetry breaking in frustrated superconductor ${\mathrm{Re}}_{2}\mathrm{Hf}$}},}\ }\href {\doibase 10.1103/PhysRevB.105.094513} {\bibfield  {journal} {\bibinfo  {journal} {Phys. Rev. B}\ }\textbf {\bibinfo {volume} {105}},\ \bibinfo {pages} {094513} (\bibinfo {year} {2022})}\BibitemShut {NoStop}%
\bibitem [{\citenamefont {Kokalj}(1999)}]{Kokalj1999}%
  \BibitemOpen
  \bibfield  {author} {\bibinfo {author} {\bibfnamefont {Anton}\ \bibnamefont {Kokalj}},\ }\bibfield  {title} {\enquote {\bibinfo {title} {{XCrySDen—a new program for displaying crystalline structures and electron densities}},}\ }\href {\doibase https://doi.org/10.1016/S1093-3263(99)00028-5} {\bibfield  {journal} {\bibinfo  {journal} {Journal of Molecular Graphics and Modelling}\ }\textbf {\bibinfo {volume} {17}},\ \bibinfo {pages} {176--179} (\bibinfo {year} {1999})}\BibitemShut {NoStop}%
\bibitem [{\citenamefont {Jiang}\ \emph {et~al.}(2021)\citenamefont {Jiang}, \citenamefont {Yin}, \citenamefont {Denner}, \citenamefont {Shumiya}, \citenamefont {Ortiz}, \citenamefont {Xu}, \citenamefont {Guguchia}, \citenamefont {He}, \citenamefont {Hossain}, \citenamefont {Liu}, \citenamefont {Ruff}, \citenamefont {Kautzsch}, \citenamefont {Zhang}, \citenamefont {Chang}, \citenamefont {Belopolski}, \citenamefont {Zhang}, \citenamefont {Cochran}, \citenamefont {Multer}, \citenamefont {Litskevich}, \citenamefont {Cheng}, \citenamefont {Yang}, \citenamefont {Wang}, \citenamefont {Thomale}, \citenamefont {Neupert}, \citenamefont {Wilson},\ and\ \citenamefont {Hasan}}]{Jiang2021}%
  \BibitemOpen
  \bibfield  {author} {\bibinfo {author} {\bibfnamefont {Yu-Xiao}\ \bibnamefont {Jiang}}, \bibinfo {author} {\bibfnamefont {Jia-Xin}\ \bibnamefont {Yin}}, \bibinfo {author} {\bibfnamefont {M~Michael}\ \bibnamefont {Denner}}, \bibinfo {author} {\bibfnamefont {Nana}\ \bibnamefont {Shumiya}}, \bibinfo {author} {\bibfnamefont {Brenden~R}\ \bibnamefont {Ortiz}}, \bibinfo {author} {\bibfnamefont {Gang}\ \bibnamefont {Xu}}, \bibinfo {author} {\bibfnamefont {Zurab}\ \bibnamefont {Guguchia}}, \bibinfo {author} {\bibfnamefont {Junyi}\ \bibnamefont {He}}, \bibinfo {author} {\bibfnamefont {Md~Shafayat}\ \bibnamefont {Hossain}}, \bibinfo {author} {\bibfnamefont {Xiaoxiong}\ \bibnamefont {Liu}}, \bibinfo {author} {\bibfnamefont {Jacob}\ \bibnamefont {Ruff}}, \bibinfo {author} {\bibfnamefont {Linus}\ \bibnamefont {Kautzsch}}, \bibinfo {author} {\bibfnamefont {Songtian~S}\ \bibnamefont {Zhang}}, \bibinfo {author} {\bibfnamefont {Guoqing}\ \bibnamefont {Chang}}, \bibinfo {author} {\bibfnamefont {Ilya}\ \bibnamefont
  {Belopolski}}, \bibinfo {author} {\bibfnamefont {Qi}~\bibnamefont {Zhang}}, \bibinfo {author} {\bibfnamefont {Tyler~A}\ \bibnamefont {Cochran}}, \bibinfo {author} {\bibfnamefont {Daniel}\ \bibnamefont {Multer}}, \bibinfo {author} {\bibfnamefont {Maksim}\ \bibnamefont {Litskevich}}, \bibinfo {author} {\bibfnamefont {Zi-Jia}\ \bibnamefont {Cheng}}, \bibinfo {author} {\bibfnamefont {Xian~P}\ \bibnamefont {Yang}}, \bibinfo {author} {\bibfnamefont {Ziqiang}\ \bibnamefont {Wang}}, \bibinfo {author} {\bibfnamefont {Ronny}\ \bibnamefont {Thomale}}, \bibinfo {author} {\bibfnamefont {Titus}\ \bibnamefont {Neupert}}, \bibinfo {author} {\bibfnamefont {Stephen~D}\ \bibnamefont {Wilson}}, \ and\ \bibinfo {author} {\bibfnamefont {M~Zahid}\ \bibnamefont {Hasan}},\ }\bibfield  {title} {\enquote {\bibinfo {title} {Unconventional chiral charge order in kagome superconductor {KV$_3$Sb$_5$}},}\ }\href@noop {} {\bibfield  {journal} {\bibinfo  {journal} {Nature Materials}\ }\textbf {\bibinfo {volume} {20}},\ \bibinfo {pages}
  {1353--1357} (\bibinfo {year} {2021})}\BibitemShut {NoStop}%
\bibitem [{\citenamefont {Ortiz}\ \emph {et~al.}(2021)\citenamefont {Ortiz}, \citenamefont {Sarte}, \citenamefont {Kenney}, \citenamefont {Graf}, \citenamefont {Teicher}, \citenamefont {Seshadri},\ and\ \citenamefont {Wilson}}]{Ortiz2021}%
  \BibitemOpen
  \bibfield  {author} {\bibinfo {author} {\bibfnamefont {Brenden~R.}\ \bibnamefont {Ortiz}}, \bibinfo {author} {\bibfnamefont {Paul~M.}\ \bibnamefont {Sarte}}, \bibinfo {author} {\bibfnamefont {Eric~M.}\ \bibnamefont {Kenney}}, \bibinfo {author} {\bibfnamefont {Michael~J.}\ \bibnamefont {Graf}}, \bibinfo {author} {\bibfnamefont {Samuel M.~L.}\ \bibnamefont {Teicher}}, \bibinfo {author} {\bibfnamefont {Ram}\ \bibnamefont {Seshadri}}, \ and\ \bibinfo {author} {\bibfnamefont {Stephen~D.}\ \bibnamefont {Wilson}},\ }\bibfield  {title} {\enquote {\bibinfo {title} {{Superconductivity in the ${\mathbb{Z}}_{2}$ kagome metal ${\mathrm{KV}}_{3}{\mathrm{Sb}}_{5}$}},}\ }\href {\doibase 10.1103/PhysRevMaterials.5.034801} {\bibfield  {journal} {\bibinfo  {journal} {Phys. Rev. Mater.}\ }\textbf {\bibinfo {volume} {5}},\ \bibinfo {pages} {034801} (\bibinfo {year} {2021})}\BibitemShut {NoStop}%
\bibitem [{\citenamefont {Yu}\ \emph {et~al.}(2021)\citenamefont {Yu}, \citenamefont {Wu}, \citenamefont {Wang}, \citenamefont {Lei}, \citenamefont {Zhuo}, \citenamefont {Ying},\ and\ \citenamefont {Chen}}]{Yu2021}%
  \BibitemOpen
  \bibfield  {author} {\bibinfo {author} {\bibfnamefont {F.~H.}\ \bibnamefont {Yu}}, \bibinfo {author} {\bibfnamefont {T.}~\bibnamefont {Wu}}, \bibinfo {author} {\bibfnamefont {Z.~Y.}\ \bibnamefont {Wang}}, \bibinfo {author} {\bibfnamefont {B.}~\bibnamefont {Lei}}, \bibinfo {author} {\bibfnamefont {W.~Z.}\ \bibnamefont {Zhuo}}, \bibinfo {author} {\bibfnamefont {J.~J.}\ \bibnamefont {Ying}}, \ and\ \bibinfo {author} {\bibfnamefont {X.~H.}\ \bibnamefont {Chen}},\ }\bibfield  {title} {\enquote {\bibinfo {title} {{Concurrence of anomalous Hall effect and charge density wave in a superconducting topological kagome metal}},}\ }\href {\doibase 10.1103/PhysRevB.104.L041103} {\bibfield  {journal} {\bibinfo  {journal} {Phys. Rev. B}\ }\textbf {\bibinfo {volume} {104}},\ \bibinfo {pages} {L041103} (\bibinfo {year} {2021})}\BibitemShut {NoStop}%
\bibitem [{\citenamefont {Neupert}\ \emph {et~al.}(2022)\citenamefont {Neupert}, \citenamefont {Denner}, \citenamefont {Yin}, \citenamefont {Thomale},\ and\ \citenamefont {Hasan}}]{Neupert2022}%
  \BibitemOpen
  \bibfield  {author} {\bibinfo {author} {\bibfnamefont {Titus}\ \bibnamefont {Neupert}}, \bibinfo {author} {\bibfnamefont {M~Michael}\ \bibnamefont {Denner}}, \bibinfo {author} {\bibfnamefont {Jia-Xin}\ \bibnamefont {Yin}}, \bibinfo {author} {\bibfnamefont {Ronny}\ \bibnamefont {Thomale}}, \ and\ \bibinfo {author} {\bibfnamefont {M~Zahid}\ \bibnamefont {Hasan}},\ }\bibfield  {title} {\enquote {\bibinfo {title} {Charge order and superconductivity in kagome materials},}\ }\href@noop {} {\bibfield  {journal} {\bibinfo  {journal} {Nature Physics}\ }\textbf {\bibinfo {volume} {18}},\ \bibinfo {pages} {137--143} (\bibinfo {year} {2022})}\BibitemShut {NoStop}%
\bibitem [{\citenamefont {Bolens}\ and\ \citenamefont {Nagaosa}(2019)}]{Bolens2019}%
  \BibitemOpen
  \bibfield  {author} {\bibinfo {author} {\bibfnamefont {Adrien}\ \bibnamefont {Bolens}}\ and\ \bibinfo {author} {\bibfnamefont {Naoto}\ \bibnamefont {Nagaosa}},\ }\bibfield  {title} {\enquote {\bibinfo {title} {{Topological states on the breathing kagome lattice}},}\ }\href {\doibase 10.1103/PhysRevB.99.165141} {\bibfield  {journal} {\bibinfo  {journal} {Phys. Rev. B}\ }\textbf {\bibinfo {volume} {99}},\ \bibinfo {pages} {165141} (\bibinfo {year} {2019})}\BibitemShut {NoStop}%
\bibitem [{\citenamefont {Peng}\ \emph {et~al.}(2021)\citenamefont {Peng}, \citenamefont {Han}, \citenamefont {Pokharel}, \citenamefont {Shen}, \citenamefont {Li}, \citenamefont {Hashimoto}, \citenamefont {Lu}, \citenamefont {Ortiz}, \citenamefont {Luo}, \citenamefont {Li}, \citenamefont {Guo}, \citenamefont {Wang}, \citenamefont {Cui}, \citenamefont {Sun}, \citenamefont {Qiao}, \citenamefont {Wilson},\ and\ \citenamefont {He}}]{Peng2021}%
  \BibitemOpen
  \bibfield  {author} {\bibinfo {author} {\bibfnamefont {Shuting}\ \bibnamefont {Peng}}, \bibinfo {author} {\bibfnamefont {Yulei}\ \bibnamefont {Han}}, \bibinfo {author} {\bibfnamefont {Ganesh}\ \bibnamefont {Pokharel}}, \bibinfo {author} {\bibfnamefont {Jianchang}\ \bibnamefont {Shen}}, \bibinfo {author} {\bibfnamefont {Zeyu}\ \bibnamefont {Li}}, \bibinfo {author} {\bibfnamefont {Makoto}\ \bibnamefont {Hashimoto}}, \bibinfo {author} {\bibfnamefont {Donghui}\ \bibnamefont {Lu}}, \bibinfo {author} {\bibfnamefont {Brenden~R.}\ \bibnamefont {Ortiz}}, \bibinfo {author} {\bibfnamefont {Yang}\ \bibnamefont {Luo}}, \bibinfo {author} {\bibfnamefont {Houchen}\ \bibnamefont {Li}}, \bibinfo {author} {\bibfnamefont {Mingyao}\ \bibnamefont {Guo}}, \bibinfo {author} {\bibfnamefont {Bingqian}\ \bibnamefont {Wang}}, \bibinfo {author} {\bibfnamefont {Shengtao}\ \bibnamefont {Cui}}, \bibinfo {author} {\bibfnamefont {Zhe}\ \bibnamefont {Sun}}, \bibinfo {author} {\bibfnamefont {Zhenhua}\ \bibnamefont {Qiao}}, \bibinfo {author}
  {\bibfnamefont {Stephen~D.}\ \bibnamefont {Wilson}}, \ and\ \bibinfo {author} {\bibfnamefont {Junfeng}\ \bibnamefont {He}},\ }\bibfield  {title} {\enquote {\bibinfo {title} {{Realizing Kagome Band Structure in Two-Dimensional Kagome Surface States of $R{\mathrm{V}}_{6}{\mathrm{Sn}}_{6}$ ($R=\mathrm{Gd}$, Ho)}},}\ }\href {\doibase 10.1103/PhysRevLett.127.266401} {\bibfield  {journal} {\bibinfo  {journal} {Phys. Rev. Lett.}\ }\textbf {\bibinfo {volume} {127}},\ \bibinfo {pages} {266401} (\bibinfo {year} {2021})}\BibitemShut {NoStop}%
\bibitem [{\citenamefont {Yin}\ \emph {et~al.}(2022)\citenamefont {Yin}, \citenamefont {Lian},\ and\ \citenamefont {Hasan}}]{Yin2022}%
  \BibitemOpen
  \bibfield  {author} {\bibinfo {author} {\bibfnamefont {Jia-Xin}\ \bibnamefont {Yin}}, \bibinfo {author} {\bibfnamefont {Biao}\ \bibnamefont {Lian}}, \ and\ \bibinfo {author} {\bibfnamefont {M~Zahid}\ \bibnamefont {Hasan}},\ }\bibfield  {title} {\enquote {\bibinfo {title} {Topological kagome magnets and superconductors},}\ }\href@noop {} {\bibfield  {journal} {\bibinfo  {journal} {Nature}\ }\textbf {\bibinfo {volume} {612}},\ \bibinfo {pages} {647--657} (\bibinfo {year} {2022})}\BibitemShut {NoStop}%
\bibitem [{\citenamefont {Jiang}\ \emph {et~al.}(2022)\citenamefont {Jiang}, \citenamefont {Wu}, \citenamefont {Yin}, \citenamefont {Wang}, \citenamefont {Hasan}, \citenamefont {Wilson}, \citenamefont {Chen},\ and\ \citenamefont {Hu}}]{Jiang2022}%
  \BibitemOpen
  \bibfield  {author} {\bibinfo {author} {\bibfnamefont {Kun}\ \bibnamefont {Jiang}}, \bibinfo {author} {\bibfnamefont {Tao}\ \bibnamefont {Wu}}, \bibinfo {author} {\bibfnamefont {Jia-Xin}\ \bibnamefont {Yin}}, \bibinfo {author} {\bibfnamefont {Zhenyu}\ \bibnamefont {Wang}}, \bibinfo {author} {\bibfnamefont {M~Zahid}\ \bibnamefont {Hasan}}, \bibinfo {author} {\bibfnamefont {Stephen~D}\ \bibnamefont {Wilson}}, \bibinfo {author} {\bibfnamefont {Xianhui}\ \bibnamefont {Chen}}, \ and\ \bibinfo {author} {\bibfnamefont {Jiangping}\ \bibnamefont {Hu}},\ }\bibfield  {title} {\enquote {\bibinfo {title} {{Kagome superconductors AV$_3$Sb$_5$ (A = K, Rb, Cs)}},}\ }\href {\doibase 10.1093/nsr/nwac199} {\bibfield  {journal} {\bibinfo  {journal} {National Science Review}\ }\textbf {\bibinfo {volume} {10}},\ \bibinfo {pages} {nwac199} (\bibinfo {year} {2022})},\ \Eprint {http://arxiv.org/abs/https://academic.oup.com/nsr/article-pdf/10/2/nwac199/49561488/nwac199.pdf}
  {https://academic.oup.com/nsr/article-pdf/10/2/nwac199/49561488/nwac199.pdf} \BibitemShut {NoStop}%
\bibitem [{\citenamefont {Zhao}\ \emph {et~al.}(2021)\citenamefont {Zhao}, \citenamefont {Li}, \citenamefont {Ortiz}, \citenamefont {Teicher}, \citenamefont {Park}, \citenamefont {Ye}, \citenamefont {Wang}, \citenamefont {Balents}, \citenamefont {Wilson},\ and\ \citenamefont {Zeljkovic}}]{Zhao2021}%
  \BibitemOpen
  \bibfield  {author} {\bibinfo {author} {\bibfnamefont {H.}~\bibnamefont {Zhao}}, \bibinfo {author} {\bibfnamefont {H.}~\bibnamefont {Li}}, \bibinfo {author} {\bibfnamefont {B.~R.}\ \bibnamefont {Ortiz}}, \bibinfo {author} {\bibfnamefont {S.~M.}\ \bibnamefont {Teicher}}, \bibinfo {author} {\bibfnamefont {T.}~\bibnamefont {Park}}, \bibinfo {author} {\bibfnamefont {M.}~\bibnamefont {Ye}}, \bibinfo {author} {\bibfnamefont {Z.}~\bibnamefont {Wang}}, \bibinfo {author} {\bibfnamefont {L.}~\bibnamefont {Balents}}, \bibinfo {author} {\bibfnamefont {S.~D.}\ \bibnamefont {Wilson}}, \ and\ \bibinfo {author} {\bibfnamefont {I.}~\bibnamefont {Zeljkovic}},\ }\bibfield  {title} {\enquote {\bibinfo {title} {{Cascade of correlated electron states in the kagome superconductor CsV$_3$Sb$_5$}},}\ }\href@noop {} {\bibfield  {journal} {\bibinfo  {journal} {Nature (London)}\ }\textbf {\bibinfo {volume} {599}},\ \bibinfo {pages} {216--221} (\bibinfo {year} {2021})}\BibitemShut {NoStop}%
\bibitem [{\citenamefont {Guguchia}\ \emph {et~al.}(2023)\citenamefont {Guguchia}, \citenamefont {Mielke}, \citenamefont {Das}, \citenamefont {Gupta}, \citenamefont {Yin}, \citenamefont {Liu}, \citenamefont {Yin}, \citenamefont {Christensen}, \citenamefont {Tu}, \citenamefont {Gong}, \citenamefont {Shumiya}, \citenamefont {Hossain}, \citenamefont {Gamsakhurdashvili}, \citenamefont {Elender}, \citenamefont {Dai}, \citenamefont {Amato}, \citenamefont {Shi}, \citenamefont {Lei}, \citenamefont {Fernandes}, \citenamefont {Hasan}, \citenamefont {Luetkens},\ and\ \citenamefont {Khasanov}}]{Guguchia2023}%
  \BibitemOpen
  \bibfield  {author} {\bibinfo {author} {\bibfnamefont {Z}~\bibnamefont {Guguchia}}, \bibinfo {author} {\bibfnamefont {C}~\bibnamefont {Mielke}}, \bibinfo {author} {\bibfnamefont {D}~\bibnamefont {Das}}, \bibinfo {author} {\bibfnamefont {R}~\bibnamefont {Gupta}}, \bibinfo {author} {\bibfnamefont {J-X}\ \bibnamefont {Yin}}, \bibinfo {author} {\bibfnamefont {H}~\bibnamefont {Liu}}, \bibinfo {author} {\bibfnamefont {Q}~\bibnamefont {Yin}}, \bibinfo {author} {\bibfnamefont {M~H}\ \bibnamefont {Christensen}}, \bibinfo {author} {\bibfnamefont {Z}~\bibnamefont {Tu}}, \bibinfo {author} {\bibfnamefont {C}~\bibnamefont {Gong}}, \bibinfo {author} {\bibfnamefont {N}~\bibnamefont {Shumiya}}, \bibinfo {author} {\bibfnamefont {Md~Shafayat}\ \bibnamefont {Hossain}}, \bibinfo {author} {\bibfnamefont {Ts}~\bibnamefont {Gamsakhurdashvili}}, \bibinfo {author} {\bibfnamefont {M}~\bibnamefont {Elender}}, \bibinfo {author} {\bibfnamefont {Pengcheng}\ \bibnamefont {Dai}}, \bibinfo {author} {\bibfnamefont {A}~\bibnamefont {Amato}},
  \bibinfo {author} {\bibfnamefont {Y}~\bibnamefont {Shi}}, \bibinfo {author} {\bibfnamefont {H~C}\ \bibnamefont {Lei}}, \bibinfo {author} {\bibfnamefont {R~M}\ \bibnamefont {Fernandes}}, \bibinfo {author} {\bibfnamefont {M~Z}\ \bibnamefont {Hasan}}, \bibinfo {author} {\bibfnamefont {H}~\bibnamefont {Luetkens}}, \ and\ \bibinfo {author} {\bibfnamefont {R}~\bibnamefont {Khasanov}},\ }\bibfield  {title} {\enquote {\bibinfo {title} {Tunable unconventional kagome superconductivity in charge ordered {RbV$_3$Sb$_5$} and {KV$_3$Sb$_5$}},}\ }\href@noop {} {\bibfield  {journal} {\bibinfo  {journal} {Nature Communications}\ }\textbf {\bibinfo {volume} {14}},\ \bibinfo {pages} {153} (\bibinfo {year} {2023})}\BibitemShut {NoStop}%
\bibitem [{\citenamefont {Chaudhary}\ \emph {et~al.}(2023)\citenamefont {Chaudhary}, \citenamefont {Shama}, \citenamefont {Singh}, \citenamefont {Consiglio}, \citenamefont {Di~Sante}, \citenamefont {Thomale},\ and\ \citenamefont {Singh}}]{Chaundhary2023}%
  \BibitemOpen
  \bibfield  {author} {\bibinfo {author} {\bibfnamefont {Savita}\ \bibnamefont {Chaudhary}}, \bibinfo {author} {\bibnamefont {Shama}}, \bibinfo {author} {\bibfnamefont {Jaskaran}\ \bibnamefont {Singh}}, \bibinfo {author} {\bibfnamefont {Armando}\ \bibnamefont {Consiglio}}, \bibinfo {author} {\bibfnamefont {Domenico}\ \bibnamefont {Di~Sante}}, \bibinfo {author} {\bibfnamefont {Ronny}\ \bibnamefont {Thomale}}, \ and\ \bibinfo {author} {\bibfnamefont {Yogesh}\ \bibnamefont {Singh}},\ }\bibfield  {title} {\enquote {\bibinfo {title} {{Role of electronic correlations in the kagome-lattice superconductor ${\mathrm{LaRh}}_{3}{\mathrm{B}}_{2}$}},}\ }\href {\doibase 10.1103/PhysRevB.107.085103} {\bibfield  {journal} {\bibinfo  {journal} {Phys. Rev. B}\ }\textbf {\bibinfo {volume} {107}},\ \bibinfo {pages} {085103} (\bibinfo {year} {2023})}\BibitemShut {NoStop}%
\bibitem [{\citenamefont {Gui}\ and\ \citenamefont {Cava}(2022)}]{Gui2022}%
  \BibitemOpen
  \bibfield  {author} {\bibinfo {author} {\bibfnamefont {Xin}\ \bibnamefont {Gui}}\ and\ \bibinfo {author} {\bibfnamefont {Robert~J.}\ \bibnamefont {Cava}},\ }\bibfield  {title} {\enquote {\bibinfo {title} {{LaIr$_3$Ga$_2$: A Superconductor Based on a Kagome Lattice of Ir}},}\ }\href {\doibase 10.1021/acs.chemmater.2c00280} {\bibfield  {journal} {\bibinfo  {journal} {Chemistry of Materials}\ }\textbf {\bibinfo {volume} {34}},\ \bibinfo {pages} {2824--2832} (\bibinfo {year} {2022})},\ \Eprint {http://arxiv.org/abs/https://doi.org/10.1021/acs.chemmater.2c00280} {https://doi.org/10.1021/acs.chemmater.2c00280} \BibitemShut {NoStop}%
\bibitem [{\citenamefont {Mielke}\ \emph {et~al.}(2021)\citenamefont {Mielke}, \citenamefont {Qin}, \citenamefont {Yin}, \citenamefont {Nakamura}, \citenamefont {Das}, \citenamefont {Guo}, \citenamefont {Khasanov}, \citenamefont {Chang}, \citenamefont {Wang}, \citenamefont {Jia}, \citenamefont {Nakatsuji}, \citenamefont {Amato}, \citenamefont {Luetkens}, \citenamefont {Xu}, \citenamefont {Hasan},\ and\ \citenamefont {Guguchia}}]{Mielke2021}%
  \BibitemOpen
  \bibfield  {author} {\bibinfo {author} {\bibfnamefont {C.}~\bibnamefont {Mielke}}, \bibinfo {author} {\bibfnamefont {Y.}~\bibnamefont {Qin}}, \bibinfo {author} {\bibfnamefont {J.-X.}\ \bibnamefont {Yin}}, \bibinfo {author} {\bibfnamefont {H.}~\bibnamefont {Nakamura}}, \bibinfo {author} {\bibfnamefont {D.}~\bibnamefont {Das}}, \bibinfo {author} {\bibfnamefont {K.}~\bibnamefont {Guo}}, \bibinfo {author} {\bibfnamefont {R.}~\bibnamefont {Khasanov}}, \bibinfo {author} {\bibfnamefont {J.}~\bibnamefont {Chang}}, \bibinfo {author} {\bibfnamefont {Z.~Q.}\ \bibnamefont {Wang}}, \bibinfo {author} {\bibfnamefont {S.}~\bibnamefont {Jia}}, \bibinfo {author} {\bibfnamefont {S.}~\bibnamefont {Nakatsuji}}, \bibinfo {author} {\bibfnamefont {A.}~\bibnamefont {Amato}}, \bibinfo {author} {\bibfnamefont {H.}~\bibnamefont {Luetkens}}, \bibinfo {author} {\bibfnamefont {G.}~\bibnamefont {Xu}}, \bibinfo {author} {\bibfnamefont {M.~Z.}\ \bibnamefont {Hasan}}, \ and\ \bibinfo {author} {\bibfnamefont {Z.}~\bibnamefont {Guguchia}},\
  }\bibfield  {title} {\enquote {\bibinfo {title} {{Nodeless kagome superconductivity in ${\mathrm{LaRu}}_{3}{\mathrm{Si}}_{2}$}},}\ }\href {\doibase 10.1103/PhysRevMaterials.5.034803} {\bibfield  {journal} {\bibinfo  {journal} {Phys. Rev. Mater.}\ }\textbf {\bibinfo {volume} {5}},\ \bibinfo {pages} {034803} (\bibinfo {year} {2021})}\BibitemShut {NoStop}%
\bibitem [{\citenamefont {Kang}\ \emph {et~al.}(2023)\citenamefont {Kang}, \citenamefont {Fang}, \citenamefont {Yoo}, \citenamefont {Ortiz}, \citenamefont {Oey}, \citenamefont {Choi}, \citenamefont {Ryu}, \citenamefont {Kim}, \citenamefont {Jozwiak}, \citenamefont {Bostwick}, \citenamefont {Rotenberg}, \citenamefont {Kaxiras}, \citenamefont {Checkelsky}, \citenamefont {Wilson}, \citenamefont {Park},\ and\ \citenamefont {Comin}}]{Kang2023}%
  \BibitemOpen
  \bibfield  {author} {\bibinfo {author} {\bibfnamefont {Mingu}\ \bibnamefont {Kang}}, \bibinfo {author} {\bibfnamefont {Shiang}\ \bibnamefont {Fang}}, \bibinfo {author} {\bibfnamefont {Jonggyu}\ \bibnamefont {Yoo}}, \bibinfo {author} {\bibfnamefont {Brenden~R}\ \bibnamefont {Ortiz}}, \bibinfo {author} {\bibfnamefont {Yuzki~M}\ \bibnamefont {Oey}}, \bibinfo {author} {\bibfnamefont {Jonghyeok}\ \bibnamefont {Choi}}, \bibinfo {author} {\bibfnamefont {Sae~Hee}\ \bibnamefont {Ryu}}, \bibinfo {author} {\bibfnamefont {Jimin}\ \bibnamefont {Kim}}, \bibinfo {author} {\bibfnamefont {Chris}\ \bibnamefont {Jozwiak}}, \bibinfo {author} {\bibfnamefont {Aaron}\ \bibnamefont {Bostwick}}, \bibinfo {author} {\bibfnamefont {Eli}\ \bibnamefont {Rotenberg}}, \bibinfo {author} {\bibfnamefont {Efthimios}\ \bibnamefont {Kaxiras}}, \bibinfo {author} {\bibfnamefont {Joseph~G}\ \bibnamefont {Checkelsky}}, \bibinfo {author} {\bibfnamefont {Stephen~D}\ \bibnamefont {Wilson}}, \bibinfo {author} {\bibfnamefont {Jae-Hoon}\ \bibnamefont
  {Park}}, \ and\ \bibinfo {author} {\bibfnamefont {Riccardo}\ \bibnamefont {Comin}},\ }\bibfield  {title} {\enquote {\bibinfo {title} {Charge order landscape and competition with superconductivity in kagome metals},}\ }\href@noop {} {\bibfield  {journal} {\bibinfo  {journal} {Nature Materials}\ }\textbf {\bibinfo {volume} {22}},\ \bibinfo {pages} {186--193} (\bibinfo {year} {2023})}\BibitemShut {NoStop}%
\bibitem [{\citenamefont {G{\'o}rnicka}\ \emph {et~al.}(2023)\citenamefont {G{\'o}rnicka}, \citenamefont {Winiarski}, \citenamefont {Walicka},\ and\ \citenamefont {Klimczuk}}]{Gornicka2023}%
  \BibitemOpen
  \bibfield  {author} {\bibinfo {author} {\bibfnamefont {Karolina}\ \bibnamefont {G{\'o}rnicka}}, \bibinfo {author} {\bibfnamefont {Micha{\l}~J}\ \bibnamefont {Winiarski}}, \bibinfo {author} {\bibfnamefont {Dorota~I}\ \bibnamefont {Walicka}}, \ and\ \bibinfo {author} {\bibfnamefont {Tomasz}\ \bibnamefont {Klimczuk}},\ }\bibfield  {title} {\enquote {\bibinfo {title} {{Superconductivity in a breathing kagome metals {ROs$_2$} (R = Sc, Y, Lu)}},}\ }\href@noop {} {\bibfield  {journal} {\bibinfo  {journal} {Scientific Reports}\ }\textbf {\bibinfo {volume} {13}},\ \bibinfo {pages} {16704} (\bibinfo {year} {2023})}\BibitemShut {NoStop}%
\bibitem [{\citenamefont {Garmroudi}\ \emph {et~al.}(2025)\citenamefont {Garmroudi}, \citenamefont {Coulter}, \citenamefont {Serhiienko}, \citenamefont {Di~Cataldo}, \citenamefont {Parzer}, \citenamefont {Riss}, \citenamefont {Grasser}, \citenamefont {Stockinger}, \citenamefont {Khmelevskyi}, \citenamefont {Pryga}, \citenamefont {Wiendlocha}, \citenamefont {Held}, \citenamefont {Mori}, \citenamefont {Bauer}, \citenamefont {Georges},\ and\ \citenamefont {Pustogow}}]{Garmroudi_ni3in}%
  \BibitemOpen
  \bibfield  {author} {\bibinfo {author} {\bibfnamefont {Fabian}\ \bibnamefont {Garmroudi}}, \bibinfo {author} {\bibfnamefont {Jennifer}\ \bibnamefont {Coulter}}, \bibinfo {author} {\bibfnamefont {Illia}\ \bibnamefont {Serhiienko}}, \bibinfo {author} {\bibfnamefont {Simone}\ \bibnamefont {Di~Cataldo}}, \bibinfo {author} {\bibfnamefont {Michael}\ \bibnamefont {Parzer}}, \bibinfo {author} {\bibfnamefont {Alexander}\ \bibnamefont {Riss}}, \bibinfo {author} {\bibfnamefont {Matthias}\ \bibnamefont {Grasser}}, \bibinfo {author} {\bibfnamefont {Simon}\ \bibnamefont {Stockinger}}, \bibinfo {author} {\bibfnamefont {Sergii}\ \bibnamefont {Khmelevskyi}}, \bibinfo {author} {\bibfnamefont {Kacper}\ \bibnamefont {Pryga}}, \bibinfo {author} {\bibfnamefont {Bartlomiej}\ \bibnamefont {Wiendlocha}}, \bibinfo {author} {\bibfnamefont {Karsten}\ \bibnamefont {Held}}, \bibinfo {author} {\bibfnamefont {Takao}\ \bibnamefont {Mori}}, \bibinfo {author} {\bibfnamefont {Ernst}\ \bibnamefont {Bauer}}, \bibinfo {author} {\bibfnamefont
  {Antoine}\ \bibnamefont {Georges}}, \ and\ \bibinfo {author} {\bibfnamefont {Andrej}\ \bibnamefont {Pustogow}},\ }\bibfield  {title} {\enquote {\bibinfo {title} {{Topological Flat-Band-Driven Metallic Thermoelectricity}},}\ }\href {\doibase 10.1103/PhysRevX.15.021054} {\bibfield  {journal} {\bibinfo  {journal} {Phys. Rev. X}\ }\textbf {\bibinfo {volume} {15}},\ \bibinfo {pages} {021054} (\bibinfo {year} {2025})}\BibitemShut {NoStop}%
\bibitem [{\citenamefont {Tan}\ \emph {et~al.}(2021)\citenamefont {Tan}, \citenamefont {Liu}, \citenamefont {Wang},\ and\ \citenamefont {Yan}}]{Tan2021}%
  \BibitemOpen
  \bibfield  {author} {\bibinfo {author} {\bibfnamefont {Hengxin}\ \bibnamefont {Tan}}, \bibinfo {author} {\bibfnamefont {Yizhou}\ \bibnamefont {Liu}}, \bibinfo {author} {\bibfnamefont {Ziqiang}\ \bibnamefont {Wang}}, \ and\ \bibinfo {author} {\bibfnamefont {Binghai}\ \bibnamefont {Yan}},\ }\bibfield  {title} {\enquote {\bibinfo {title} {{Charge Density Waves and Electronic Properties of Superconducting Kagome Metals}},}\ }\href {\doibase 10.1103/PhysRevLett.127.046401} {\bibfield  {journal} {\bibinfo  {journal} {Phys. Rev. Lett.}\ }\textbf {\bibinfo {volume} {127}},\ \bibinfo {pages} {046401} (\bibinfo {year} {2021})}\BibitemShut {NoStop}%
\bibitem [{\citenamefont {Watkins}\ \emph {et~al.}(2024)\citenamefont {Watkins}, \citenamefont {Johrendt}, \citenamefont {Vlcek}, \citenamefont {Wilson},\ and\ \citenamefont {Seshadri}}]{Watkins2024}%
  \BibitemOpen
  \bibfield  {author} {\bibinfo {author} {\bibfnamefont {Aurland~K.}\ \bibnamefont {Watkins}}, \bibinfo {author} {\bibfnamefont {Dirk}\ \bibnamefont {Johrendt}}, \bibinfo {author} {\bibfnamefont {Vojtech}\ \bibnamefont {Vlcek}}, \bibinfo {author} {\bibfnamefont {Stephen~D.}\ \bibnamefont {Wilson}}, \ and\ \bibinfo {author} {\bibfnamefont {Ram}\ \bibnamefont {Seshadri}},\ }\bibfield  {title} {\enquote {\bibinfo {title} {{Fidelity and variability in the interlayer electronic structure of the kagome superconductor ${\mathrm{CsV}}_{3}{\mathrm{Sb}}_{5}$}},}\ }\href {\doibase 10.1103/PhysRevMaterials.8.054204} {\bibfield  {journal} {\bibinfo  {journal} {Phys. Rev. Mater.}\ }\textbf {\bibinfo {volume} {8}},\ \bibinfo {pages} {054204} (\bibinfo {year} {2024})}\BibitemShut {NoStop}%
\bibitem [{\citenamefont {Yi}\ \emph {et~al.}(2022)\citenamefont {Yi}, \citenamefont {Ma}, \citenamefont {Zhang}, \citenamefont {Liao}, \citenamefont {You},\ and\ \citenamefont {Su}}]{Yi2022}%
  \BibitemOpen
  \bibfield  {author} {\bibinfo {author} {\bibfnamefont {Xin-Wei}\ \bibnamefont {Yi}}, \bibinfo {author} {\bibfnamefont {Xing-Yu}\ \bibnamefont {Ma}}, \bibinfo {author} {\bibfnamefont {Zhen}\ \bibnamefont {Zhang}}, \bibinfo {author} {\bibfnamefont {Zheng-Wei}\ \bibnamefont {Liao}}, \bibinfo {author} {\bibfnamefont {Jing-Yang}\ \bibnamefont {You}}, \ and\ \bibinfo {author} {\bibfnamefont {Gang}\ \bibnamefont {Su}},\ }\bibfield  {title} {\enquote {\bibinfo {title} {{Large kagome family candidates with topological superconductivity and charge density waves}},}\ }\href {\doibase 10.1103/PhysRevB.106.L220505} {\bibfield  {journal} {\bibinfo  {journal} {Phys. Rev. B}\ }\textbf {\bibinfo {volume} {106}},\ \bibinfo {pages} {L220505} (\bibinfo {year} {2022})}\BibitemShut {NoStop}%
\bibitem [{\citenamefont {N.}\ \emph {et~al.}(1988)\citenamefont {N.}, \citenamefont {U.},\ and\ \citenamefont {W.}}]{Oliveira1988}%
  \BibitemOpen
  \bibfield  {author} {\bibinfo {author} {\bibfnamefont {Oliveira~L.}\ \bibnamefont {N.}}, \bibinfo {author} {\bibfnamefont {Gross E.~K.}\ \bibnamefont {U.}}, \ and\ \bibinfo {author} {\bibfnamefont {Kohn}\ \bibnamefont {W.}},\ }\bibfield  {title} {\enquote {\bibinfo {title} {{Density-Functional Theory for Superconductors}},}\ }\href {\doibase 10.1103/PhysRevLett.60.2430} {\bibfield  {journal} {\bibinfo  {journal} {Phys. Rev. Lett.}\ }\textbf {\bibinfo {volume} {60}},\ \bibinfo {pages} {2430--2433} (\bibinfo {year} {1988})}\BibitemShut {NoStop}%
\bibitem [{\citenamefont {M.}\ \emph {et~al.}(2005)\citenamefont {M.}, \citenamefont {L.}, \citenamefont {N.}, \citenamefont {A.}, \citenamefont {G.}, \citenamefont {L.}, \citenamefont {A.}, \citenamefont {S.},\ and\ \citenamefont {U.}}]{Luders2005}%
  \BibitemOpen
  \bibfield  {author} {\bibinfo {author} {\bibfnamefont {L\"uders}\ \bibnamefont {M.}}, \bibinfo {author} {\bibfnamefont {Marques M.~A.}\ \bibnamefont {L.}}, \bibinfo {author} {\bibfnamefont {Lathiotakis~N.}\ \bibnamefont {N.}}, \bibinfo {author} {\bibfnamefont {Floris}\ \bibnamefont {A.}}, \bibinfo {author} {\bibfnamefont {Profeta}\ \bibnamefont {G.}}, \bibinfo {author} {\bibfnamefont {Fast}\ \bibnamefont {L.}}, \bibinfo {author} {\bibfnamefont {Continenza}\ \bibnamefont {A.}}, \bibinfo {author} {\bibfnamefont {Massidda}\ \bibnamefont {S.}}, \ and\ \bibinfo {author} {\bibfnamefont {Gross E.~K.}\ \bibnamefont {U.}},\ }\bibfield  {title} {\enquote {\bibinfo {title} {{Ab initio theory of superconductivity. I. Density functional formalism and approximate functionals}},}\ }\href {\doibase 10.1103/PhysRevB.72.024545} {\bibfield  {journal} {\bibinfo  {journal} {Phys. Rev. B}\ }\textbf {\bibinfo {volume} {72}},\ \bibinfo {pages} {024545} (\bibinfo {year} {2005})}\BibitemShut {NoStop}%
\bibitem [{\citenamefont {Kawamura}\ \emph {et~al.}(2017)\citenamefont {Kawamura}, \citenamefont {Akashi},\ and\ \citenamefont {Tsuneyuki}}]{Kawamura2017}%
  \BibitemOpen
  \bibfield  {author} {\bibinfo {author} {\bibfnamefont {Mitsuaki}\ \bibnamefont {Kawamura}}, \bibinfo {author} {\bibfnamefont {Ryosuke}\ \bibnamefont {Akashi}}, \ and\ \bibinfo {author} {\bibfnamefont {Shinji}\ \bibnamefont {Tsuneyuki}},\ }\bibfield  {title} {\enquote {\bibinfo {title} {{Anisotropic superconducting gaps in ${\mathrm{YNi}}_{2}{\mathrm{B}}_{2}\mathrm{C}$: A first-principles investigation}},}\ }\href {\doibase 10.1103/PhysRevB.95.054506} {\bibfield  {journal} {\bibinfo  {journal} {Phys. Rev. B}\ }\textbf {\bibinfo {volume} {95}},\ \bibinfo {pages} {054506} (\bibinfo {year} {2017})}\BibitemShut {NoStop}%
\bibitem [{\citenamefont {Kawamura}\ \emph {et~al.}(2020)\citenamefont {Kawamura}, \citenamefont {Hizume},\ and\ \citenamefont {Ozaki}}]{Kawamura2020}%
  \BibitemOpen
  \bibfield  {author} {\bibinfo {author} {\bibfnamefont {Mitsuaki}\ \bibnamefont {Kawamura}}, \bibinfo {author} {\bibfnamefont {Yuma}\ \bibnamefont {Hizume}}, \ and\ \bibinfo {author} {\bibfnamefont {Taisuke}\ \bibnamefont {Ozaki}},\ }\bibfield  {title} {\enquote {\bibinfo {title} {{Benchmark of density functional theory for superconductors in elemental materials}},}\ }\href {\doibase 10.1103/PhysRevB.101.134511} {\bibfield  {journal} {\bibinfo  {journal} {Phys. Rev. B}\ }\textbf {\bibinfo {volume} {101}},\ \bibinfo {pages} {134511} (\bibinfo {year} {2020})}\BibitemShut {NoStop}%
\bibitem [{\citenamefont {Barker}\ \emph {et~al.}(2015)\citenamefont {Barker}, \citenamefont {Singh}, \citenamefont {Thamizhavel}, \citenamefont {Hillier}, \citenamefont {Lees}, \citenamefont {Balakrishnan}, \citenamefont {Paul},\ and\ \citenamefont {Singh}}]{Barker2015}%
  \BibitemOpen
  \bibfield  {author} {\bibinfo {author} {\bibfnamefont {J.~A.~T.}\ \bibnamefont {Barker}}, \bibinfo {author} {\bibfnamefont {D.}~\bibnamefont {Singh}}, \bibinfo {author} {\bibfnamefont {A.}~\bibnamefont {Thamizhavel}}, \bibinfo {author} {\bibfnamefont {A.~D.}\ \bibnamefont {Hillier}}, \bibinfo {author} {\bibfnamefont {M.~R.}\ \bibnamefont {Lees}}, \bibinfo {author} {\bibfnamefont {G.}~\bibnamefont {Balakrishnan}}, \bibinfo {author} {\bibfnamefont {D.~McK.}\ \bibnamefont {Paul}}, \ and\ \bibinfo {author} {\bibfnamefont {R.~P.}\ \bibnamefont {Singh}},\ }\bibfield  {title} {\enquote {\bibinfo {title} {{Unconventional Superconductivity in ${\mathrm{La}}_{7}{\mathrm{Ir}}_{3}$ Revealed by Muon Spin Relaxation: Introducing a New Family of Noncentrosymmetric Superconductor That Breaks Time-Reversal Symmetry}},}\ }\href {\doibase 10.1103/PhysRevLett.115.267001} {\bibfield  {journal} {\bibinfo  {journal} {Phys. Rev. Lett.}\ }\textbf {\bibinfo {volume} {115}},\ \bibinfo {pages} {267001} (\bibinfo {year}
  {2015})}\BibitemShut {NoStop}%
\bibitem [{\citenamefont {Mayoh}\ \emph {et~al.}(2021)\citenamefont {Mayoh}, \citenamefont {Hillier}, \citenamefont {Balakrishnan},\ and\ \citenamefont {Lees}}]{Mayoh2021}%
  \BibitemOpen
  \bibfield  {author} {\bibinfo {author} {\bibfnamefont {D.~A.}\ \bibnamefont {Mayoh}}, \bibinfo {author} {\bibfnamefont {A.~D.}\ \bibnamefont {Hillier}}, \bibinfo {author} {\bibfnamefont {G.}~\bibnamefont {Balakrishnan}}, \ and\ \bibinfo {author} {\bibfnamefont {M.~R.}\ \bibnamefont {Lees}},\ }\bibfield  {title} {\enquote {\bibinfo {title} {{Evidence for the coexistence of time-reversal symmetry breaking and Bardeen-Cooper-Schrieffer-like superconductivity in ${\mathrm{La}}_{7}{\mathrm{Pd}}_{3}$}},}\ }\href {\doibase 10.1103/PhysRevB.103.024507} {\bibfield  {journal} {\bibinfo  {journal} {Phys. Rev. B}\ }\textbf {\bibinfo {volume} {103}},\ \bibinfo {pages} {024507} (\bibinfo {year} {2021})}\BibitemShut {NoStop}%
\bibitem [{\citenamefont {Singh}\ \emph {et~al.}(2020)\citenamefont {Singh}, \citenamefont {Scheurer}, \citenamefont {Hillier}, \citenamefont {Adroja},\ and\ \citenamefont {Singh}}]{Singh2020}%
  \BibitemOpen
  \bibfield  {author} {\bibinfo {author} {\bibfnamefont {D.}~\bibnamefont {Singh}}, \bibinfo {author} {\bibfnamefont {M.~S.}\ \bibnamefont {Scheurer}}, \bibinfo {author} {\bibfnamefont {A.~D.}\ \bibnamefont {Hillier}}, \bibinfo {author} {\bibfnamefont {D.~T.}\ \bibnamefont {Adroja}}, \ and\ \bibinfo {author} {\bibfnamefont {R.~P.}\ \bibnamefont {Singh}},\ }\bibfield  {title} {\enquote {\bibinfo {title} {{Time-reversal-symmetry breaking and unconventional pairing in the noncentrosymmetric superconductor ${\mathrm{La}}_{7}{\mathrm{Rh}}_{3}$}},}\ }\href {\doibase 10.1103/PhysRevB.102.134511} {\bibfield  {journal} {\bibinfo  {journal} {Phys. Rev. B}\ }\textbf {\bibinfo {volume} {102}},\ \bibinfo {pages} {134511} (\bibinfo {year} {2020})}\BibitemShut {NoStop}%
\bibitem [{\citenamefont {Shu}\ \emph {et~al.}(2011)\citenamefont {Shu}, \citenamefont {Higemoto}, \citenamefont {Aoki}, \citenamefont {Hillier}, \citenamefont {Ohishi}, \citenamefont {Ishida}, \citenamefont {Kadono}, \citenamefont {Koda}, \citenamefont {Bernal}, \citenamefont {MacLaughlin}, \citenamefont {Tunashima}, \citenamefont {Yonezawa}, \citenamefont {Sanada}, \citenamefont {Kikuchi}, \citenamefont {Sato}, \citenamefont {Sugawara}, \citenamefont {Ito},\ and\ \citenamefont {Maple}}]{Shu2021}%
  \BibitemOpen
  \bibfield  {author} {\bibinfo {author} {\bibfnamefont {Lei}\ \bibnamefont {Shu}}, \bibinfo {author} {\bibfnamefont {W.}~\bibnamefont {Higemoto}}, \bibinfo {author} {\bibfnamefont {Y.}~\bibnamefont {Aoki}}, \bibinfo {author} {\bibfnamefont {A.~D.}\ \bibnamefont {Hillier}}, \bibinfo {author} {\bibfnamefont {K.}~\bibnamefont {Ohishi}}, \bibinfo {author} {\bibfnamefont {K.}~\bibnamefont {Ishida}}, \bibinfo {author} {\bibfnamefont {R.}~\bibnamefont {Kadono}}, \bibinfo {author} {\bibfnamefont {A.}~\bibnamefont {Koda}}, \bibinfo {author} {\bibfnamefont {O.~O.}\ \bibnamefont {Bernal}}, \bibinfo {author} {\bibfnamefont {D.~E.}\ \bibnamefont {MacLaughlin}}, \bibinfo {author} {\bibfnamefont {Y.}~\bibnamefont {Tunashima}}, \bibinfo {author} {\bibfnamefont {Y.}~\bibnamefont {Yonezawa}}, \bibinfo {author} {\bibfnamefont {S.}~\bibnamefont {Sanada}}, \bibinfo {author} {\bibfnamefont {D.}~\bibnamefont {Kikuchi}}, \bibinfo {author} {\bibfnamefont {H.}~\bibnamefont {Sato}}, \bibinfo {author} {\bibfnamefont {H.}~\bibnamefont
  {Sugawara}}, \bibinfo {author} {\bibfnamefont {T.~U.}\ \bibnamefont {Ito}}, \ and\ \bibinfo {author} {\bibfnamefont {M.~B.}\ \bibnamefont {Maple}},\ }\bibfield  {title} {\enquote {\bibinfo {title} {{Suppression of time-reversal symmetry breaking superconductivity in Pr(Os${}_{1\ensuremath{-}x}$Ru${}_{x}$)${}_{4}$Sb${}_{12}$ and Pr${}_{1\ensuremath{-}y}$La${}_{y}$Os${}_{4}$Sb${}_{12}$}},}\ }\href {\doibase 10.1103/PhysRevB.83.100504} {\bibfield  {journal} {\bibinfo  {journal} {Phys. Rev. B}\ }\textbf {\bibinfo {volume} {83}},\ \bibinfo {pages} {100504} (\bibinfo {year} {2011})}\BibitemShut {NoStop}%
\bibitem [{\citenamefont {Zhang}\ \emph {et~al.}(2019)\citenamefont {Zhang}, \citenamefont {Ding}, \citenamefont {Huang}, \citenamefont {Tan}, \citenamefont {Hillier}, \citenamefont {Biswas}, \citenamefont {MacLaughlin},\ and\ \citenamefont {Shu}}]{Zhang2019}%
  \BibitemOpen
  \bibfield  {author} {\bibinfo {author} {\bibfnamefont {J.}~\bibnamefont {Zhang}}, \bibinfo {author} {\bibfnamefont {Z.~F.}\ \bibnamefont {Ding}}, \bibinfo {author} {\bibfnamefont {K.}~\bibnamefont {Huang}}, \bibinfo {author} {\bibfnamefont {C.}~\bibnamefont {Tan}}, \bibinfo {author} {\bibfnamefont {A.~D.}\ \bibnamefont {Hillier}}, \bibinfo {author} {\bibfnamefont {P.~K.}\ \bibnamefont {Biswas}}, \bibinfo {author} {\bibfnamefont {D.~E.}\ \bibnamefont {MacLaughlin}}, \ and\ \bibinfo {author} {\bibfnamefont {L.}~\bibnamefont {Shu}},\ }\bibfield  {title} {\enquote {\bibinfo {title} {{Broken time-reversal symmetry in superconducting ${\mathrm{Pr}}_{1\ensuremath{-}x}{\mathrm{La}}_{x}{\mathrm{Pt}}_{4}{\mathrm{Ge}}_{12}$}},}\ }\href {\doibase 10.1103/PhysRevB.100.024508} {\bibfield  {journal} {\bibinfo  {journal} {Phys. Rev. B}\ }\textbf {\bibinfo {volume} {100}},\ \bibinfo {pages} {024508} (\bibinfo {year} {2019})}\BibitemShut {NoStop}%
\bibitem [{\citenamefont {Kataria}\ \emph {et~al.}(2023)\citenamefont {Kataria}, \citenamefont {Verezhak}, \citenamefont {Prakash}, \citenamefont {Kushwaha}, \citenamefont {Thamizhavel}, \citenamefont {Ramakrishnan}, \citenamefont {Scheurer}, \citenamefont {Hillier},\ and\ \citenamefont {Singh}}]{Kataria2023}%
  \BibitemOpen
  \bibfield  {author} {\bibinfo {author} {\bibfnamefont {A.}~\bibnamefont {Kataria}}, \bibinfo {author} {\bibfnamefont {J.~A.~T.}\ \bibnamefont {Verezhak}}, \bibinfo {author} {\bibfnamefont {O.}~\bibnamefont {Prakash}}, \bibinfo {author} {\bibfnamefont {R.~K.}\ \bibnamefont {Kushwaha}}, \bibinfo {author} {\bibfnamefont {A.}~\bibnamefont {Thamizhavel}}, \bibinfo {author} {\bibfnamefont {S.}~\bibnamefont {Ramakrishnan}}, \bibinfo {author} {\bibfnamefont {M.~S.}\ \bibnamefont {Scheurer}}, \bibinfo {author} {\bibfnamefont {A.~D.}\ \bibnamefont {Hillier}}, \ and\ \bibinfo {author} {\bibfnamefont {R.~P.}\ \bibnamefont {Singh}},\ }\bibfield  {title} {\enquote {\bibinfo {title} {{Time-reversal symmetry breaking in the superconducting low carrier density quasiskutterudite ${\mathrm{Lu}}_{3}{\mathrm{Os}}_{4}{\mathrm{Ge}}_{13}$}},}\ }\href {\doibase 10.1103/PhysRevB.107.L100506} {\bibfield  {journal} {\bibinfo  {journal} {Phys. Rev. B}\ }\textbf {\bibinfo {volume} {107}},\ \bibinfo {pages} {L100506} (\bibinfo {year}
  {2023})}\BibitemShut {NoStop}%
\bibitem [{\citenamefont {Giannozzi}\ \emph {et~al.}(2009)\citenamefont {Giannozzi}, \citenamefont {Baroni}, \citenamefont {Bonini}, \citenamefont {Calandra}, \citenamefont {Car}, \citenamefont {Cavazzoni}, \citenamefont {Ceresoli}, \citenamefont {Chiarotti}, \citenamefont {Cococcioni}, \citenamefont {Dabo}, \citenamefont {Corso}, \citenamefont {de~Gironcoli}, \citenamefont {Fabris}, \citenamefont {Fratesi}, \citenamefont {Gebauer}, \citenamefont {Gerstmann}, \citenamefont {Gougoussis}, \citenamefont {Kokalj}, \citenamefont {Lazzeri}, \citenamefont {Martin-Samos}, \citenamefont {Marzari}, \citenamefont {Mauri}, \citenamefont {Mazzarello}, \citenamefont {Paolini}, \citenamefont {Pasquarello}, \citenamefont {Paulatto}, \citenamefont {Sbraccia}, \citenamefont {Scandolo}, \citenamefont {Sclauzero}, \citenamefont {Seitsonen}, \citenamefont {Smogunov}, \citenamefont {Umari},\ and\ \citenamefont {Wentzcovitch}}]{Giannozzi2009}%
  \BibitemOpen
  \bibfield  {author} {\bibinfo {author} {\bibfnamefont {Paolo}\ \bibnamefont {Giannozzi}}, \bibinfo {author} {\bibfnamefont {Stefano}\ \bibnamefont {Baroni}}, \bibinfo {author} {\bibfnamefont {Nicola}\ \bibnamefont {Bonini}}, \bibinfo {author} {\bibfnamefont {Matteo}\ \bibnamefont {Calandra}}, \bibinfo {author} {\bibfnamefont {Roberto}\ \bibnamefont {Car}}, \bibinfo {author} {\bibfnamefont {Carlo}\ \bibnamefont {Cavazzoni}}, \bibinfo {author} {\bibfnamefont {Davide}\ \bibnamefont {Ceresoli}}, \bibinfo {author} {\bibfnamefont {Guido~L}\ \bibnamefont {Chiarotti}}, \bibinfo {author} {\bibfnamefont {Matteo}\ \bibnamefont {Cococcioni}}, \bibinfo {author} {\bibfnamefont {Ismaila}\ \bibnamefont {Dabo}}, \bibinfo {author} {\bibfnamefont {Andrea~Dal}\ \bibnamefont {Corso}}, \bibinfo {author} {\bibfnamefont {Stefano}\ \bibnamefont {de~Gironcoli}}, \bibinfo {author} {\bibfnamefont {Stefano}\ \bibnamefont {Fabris}}, \bibinfo {author} {\bibfnamefont {Guido}\ \bibnamefont {Fratesi}}, \bibinfo {author} {\bibfnamefont
  {Ralph}\ \bibnamefont {Gebauer}}, \bibinfo {author} {\bibfnamefont {Uwe}\ \bibnamefont {Gerstmann}}, \bibinfo {author} {\bibfnamefont {Christos}\ \bibnamefont {Gougoussis}}, \bibinfo {author} {\bibfnamefont {Anton}\ \bibnamefont {Kokalj}}, \bibinfo {author} {\bibfnamefont {Michele}\ \bibnamefont {Lazzeri}}, \bibinfo {author} {\bibfnamefont {Layla}\ \bibnamefont {Martin-Samos}}, \bibinfo {author} {\bibfnamefont {Nicola}\ \bibnamefont {Marzari}}, \bibinfo {author} {\bibfnamefont {Francesco}\ \bibnamefont {Mauri}}, \bibinfo {author} {\bibfnamefont {Riccardo}\ \bibnamefont {Mazzarello}}, \bibinfo {author} {\bibfnamefont {Stefano}\ \bibnamefont {Paolini}}, \bibinfo {author} {\bibfnamefont {Alfredo}\ \bibnamefont {Pasquarello}}, \bibinfo {author} {\bibfnamefont {Lorenzo}\ \bibnamefont {Paulatto}}, \bibinfo {author} {\bibfnamefont {Carlo}\ \bibnamefont {Sbraccia}}, \bibinfo {author} {\bibfnamefont {Sandro}\ \bibnamefont {Scandolo}}, \bibinfo {author} {\bibfnamefont {Gabriele}\ \bibnamefont {Sclauzero}}, \bibinfo
  {author} {\bibfnamefont {Ari~P}\ \bibnamefont {Seitsonen}}, \bibinfo {author} {\bibfnamefont {Alexander}\ \bibnamefont {Smogunov}}, \bibinfo {author} {\bibfnamefont {Paolo}\ \bibnamefont {Umari}}, \ and\ \bibinfo {author} {\bibfnamefont {Renata~M}\ \bibnamefont {Wentzcovitch}},\ }\bibfield  {title} {\enquote {\bibinfo {title} {{{QUANTUM} {ESPRESSO}: a modular and open-source software project for quantum simulations of materials}},}\ }\href {\doibase 10.1088/0953-8984/21/39/395502} {\bibfield  {journal} {\bibinfo  {journal} {Journal of Physics: Condensed Matter}\ }\textbf {\bibinfo {volume} {21}},\ \bibinfo {pages} {395502} (\bibinfo {year} {2009})}\BibitemShut {NoStop}%
\bibitem [{\citenamefont {Giannozzi}\ \emph {et~al.}(2017)\citenamefont {Giannozzi}, \citenamefont {Andreussi}, \citenamefont {Brumme}, \citenamefont {Bunau}, \citenamefont {Nardelli}, \citenamefont {Calandra}, \citenamefont {Car}, \citenamefont {Cavazzoni}, \citenamefont {Ceresoli}, \citenamefont {Cococcioni}, \citenamefont {Colonna}, \citenamefont {Carnimeo}, \citenamefont {Corso}, \citenamefont {de~Gironcoli}, \citenamefont {Delugas}, \citenamefont {Jr}, \citenamefont {Ferretti}, \citenamefont {Floris}, \citenamefont {Fratesi}, \citenamefont {Fugallo}, \citenamefont {Gebauer}, \citenamefont {Gerstmann}, \citenamefont {Giustino}, \citenamefont {Gorni}, \citenamefont {Jia}, \citenamefont {Kawamura}, \citenamefont {Ko}, \citenamefont {Kokalj}, \citenamefont {Kucukbenli}, \citenamefont {Lazzeri}, \citenamefont {Marsili}, \citenamefont {Marzari}, \citenamefont {Mauri}, \citenamefont {Nguyen}, \citenamefont {Nguyen}, \citenamefont {de-la Roza}, \citenamefont {Paulatto}, \citenamefont {Ponce}, \citenamefont
  {Rocca}, \citenamefont {Sabatini}, \citenamefont {Santra}, \citenamefont {Schlipf}, \citenamefont {Seitsonen}, \citenamefont {Smogunov}, \citenamefont {Timrov}, \citenamefont {Thonhauser}, \citenamefont {Umari}, \citenamefont {Vast}, \citenamefont {Wu},\ and\ \citenamefont {Baroni}}]{Giannozzi2017}%
  \BibitemOpen
  \bibfield  {author} {\bibinfo {author} {\bibfnamefont {P}~\bibnamefont {Giannozzi}}, \bibinfo {author} {\bibfnamefont {O}~\bibnamefont {Andreussi}}, \bibinfo {author} {\bibfnamefont {T}~\bibnamefont {Brumme}}, \bibinfo {author} {\bibfnamefont {O}~\bibnamefont {Bunau}}, \bibinfo {author} {\bibfnamefont {M~Buongiorno}\ \bibnamefont {Nardelli}}, \bibinfo {author} {\bibfnamefont {M}~\bibnamefont {Calandra}}, \bibinfo {author} {\bibfnamefont {R}~\bibnamefont {Car}}, \bibinfo {author} {\bibfnamefont {C}~\bibnamefont {Cavazzoni}}, \bibinfo {author} {\bibfnamefont {D}~\bibnamefont {Ceresoli}}, \bibinfo {author} {\bibfnamefont {M}~\bibnamefont {Cococcioni}}, \bibinfo {author} {\bibfnamefont {N}~\bibnamefont {Colonna}}, \bibinfo {author} {\bibfnamefont {I}~\bibnamefont {Carnimeo}}, \bibinfo {author} {\bibfnamefont {A~Dal}\ \bibnamefont {Corso}}, \bibinfo {author} {\bibfnamefont {S}~\bibnamefont {de~Gironcoli}}, \bibinfo {author} {\bibfnamefont {P}~\bibnamefont {Delugas}}, \bibinfo {author} {\bibfnamefont
  {R~A~DiStasio}\ \bibnamefont {Jr}}, \bibinfo {author} {\bibfnamefont {A}~\bibnamefont {Ferretti}}, \bibinfo {author} {\bibfnamefont {A}~\bibnamefont {Floris}}, \bibinfo {author} {\bibfnamefont {G}~\bibnamefont {Fratesi}}, \bibinfo {author} {\bibfnamefont {G}~\bibnamefont {Fugallo}}, \bibinfo {author} {\bibfnamefont {R}~\bibnamefont {Gebauer}}, \bibinfo {author} {\bibfnamefont {U}~\bibnamefont {Gerstmann}}, \bibinfo {author} {\bibfnamefont {F}~\bibnamefont {Giustino}}, \bibinfo {author} {\bibfnamefont {T}~\bibnamefont {Gorni}}, \bibinfo {author} {\bibfnamefont {J}~\bibnamefont {Jia}}, \bibinfo {author} {\bibfnamefont {M}~\bibnamefont {Kawamura}}, \bibinfo {author} {\bibfnamefont {H-Y}\ \bibnamefont {Ko}}, \bibinfo {author} {\bibfnamefont {A}~\bibnamefont {Kokalj}}, \bibinfo {author} {\bibfnamefont {E}~\bibnamefont {Kucukbenli}}, \bibinfo {author} {\bibfnamefont {M}~\bibnamefont {Lazzeri}}, \bibinfo {author} {\bibfnamefont {M}~\bibnamefont {Marsili}}, \bibinfo {author} {\bibfnamefont {N}~\bibnamefont
  {Marzari}}, \bibinfo {author} {\bibfnamefont {F}~\bibnamefont {Mauri}}, \bibinfo {author} {\bibfnamefont {N~L}\ \bibnamefont {Nguyen}}, \bibinfo {author} {\bibfnamefont {H-V}\ \bibnamefont {Nguyen}}, \bibinfo {author} {\bibfnamefont {A~Otero}\ \bibnamefont {de-la Roza}}, \bibinfo {author} {\bibfnamefont {L}~\bibnamefont {Paulatto}}, \bibinfo {author} {\bibfnamefont {S}~\bibnamefont {Ponce}}, \bibinfo {author} {\bibfnamefont {D}~\bibnamefont {Rocca}}, \bibinfo {author} {\bibfnamefont {R}~\bibnamefont {Sabatini}}, \bibinfo {author} {\bibfnamefont {B}~\bibnamefont {Santra}}, \bibinfo {author} {\bibfnamefont {M}~\bibnamefont {Schlipf}}, \bibinfo {author} {\bibfnamefont {A~P}\ \bibnamefont {Seitsonen}}, \bibinfo {author} {\bibfnamefont {A}~\bibnamefont {Smogunov}}, \bibinfo {author} {\bibfnamefont {I}~\bibnamefont {Timrov}}, \bibinfo {author} {\bibfnamefont {T}~\bibnamefont {Thonhauser}}, \bibinfo {author} {\bibfnamefont {P}~\bibnamefont {Umari}}, \bibinfo {author} {\bibfnamefont {N}~\bibnamefont {Vast}},
  \bibinfo {author} {\bibfnamefont {X}~\bibnamefont {Wu}}, \ and\ \bibinfo {author} {\bibfnamefont {S}~\bibnamefont {Baroni}},\ }\bibfield  {title} {\enquote {\bibinfo {title} {{Advanced capabilities for materials modelling with QUANTUM ESPRESSO}},}\ }\href {\doibase 10.1088/1361-648X/aa8f79} {\bibfield  {journal} {\bibinfo  {journal} {Journal of Physics: Condensed Matter}\ }\textbf {\bibinfo {volume} {29}},\ \bibinfo {pages} {465901} (\bibinfo {year} {2017})}\BibitemShut {NoStop}%
\bibitem [{\citenamefont {P.}\ \emph {et~al.}(1996)\citenamefont {P.}, \citenamefont {Kieron},\ and\ \citenamefont {Matthias}}]{Perdew1996}%
  \BibitemOpen
  \bibfield  {author} {\bibinfo {author} {\bibfnamefont {Perdew~John}\ \bibnamefont {P.}}, \bibinfo {author} {\bibfnamefont {Burke}\ \bibnamefont {Kieron}}, \ and\ \bibinfo {author} {\bibfnamefont {Ernzerhof}\ \bibnamefont {Matthias}},\ }\bibfield  {title} {\enquote {\bibinfo {title} {{Generalized Gradient Approximation Made Simple}},}\ }\href {\doibase 10.1103/PhysRevLett.77.3865} {\bibfield  {journal} {\bibinfo  {journal} {Phys. Rev. Lett.}\ }\textbf {\bibinfo {volume} {77}},\ \bibinfo {pages} {3865--3868} (\bibinfo {year} {1996})}\BibitemShut {NoStop}%
\bibitem [{\citenamefont {{Dal Corso}}(2014)}]{DalCorso2014}%
  \BibitemOpen
  \bibfield  {author} {\bibinfo {author} {\bibfnamefont {Andrea}\ \bibnamefont {{Dal Corso}}},\ }\bibfield  {title} {\enquote {\bibinfo {title} {{Pseudopotentials periodic table: From H to Pu}},}\ }\href {\doibase https://doi.org/10.1016/j.commatsci.2014.07.043} {\bibfield  {journal} {\bibinfo  {journal} {Computational Materials Science}\ }\textbf {\bibinfo {volume} {95}},\ \bibinfo {pages} {337--350} (\bibinfo {year} {2014})}\BibitemShut {NoStop}%
\bibitem [{psn()}]{psnames}%
  \BibitemOpen
  \href@noop {} {}\bibinfo {note} {Pseudopotential files: Re.pbe-spn-kjpaw\_psl.1.0.0.UPF, Re.rel-pbe-spn-kjpaw\_psl.1.0.0.UPF, Zr.pbe-spn-kjpaw\_psl.1.0.0.UPF, Zr.rel-be-spn-kjpaw\_psl.1.0.0.UPF}\BibitemShut {NoStop}%
\bibitem [{\citenamefont {Baroni}\ \emph {et~al.}(2001)\citenamefont {Baroni}, \citenamefont {de~Gironcoli}, \citenamefont {Dal~Corso},\ and\ \citenamefont {Giannozzi}}]{dfpt}%
  \BibitemOpen
  \bibfield  {author} {\bibinfo {author} {\bibfnamefont {Stefano}\ \bibnamefont {Baroni}}, \bibinfo {author} {\bibfnamefont {Stefano}\ \bibnamefont {de~Gironcoli}}, \bibinfo {author} {\bibfnamefont {Andrea}\ \bibnamefont {Dal~Corso}}, \ and\ \bibinfo {author} {\bibfnamefont {Paolo}\ \bibnamefont {Giannozzi}},\ }\bibfield  {title} {\enquote {\bibinfo {title} {{Phonons and related crystal properties from density-functional perturbation theory}},}\ }\href {\doibase 10.1103/RevModPhys.73.515} {\bibfield  {journal} {\bibinfo  {journal} {Rev. Mod. Phys.}\ }\textbf {\bibinfo {volume} {73}},\ \bibinfo {pages} {515--562} (\bibinfo {year} {2001})}\BibitemShut {NoStop}%
\bibitem [{sup()}]{suppl}%
  \BibitemOpen
  \href@noop {} {}\bibinfo {note} {{See Supplemental Material for Figure S1 with the Brillouin zone and high-symmetry points; Figure S2 with electronic band structure on a longer path and in a wide energy range; Figure S3 with the electronic densities of states calculated without SOC; Figure S4 with the orbital character of the Fermi surface; Figure S5 with the orbital character of the band structure; Figure S6 with the summed distribution of the superconducting gap; Figure S7 with Fermi surface cross sections for the first three bands; Figure S8 with the scalar-relativistic Fermi surface; Figure S9 with phonon grid convergence test; Figure S10 with DOS at different hydrostatic pressures; Table I with the relaxed atomic positions (with and without SOC). Table II with atom displacements in a relaxed structure at 10 GPa. Animations of atomic displacements for selected phonon modes. }}\BibitemShut {NoStop}%
\bibitem [{\citenamefont {Stewart}(2011)}]{iron_sc}%
  \BibitemOpen
  \bibfield  {author} {\bibinfo {author} {\bibfnamefont {G.~R.}\ \bibnamefont {Stewart}},\ }\bibfield  {title} {\enquote {\bibinfo {title} {{Superconductivity in iron compounds}},}\ }\href {\doibase 10.1103/RevModPhys.83.1589} {\bibfield  {journal} {\bibinfo  {journal} {Rev. Mod. Phys.}\ }\textbf {\bibinfo {volume} {83}},\ \bibinfo {pages} {1589--1652} (\bibinfo {year} {2011})}\BibitemShut {NoStop}%
\bibitem [{\citenamefont {Carrington}(2011)}]{Carrington_2011}%
  \BibitemOpen
  \bibfield  {author} {\bibinfo {author} {\bibfnamefont {A}~\bibnamefont {Carrington}},\ }\bibfield  {title} {\enquote {\bibinfo {title} {{Quantum oscillation studies of the Fermi surface of iron-pnictide superconductors}},}\ }\href {\doibase 10.1088/0034-4885/74/12/124507} {\bibfield  {journal} {\bibinfo  {journal} {Reports on Progress in Physics}\ }\textbf {\bibinfo {volume} {74}},\ \bibinfo {pages} {124507} (\bibinfo {year} {2011})}\BibitemShut {NoStop}%
\bibitem [{\citenamefont {Kawamura}(2019)}]{Kawamura2019}%
  \BibitemOpen
  \bibfield  {author} {\bibinfo {author} {\bibfnamefont {Mitsuaki}\ \bibnamefont {Kawamura}},\ }\bibfield  {title} {\enquote {\bibinfo {title} {{FermiSurfer: Fermi-surface viewer providing multiple representation schemes}},}\ }\href {\doibase https://doi.org/10.1016/j.cpc.2019.01.017} {\bibfield  {journal} {\bibinfo  {journal} {Computer Physics Communications}\ }\textbf {\bibinfo {volume} {239}},\ \bibinfo {pages} {197--203} (\bibinfo {year} {2019})}\BibitemShut {NoStop}%
\bibitem [{\citenamefont {Gutowska}\ \emph {et~al.}(2021)\citenamefont {Gutowska}, \citenamefont {G\'ornicka}, \citenamefont {W\'ojcik}, \citenamefont {Klimczuk},\ and\ \citenamefont {Wiendlocha}}]{gutowska2021}%
  \BibitemOpen
  \bibfield  {author} {\bibinfo {author} {\bibfnamefont {Sylwia}\ \bibnamefont {Gutowska}}, \bibinfo {author} {\bibfnamefont {Karolina}\ \bibnamefont {G\'ornicka}}, \bibinfo {author} {\bibfnamefont {Pawe\l{}}\ \bibnamefont {W\'ojcik}}, \bibinfo {author} {\bibfnamefont {Tomasz}\ \bibnamefont {Klimczuk}}, \ and\ \bibinfo {author} {\bibfnamefont {Bartlomiej}\ \bibnamefont {Wiendlocha}},\ }\bibfield  {title} {\enquote {\bibinfo {title} {{Strong-coupling superconductivity of ${\mathrm{SrIr}}_{2}$ and ${\mathrm{SrRh}}_{2}$: Phonon engineering of metallic Ir and Rh}},}\ }\href {\doibase 10.1103/PhysRevB.104.054505} {\bibfield  {journal} {\bibinfo  {journal} {Phys. Rev. B}\ }\textbf {\bibinfo {volume} {104}},\ \bibinfo {pages} {054505} (\bibinfo {year} {2021})}\BibitemShut {NoStop}%
\bibitem [{\citenamefont {Grimvall}(1981)}]{Grimvall1981}%
  \BibitemOpen
  \bibfield  {author} {\bibinfo {author} {\bibfnamefont {G.}~\bibnamefont {Grimvall}},\ }\href@noop {} {\emph {\bibinfo {title} {The electron-phonon interaction in metals}}}\ (\bibinfo  {publisher} {North-Holland, Amsterdam},\ \bibinfo {year} {1981})\BibitemShut {NoStop}%
\bibitem [{\citenamefont {Wierzbowska}\ \emph {et~al.}(2005)\citenamefont {Wierzbowska}, \citenamefont {de~Gironcoli},\ and\ \citenamefont {Giannozzi}}]{Wierzbowska2005}%
  \BibitemOpen
  \bibfield  {author} {\bibinfo {author} {\bibfnamefont {Malgorzata}\ \bibnamefont {Wierzbowska}}, \bibinfo {author} {\bibfnamefont {Stefano}\ \bibnamefont {de~Gironcoli}}, \ and\ \bibinfo {author} {\bibfnamefont {Paolo}\ \bibnamefont {Giannozzi}},\ }\bibfield  {title} {\enquote {\bibinfo {title} {{Origins of low-and high-pressure discontinuities of {T}$_c$ in niobium}},}\ }\href@noop {} {\bibfield  {journal} {\bibinfo  {journal} {arXiv preprint cond-mat/0504077}\ } (\bibinfo {year} {2005})}\BibitemShut {NoStop}%
\bibitem [{\citenamefont {Heid}\ \emph {et~al.}(2010)\citenamefont {Heid}, \citenamefont {Bohnen}, \citenamefont {Sklyadneva},\ and\ \citenamefont {Chulkov}}]{heid-pb}%
  \BibitemOpen
  \bibfield  {author} {\bibinfo {author} {\bibfnamefont {R.}~\bibnamefont {Heid}}, \bibinfo {author} {\bibfnamefont {K.P.}\ \bibnamefont {Bohnen}}, \bibinfo {author} {\bibfnamefont {I.~Yu.}\ \bibnamefont {Sklyadneva}}, \ and\ \bibinfo {author} {\bibfnamefont {E.~V.}\ \bibnamefont {Chulkov}},\ }\bibfield  {title} {\enquote {\bibinfo {title} {{Effect of spin-orbit coupling on the electron-phonon interaction of the superconductors {P}b and {T}l}},}\ }\href {\doibase 10.1103/PhysRevB.81.174527} {\bibfield  {journal} {\bibinfo  {journal} {Phys. Rev. B}\ }\textbf {\bibinfo {volume} {81}},\ \bibinfo {pages} {174527} (\bibinfo {year} {2010})}\BibitemShut {NoStop}%
\bibitem [{\citenamefont {B.}\ and\ \citenamefont {C.}(1975)}]{Allen1975}%
  \BibitemOpen
  \bibfield  {author} {\bibinfo {author} {\bibfnamefont {Allen~P.}\ \bibnamefont {B.}}\ and\ \bibinfo {author} {\bibfnamefont {Dynes~R.}\ \bibnamefont {C.}},\ }\bibfield  {title} {\enquote {\bibinfo {title} {{Transition temperature of strong-coupled superconductors reanalyzed}},}\ }\href {\doibase 10.1103/PhysRevB.12.905} {\bibfield  {journal} {\bibinfo  {journal} {Phys. Rev. B}\ }\textbf {\bibinfo {volume} {12}},\ \bibinfo {pages} {905--922} (\bibinfo {year} {1975})}\BibitemShut {NoStop}%
\bibitem [{\citenamefont {Tsutsumi}\ \emph {et~al.}(2020)\citenamefont {Tsutsumi}, \citenamefont {Hizume}, \citenamefont {Kawamura}, \citenamefont {Akashi},\ and\ \citenamefont {Tsuneyuki}}]{Tsutsumi2020}%
  \BibitemOpen
  \bibfield  {author} {\bibinfo {author} {\bibfnamefont {Kentaro}\ \bibnamefont {Tsutsumi}}, \bibinfo {author} {\bibfnamefont {Yuma}\ \bibnamefont {Hizume}}, \bibinfo {author} {\bibfnamefont {Mitsuaki}\ \bibnamefont {Kawamura}}, \bibinfo {author} {\bibfnamefont {Ryosuke}\ \bibnamefont {Akashi}}, \ and\ \bibinfo {author} {\bibfnamefont {Shinji}\ \bibnamefont {Tsuneyuki}},\ }\bibfield  {title} {\enquote {\bibinfo {title} {{Effect of spin fluctuations on superconductivity in V and Nb: A first-principles study}},}\ }\href {\doibase 10.1103/PhysRevB.102.214515} {\bibfield  {journal} {\bibinfo  {journal} {Phys. Rev. B}\ }\textbf {\bibinfo {volume} {102}},\ \bibinfo {pages} {214515} (\bibinfo {year} {2020})}\BibitemShut {NoStop}%
\bibitem [{\citenamefont {Marzari}\ \emph {et~al.}(1999)\citenamefont {Marzari}, \citenamefont {Vanderbilt}, \citenamefont {De~Vita},\ and\ \citenamefont {Payne}}]{Marzari1999}%
  \BibitemOpen
  \bibfield  {author} {\bibinfo {author} {\bibfnamefont {Nicola}\ \bibnamefont {Marzari}}, \bibinfo {author} {\bibfnamefont {David}\ \bibnamefont {Vanderbilt}}, \bibinfo {author} {\bibfnamefont {Alessandro}\ \bibnamefont {De~Vita}}, \ and\ \bibinfo {author} {\bibfnamefont {M.~C.}\ \bibnamefont {Payne}},\ }\bibfield  {title} {\enquote {\bibinfo {title} {{Thermal Contraction and Disordering of the Al(110) Surface}},}\ }\href {\doibase 10.1103/PhysRevLett.82.3296} {\bibfield  {journal} {\bibinfo  {journal} {Phys. Rev. Lett.}\ }\textbf {\bibinfo {volume} {82}},\ \bibinfo {pages} {3296--3299} (\bibinfo {year} {1999})}\BibitemShut {NoStop}%
\bibitem [{\citenamefont {dos Santos}\ and\ \citenamefont {Marzari}(2023)}]{Marzari2023}%
  \BibitemOpen
  \bibfield  {author} {\bibinfo {author} {\bibfnamefont {Flaviano~Jos\'e}\ \bibnamefont {dos Santos}}\ and\ \bibinfo {author} {\bibfnamefont {Nicola}\ \bibnamefont {Marzari}},\ }\bibfield  {title} {\enquote {\bibinfo {title} {{Fermi energy determination for advanced smearing techniques}},}\ }\href {\doibase 10.1103/PhysRevB.107.195122} {\bibfield  {journal} {\bibinfo  {journal} {Phys. Rev. B}\ }\textbf {\bibinfo {volume} {107}},\ \bibinfo {pages} {195122} (\bibinfo {year} {2023})}\BibitemShut {NoStop}%
\bibitem [{\citenamefont {Gutowska}\ \emph {et~al.}(2024)\citenamefont {Gutowska}, \citenamefont {G\'ornicka}, \citenamefont {W\'ojcik}, \citenamefont {Klimczuk},\ and\ \citenamefont {Wiendlocha}}]{Gutowska2024}%
  \BibitemOpen
  \bibfield  {author} {\bibinfo {author} {\bibfnamefont {Sylwia}\ \bibnamefont {Gutowska}}, \bibinfo {author} {\bibfnamefont {Karolina}\ \bibnamefont {G\'ornicka}}, \bibinfo {author} {\bibfnamefont {Pawe\l{}}\ \bibnamefont {W\'ojcik}}, \bibinfo {author} {\bibfnamefont {Tomasz}\ \bibnamefont {Klimczuk}}, \ and\ \bibinfo {author} {\bibfnamefont {Bartlomiej}\ \bibnamefont {Wiendlocha}},\ }\bibfield  {title} {\enquote {\bibinfo {title} {{Anisotropic, multiband, and strong-coupling superconductivity of the ${\mathrm{Pb}}_{0.64}{\mathrm{Bi}}_{0.36}$ alloy}},}\ }\href {\doibase 10.1103/PhysRevB.110.214510} {\bibfield  {journal} {\bibinfo  {journal} {Phys. Rev. B}\ }\textbf {\bibinfo {volume} {110}},\ \bibinfo {pages} {214510} (\bibinfo {year} {2024})}\BibitemShut {NoStop}%
\bibitem [{\citenamefont {Kuderowicz}\ and\ \citenamefont {Wiendlocha}(2023)}]{Kuderowicz2023}%
  \BibitemOpen
  \bibfield  {author} {\bibinfo {author} {\bibfnamefont {Gabriel}\ \bibnamefont {Kuderowicz}}\ and\ \bibinfo {author} {\bibfnamefont {Bartlomiej}\ \bibnamefont {Wiendlocha}},\ }\bibfield  {title} {\enquote {\bibinfo {title} {{Strong-coupling superconductivity of the Heusler-type compound ${\mathrm{ScAu}}_{2}\mathrm{Al}$: Ab initio studies}},}\ }\href {\doibase 10.1103/PhysRevB.108.224501} {\bibfield  {journal} {\bibinfo  {journal} {Phys. Rev. B}\ }\textbf {\bibinfo {volume} {108}},\ \bibinfo {pages} {224501} (\bibinfo {year} {2023})}\BibitemShut {NoStop}%
\bibitem [{\citenamefont {Marques}\ \emph {et~al.}(2005)\citenamefont {Marques}, \citenamefont {L\"uders}, \citenamefont {Lathiotakis}, \citenamefont {Profeta}, \citenamefont {Floris}, \citenamefont {Fast}, \citenamefont {Continenza}, \citenamefont {Gross},\ and\ \citenamefont {Massidda}}]{scdft-metals}%
  \BibitemOpen
  \bibfield  {author} {\bibinfo {author} {\bibfnamefont {M.~A.~L.}\ \bibnamefont {Marques}}, \bibinfo {author} {\bibfnamefont {M.}~\bibnamefont {L\"uders}}, \bibinfo {author} {\bibfnamefont {N.~N.}\ \bibnamefont {Lathiotakis}}, \bibinfo {author} {\bibfnamefont {G.}~\bibnamefont {Profeta}}, \bibinfo {author} {\bibfnamefont {A.}~\bibnamefont {Floris}}, \bibinfo {author} {\bibfnamefont {L.}~\bibnamefont {Fast}}, \bibinfo {author} {\bibfnamefont {A.}~\bibnamefont {Continenza}}, \bibinfo {author} {\bibfnamefont {E.~K.~U.}\ \bibnamefont {Gross}}, \ and\ \bibinfo {author} {\bibfnamefont {S.}~\bibnamefont {Massidda}},\ }\bibfield  {title} {\enquote {\bibinfo {title} {{Ab initio theory of superconductivity. II. Application to elemental metals}},}\ }\href {\doibase 10.1103/PhysRevB.72.024546} {\bibfield  {journal} {\bibinfo  {journal} {Phys. Rev. B}\ }\textbf {\bibinfo {volume} {72}},\ \bibinfo {pages} {024546} (\bibinfo {year} {2005})}\BibitemShut {NoStop}%
\bibitem [{\citenamefont {Bogoljubov}(1958)}]{Bogoljubov1958}%
  \BibitemOpen
  \bibfield  {author} {\bibinfo {author} {\bibfnamefont {N.~N.}\ \bibnamefont {Bogoljubov}},\ }\bibfield  {title} {\enquote {\bibinfo {title} {{On a new method in the theory of superconductivity}},}\ }\href {\doibase 10.1007/BF02745585} {\bibfield  {journal} {\bibinfo  {journal} {Il Nuovo Cimento}\ }\textbf {\bibinfo {volume} {7}} (\bibinfo {year} {1958}),\ 10.1007/BF02745585}\BibitemShut {NoStop}%
\bibitem [{\citenamefont {Morel}\ and\ \citenamefont {Anderson}(1962)}]{Morel1962}%
  \BibitemOpen
  \bibfield  {author} {\bibinfo {author} {\bibfnamefont {P.}~\bibnamefont {Morel}}\ and\ \bibinfo {author} {\bibfnamefont {P.~W.}\ \bibnamefont {Anderson}},\ }\bibfield  {title} {\enquote {\bibinfo {title} {{Calculation of the Superconducting State Parameters with Retarded Electron-Phonon Interaction}},}\ }\href {\doibase 10.1103/PhysRev.125.1263} {\bibfield  {journal} {\bibinfo  {journal} {Phys. Rev.}\ }\textbf {\bibinfo {volume} {125}},\ \bibinfo {pages} {1263--1271} (\bibinfo {year} {1962})}\BibitemShut {NoStop}%
\bibitem [{\citenamefont {McMillan}(1968)}]{mcmillan}%
  \BibitemOpen
  \bibfield  {author} {\bibinfo {author} {\bibfnamefont {W.~L.}\ \bibnamefont {McMillan}},\ }\bibfield  {title} {\enquote {\bibinfo {title} {{Transition Temperature of Strong-Coupled Superconductors}},}\ }\href {\doibase 10.1103/PhysRev.167.331} {\bibfield  {journal} {\bibinfo  {journal} {Phys. Rev.}\ }\textbf {\bibinfo {volume} {167}},\ \bibinfo {pages} {331--344} (\bibinfo {year} {1968})}\BibitemShut {NoStop}%
\bibitem [{\citenamefont {Carbotte}(1990)}]{Carbotte1990}%
  \BibitemOpen
  \bibfield  {author} {\bibinfo {author} {\bibfnamefont {J~P}\ \bibnamefont {Carbotte}},\ }\bibfield  {title} {\enquote {\bibinfo {title} {Properties of boson-exchange superconductors},}\ }\href@noop {} {\bibfield  {journal} {\bibinfo  {journal} {Review of Modern Physics}\ }\textbf {\bibinfo {volume} {62}},\ \bibinfo {pages} {1027} (\bibinfo {year} {1990})}\BibitemShut {NoStop}%
\bibitem [{\citenamefont {Lee}\ and\ \citenamefont {Chang}(1996)}]{chang_nb}%
  \BibitemOpen
  \bibfield  {author} {\bibinfo {author} {\bibfnamefont {Keun-Ho}\ \bibnamefont {Lee}}\ and\ \bibinfo {author} {\bibfnamefont {K.~J.}\ \bibnamefont {Chang}},\ }\bibfield  {title} {\enquote {\bibinfo {title} {{Linear-response calculation of the Coulomb pseudopotential ${\mathrm{\ensuremath{\mu}}}^{\mathrm{*}}$ for Nb}},}\ }\href {\doibase 10.1103/PhysRevB.54.1419} {\bibfield  {journal} {\bibinfo  {journal} {Phys. Rev. B}\ }\textbf {\bibinfo {volume} {54}},\ \bibinfo {pages} {1419--1422} (\bibinfo {year} {1996})}\BibitemShut {NoStop}%
\bibitem [{\citenamefont {Bauer}\ \emph {et~al.}(2013)\citenamefont {Bauer}, \citenamefont {Han},\ and\ \citenamefont {Gunnarsson}}]{gunnarsson2013}%
  \BibitemOpen
  \bibfield  {author} {\bibinfo {author} {\bibfnamefont {Johannes}\ \bibnamefont {Bauer}}, \bibinfo {author} {\bibfnamefont {Jong~E.}\ \bibnamefont {Han}}, \ and\ \bibinfo {author} {\bibfnamefont {Olle}\ \bibnamefont {Gunnarsson}},\ }\bibfield  {title} {\enquote {\bibinfo {title} {{Retardation effects and the Coulomb pseudopotential in the theory of superconductivity}},}\ }\href {\doibase 10.1103/PhysRevB.87.054507} {\bibfield  {journal} {\bibinfo  {journal} {Phys. Rev. B}\ }\textbf {\bibinfo {volume} {87}},\ \bibinfo {pages} {054507} (\bibinfo {year} {2013})}\BibitemShut {NoStop}%
\bibitem [{\citenamefont {Floris}\ \emph {et~al.}(2005)\citenamefont {Floris}, \citenamefont {Profeta}, \citenamefont {Lathiotakis}, \citenamefont {L\"uders}, \citenamefont {Marques}, \citenamefont {Franchini}, \citenamefont {Gross}, \citenamefont {Continenza},\ and\ \citenamefont {Massidda}}]{mgb2}%
  \BibitemOpen
  \bibfield  {author} {\bibinfo {author} {\bibfnamefont {A.}~\bibnamefont {Floris}}, \bibinfo {author} {\bibfnamefont {G.}~\bibnamefont {Profeta}}, \bibinfo {author} {\bibfnamefont {N.~N.}\ \bibnamefont {Lathiotakis}}, \bibinfo {author} {\bibfnamefont {M.}~\bibnamefont {L\"uders}}, \bibinfo {author} {\bibfnamefont {M.~A.~L.}\ \bibnamefont {Marques}}, \bibinfo {author} {\bibfnamefont {C.}~\bibnamefont {Franchini}}, \bibinfo {author} {\bibfnamefont {E.~K.~U.}\ \bibnamefont {Gross}}, \bibinfo {author} {\bibfnamefont {A.}~\bibnamefont {Continenza}}, \ and\ \bibinfo {author} {\bibfnamefont {S.}~\bibnamefont {Massidda}},\ }\bibfield  {title} {\enquote {\bibinfo {title} {{Superconducting Properties of ${\mathrm{M}\mathrm{g}\mathrm{B}}_{2}$ from First Principles}},}\ }\href {\doibase 10.1103/PhysRevLett.94.037004} {\bibfield  {journal} {\bibinfo  {journal} {Phys. Rev. Lett.}\ }\textbf {\bibinfo {volume} {94}},\ \bibinfo {pages} {037004} (\bibinfo {year} {2005})}\BibitemShut {NoStop}%
\bibitem [{\citenamefont {Saunderson}\ \emph {et~al.}(2020)\citenamefont {Saunderson}, \citenamefont {Annett}, \citenamefont {\'Ujfalussy}, \citenamefont {Csire},\ and\ \citenamefont {Gradhand}}]{pb-gap1}%
  \BibitemOpen
  \bibfield  {author} {\bibinfo {author} {\bibfnamefont {Tom~G.}\ \bibnamefont {Saunderson}}, \bibinfo {author} {\bibfnamefont {James~F.}\ \bibnamefont {Annett}}, \bibinfo {author} {\bibfnamefont {Bal\'azs}\ \bibnamefont {\'Ujfalussy}}, \bibinfo {author} {\bibfnamefont {G\'abor}\ \bibnamefont {Csire}}, \ and\ \bibinfo {author} {\bibfnamefont {Martin}\ \bibnamefont {Gradhand}},\ }\bibfield  {title} {\enquote {\bibinfo {title} {{Gap anisotropy in multiband superconductors based on multiple scattering theory}},}\ }\href {\doibase 10.1103/PhysRevB.101.064510} {\bibfield  {journal} {\bibinfo  {journal} {Phys. Rev. B}\ }\textbf {\bibinfo {volume} {101}},\ \bibinfo {pages} {064510} (\bibinfo {year} {2020})}\BibitemShut {NoStop}%
\bibitem [{\citenamefont {Kuderowicz}\ \emph {et~al.}(2021)\citenamefont {Kuderowicz}, \citenamefont {W\'ojcik},\ and\ \citenamefont {Wiendlocha}}]{Kuderowicz2021}%
  \BibitemOpen
  \bibfield  {author} {\bibinfo {author} {\bibfnamefont {Gabriel}\ \bibnamefont {Kuderowicz}}, \bibinfo {author} {\bibfnamefont {Pawe\l{}}\ \bibnamefont {W\'ojcik}}, \ and\ \bibinfo {author} {\bibfnamefont {Bartlomiej}\ \bibnamefont {Wiendlocha}},\ }\bibfield  {title} {\enquote {\bibinfo {title} {{Electronic structure, electron-phonon coupling, and superconductivity in noncentrosymmetric ${\mathrm{ThCoC}}_{2}$ from ab initio calculations}},}\ }\href {\doibase 10.1103/PhysRevB.104.094502} {\bibfield  {journal} {\bibinfo  {journal} {Phys. Rev. B}\ }\textbf {\bibinfo {volume} {104}},\ \bibinfo {pages} {094502} (\bibinfo {year} {2021})}\BibitemShut {NoStop}%
\end{thebibliography}%

\vspace*{600pt}

\newpage
\onecolumngrid
\section*{Supplemental Material}

\renewcommand{\thefigure}{{S\arabic{figure}}}
\setcounter{figure} 0

\vspace*{24pt}
\noindent
Supplemental Material contains:\\
Figure S1 with the Brillouin zone and high-symmetry points.\\
Figure S2 with electronic band structure on a longer path and in a wide energy range.\\
Figure S3 with the electronic densities of states calculated without SOC.\\
Figure S4 with the orbital character of the Fermi surface.\\
Figure S5 with the orbital character of the band structure.\\
Figure S6 with the summed distribution of the superconducting gap.\\
Figure S7 with Fermi surface cross sections for the first three bands.\\
Figure S8 with the scalar-relativistic Fermi surface.\\
Figure S9 with phonon grid convergence test.\\
Figure S10 with DOS at different hydrostatic pressures.\\
Table I with relaxed atomic positions (with and without SOC).\\
Table II with atom displacements in a relaxed structure at 10 GPa.\\
As separate files, animations of the atomic vibrations are included.
Animations of the phonon modes at high symmetry points $\Gamma$, K, M, H, L, A and for the three highest optical modes.
Re-1 atoms are gray, Re-2 red and Zr green.
The files are following:
\begin{itemize}
    \item video{\textunderscore}GKMHLA.avi, compilation of all 18 atomic vibrations in one window,
    \item video{\textunderscore}LabelN{\textunderscore}all.avi - Label=(Gamma,K,M,H,L,A) and N=(34,35,36), atomic vibrations in separate videos,
    \item video{\textunderscore}Gamma36.avi - the breathing mode with highest frequency showing only Re-2 atoms in a kagome lattice.
\end{itemize}

\begin{figure}[htb]
    \centering
    \includegraphics[width=0.5\linewidth]{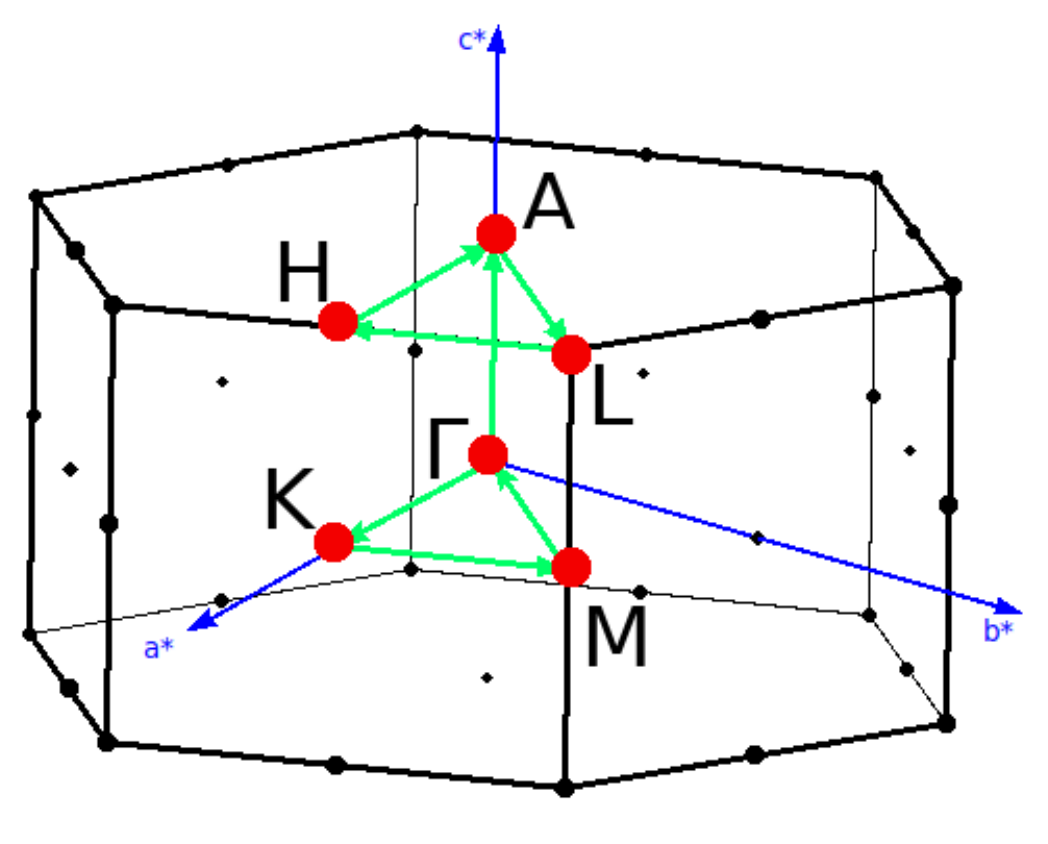}
    \caption{Brillouin zone of Re$_2$Zr with high symmetry points.}
    \label{fig_BS_sympoints}
\end{figure}

\begin{table}
\caption{Re$_2$Zr relaxed atomic positions.}
    \centering
    \begin{ruledtabular}
    \begin{tabular}{ccccccc}
     & \multicolumn{2}{c}{$x$} & \multicolumn{2}{c}{$y$} & \multicolumn{2}{c}{$z$} \\
     & with SOC & w/o SOC & with SOC & w/o SOC & with SOC & w/o SOC \\
    \hline
    \multirow{2}{*}{Re-1 (2a)} & 0.0 & 0.0 & 0.0 & 0.0 & 0.0 & 0.0 \\
     & 0.0 & 0.0 & 0.0 & 0.0 & 0.5 & 0.5 \\
    \hline
    \multirow{6}{*}{Re-2 (6h)} & 0.171048 & 0.171330 & 0.342096 & 0.342660 & 0.75 & 0.75 \\
     & 0.828952 & 0.828670 & 0.657904 & 0.657340 & 0.25 & 0.25 \\
     & 0.657904 & 0.657340 & 0.828952 & 0.828670	& 0.75 & 0.75 \\
     & 0.342096 & 0.342660 & 0.171048 & 0.171330 & 0.25 & 0.25 \\
     & 0.171048 & 0.171330 & 0.828952 & 0.828670 & 0.75 & 0.75 \\
     & 0.828952 & 0.828670 & 0.171048 & 0.171330 & 0.25 & 0.25 \\
    \hline
    \multirow{4}{*}{Zr (4f)} & 1/3 & 1/3 & 2/3 & 2/3 & 0.437628 & 0.438240 \\
     & 2/3 & 2/3 & 1/3 & 1/3 & 0.562372 & 0.561760 \\
     & 2/3 & 2/3 & 1/3 & 1/3 & 0.937628 & 0.938240 \\
     & 1/3 & 1/3 & 2/3 & 2/3 & 0.062372 & 0.061760 \\
    \end{tabular}
    \end{ruledtabular}
    \label{tab_atpos}
\end{table}

\begin{figure}[htb]
    \centering
    \includegraphics[width=0.99\linewidth]{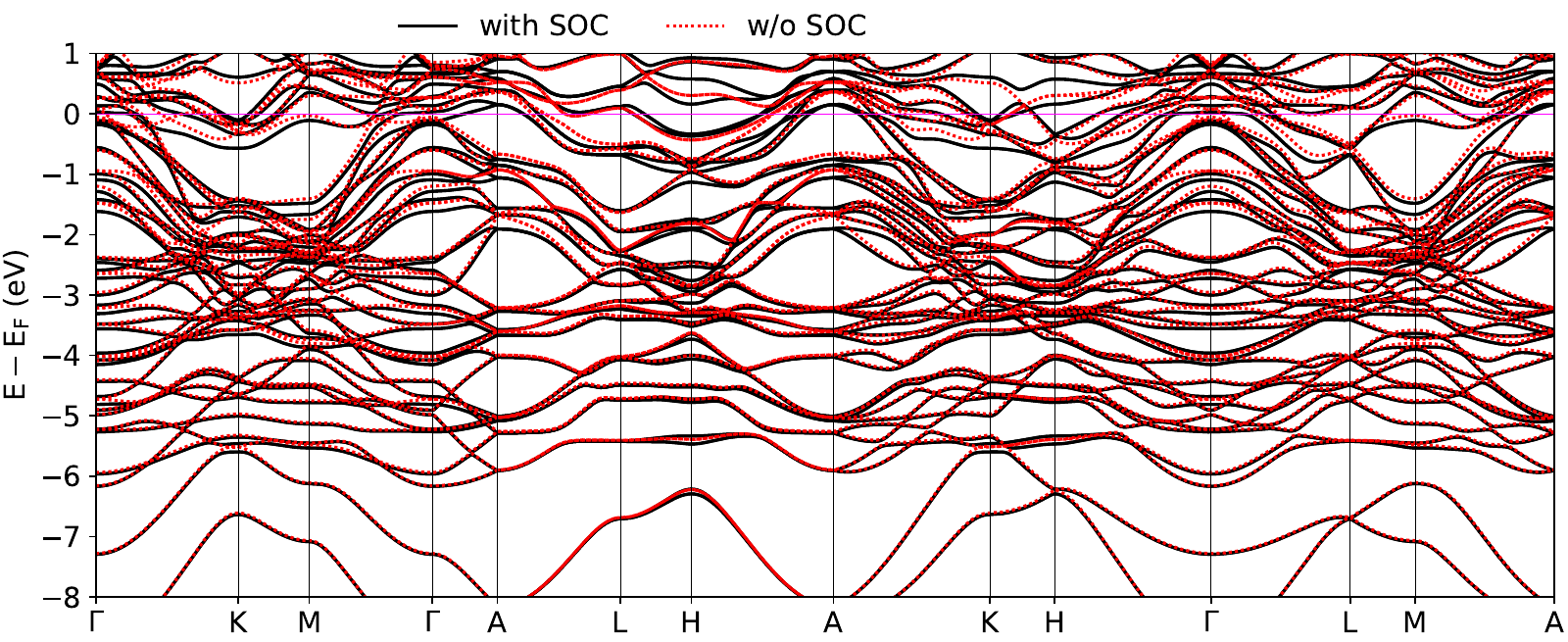}
    \caption{Re$_2$Zr electronic band structure on a longer path and in wider energy range.}
    \label{fig_elbands_large}
\end{figure}

\begin{figure}[htb]
    \centering
    \includegraphics[width=0.99\linewidth]{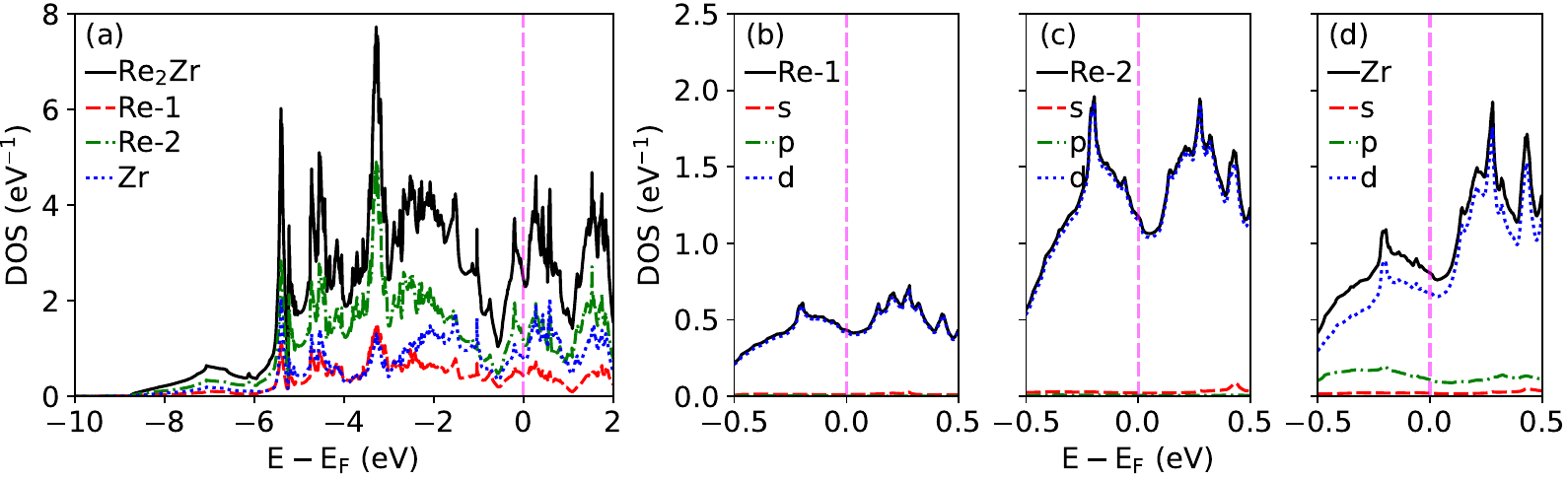}
    \caption{Re$_2$Zr electronic density of states projected onto atomic orbitals calculated without SOC.}
    \label{fig_eldos_nosoc}
\end{figure}

\begin{figure}[htb]
    \centering
    \includegraphics[width=0.99\linewidth]{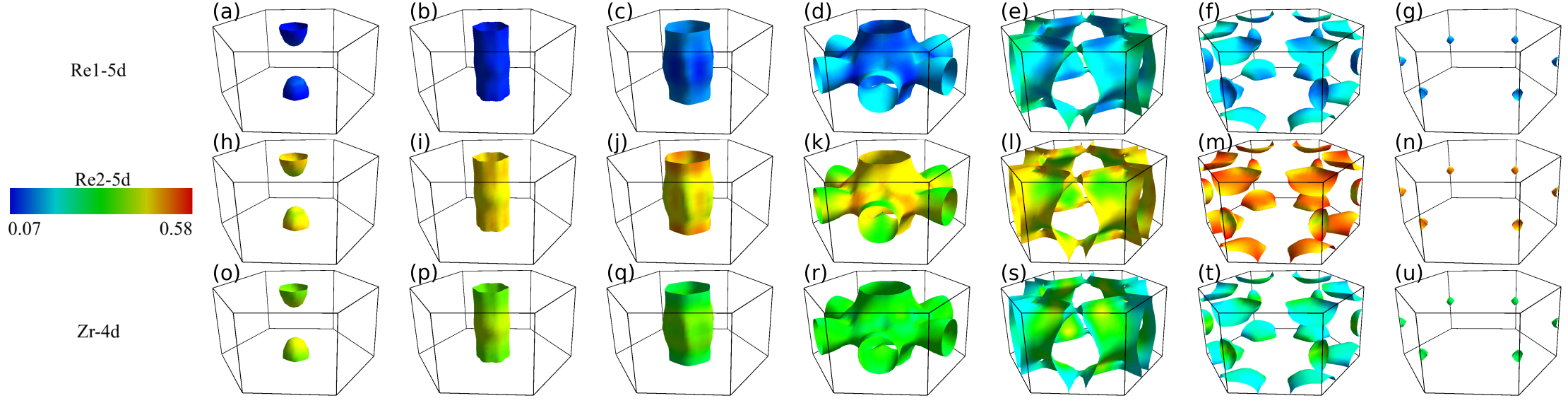}
    \caption{Re$_2$Zr orbital character of the Fermi surface projected on Re1-5d, Re2-5d and Zr-4d orbitals.}
    \label{fig_fschar}
\end{figure}

\begin{figure}[htb]
    \centering
    \includegraphics[width=0.99\linewidth]{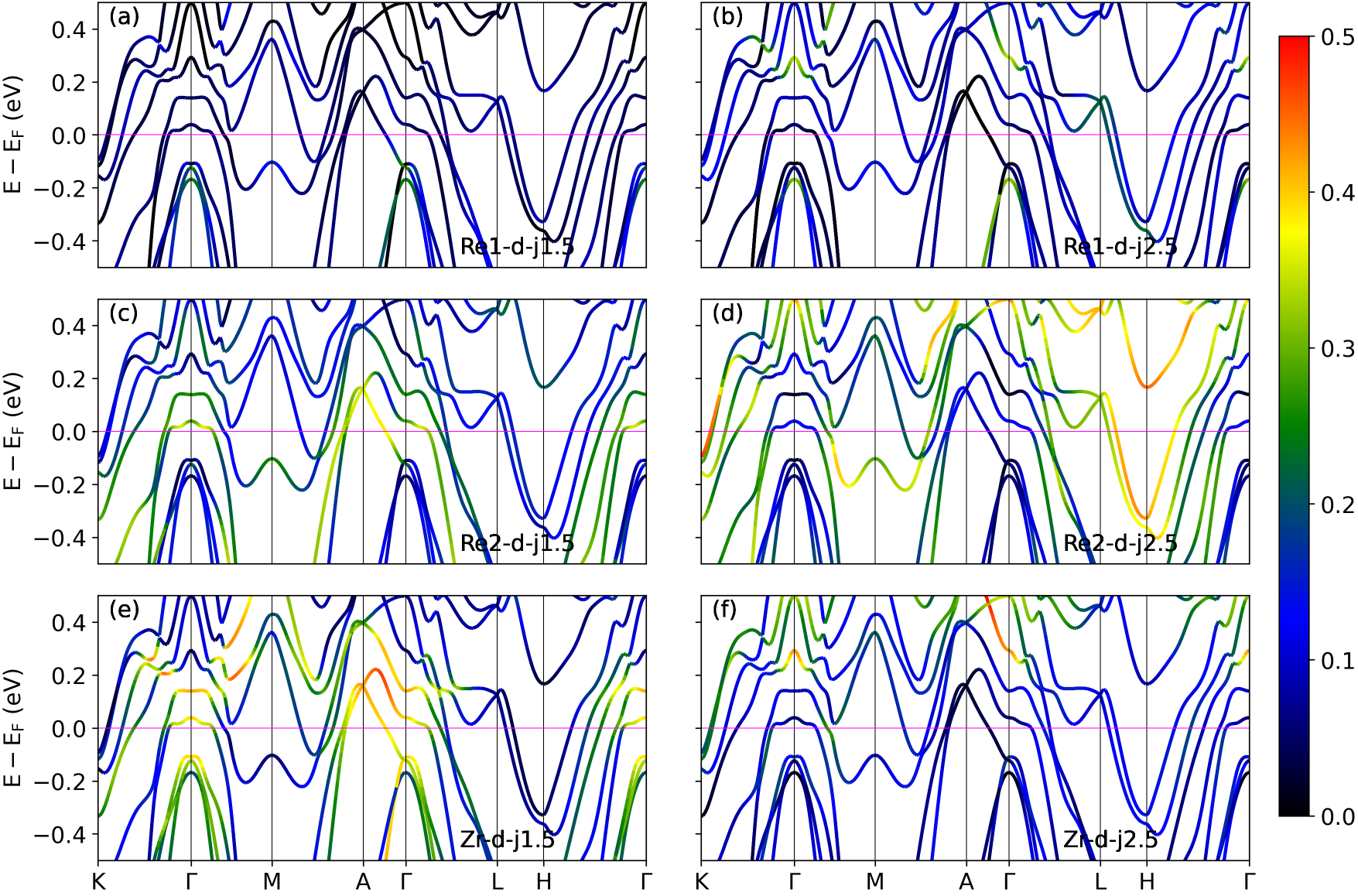}
    \caption{Re$_2$Zr orbital character of the band structure projected on Re1-d, Re2-d and Zr-d orbitals with J=(1.5,2.5).}
    \label{fig_bandschar}
\end{figure}

\begin{figure}[htb]
    \centering
    \includegraphics[width=0.7\linewidth]{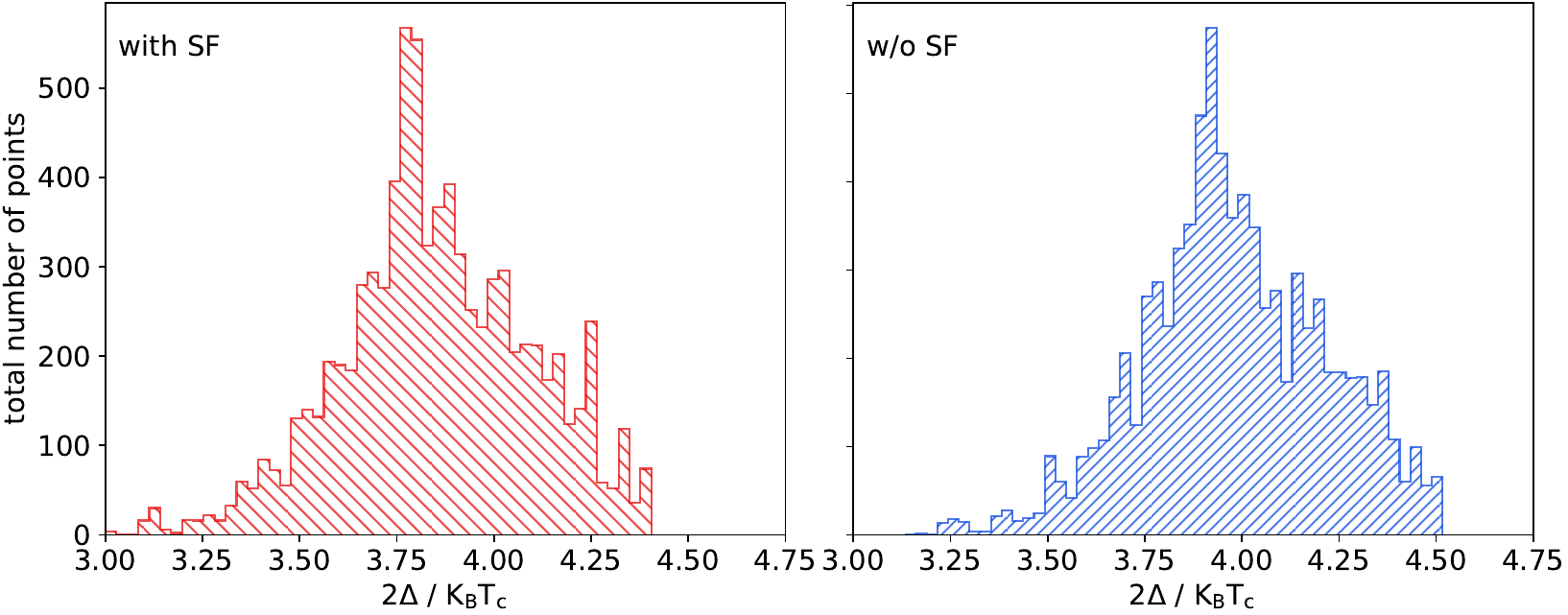}
    \caption{Superconducting gap distribution from all bands.}
    \label{fig_gapdist_all}
\end{figure}

\begin{figure}[htb]
    \centering
    \includegraphics[width=0.99\linewidth]{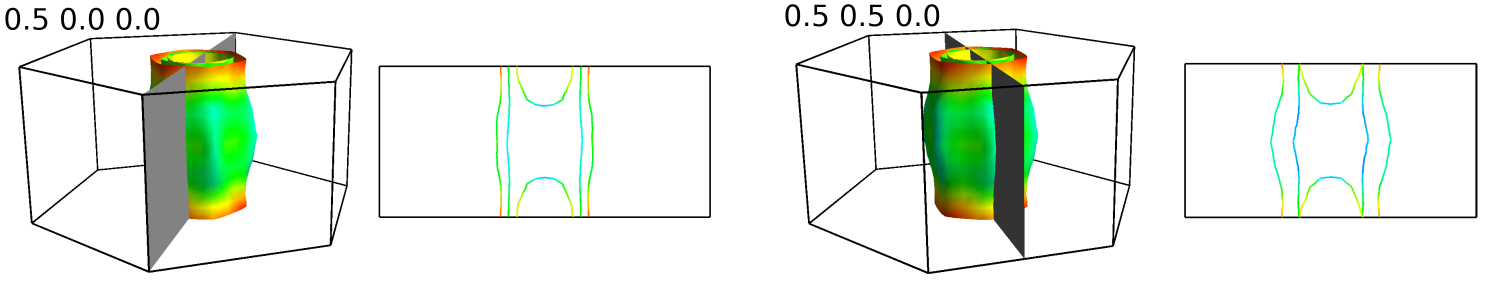}
    \caption{Re$_2$Zr Fermi surface cross sections for the first three bands.}
    \label{fig_fscuts}
\end{figure}

\begin{figure}[htb]
    \centering
    \includegraphics[width=0.5\linewidth]{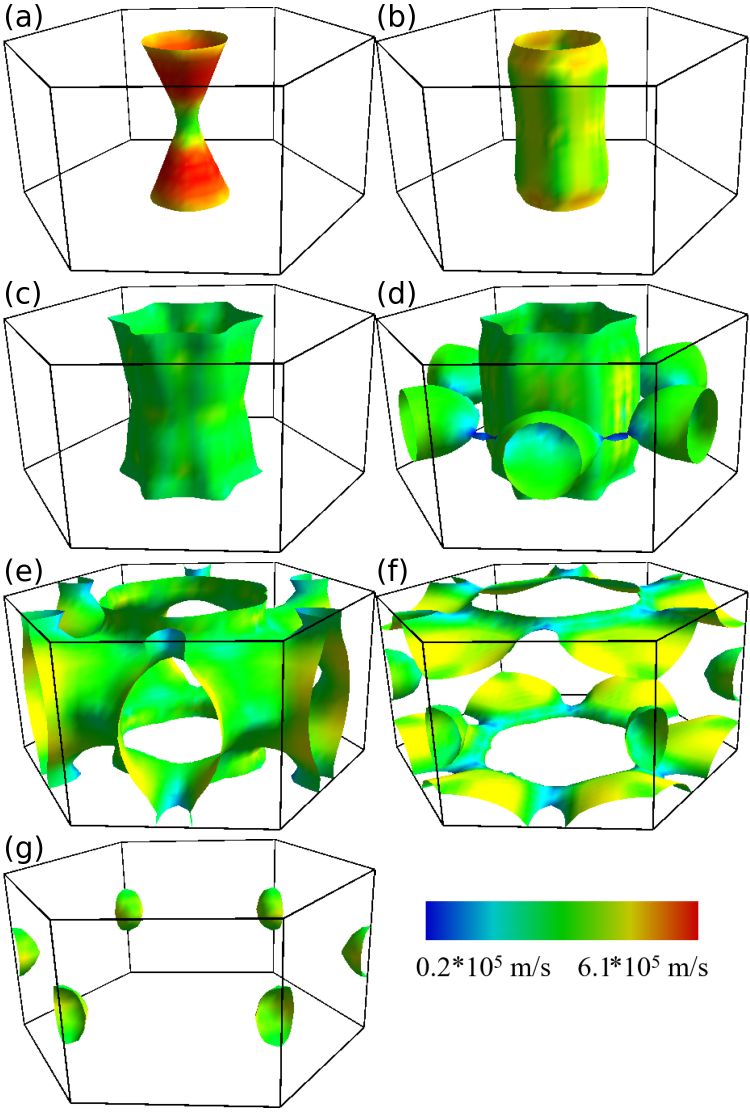}
    \caption{Re$_2$Zr Fermi surface calculated w/o SOC.}
    \label{fig_fs_nosoc}
\end{figure}


\newpage \pagebreak \cleardoublepage
\section*{Convergence analysis}

\begin{figure}[htb]
    \centering
    \includegraphics[width=0.8\linewidth]{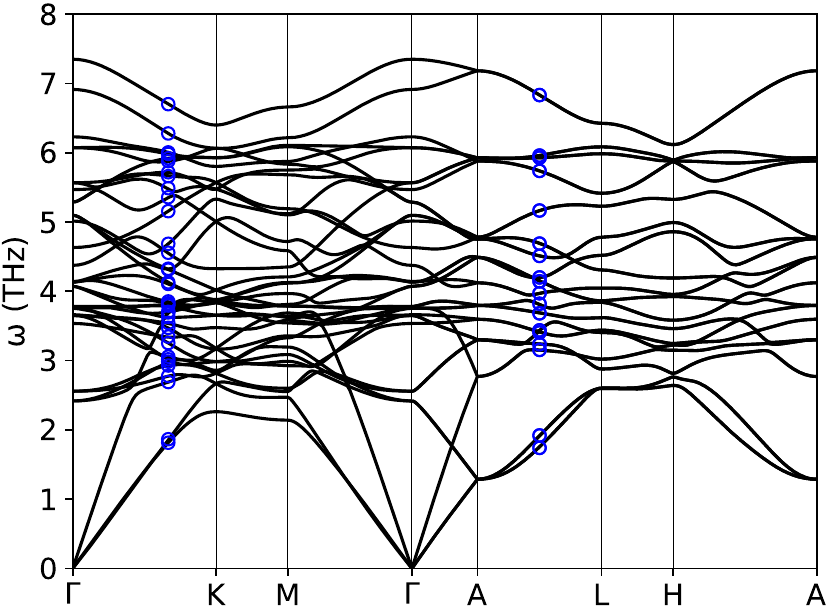}
    \caption{Convergence test: Re$_2$Zr phonon dispersion interpolated from a 7x7x3 $\mathbf{q}$-grid with exact frequencies calculated for two $\mathbf{q}$ points.}
    \label{fig_phqtest}
\end{figure}

The electronic structure has been calculated on two k-point grids, 14x14x6 and 28x28x12, as both grids lead to practically the same Fermi surface sheets and densities of states, it confirms convergence of the electronic structure. 


Convergence of the phonon structure: to test the convergence of the q-mesh, the dynamical matrices and phonon frequencies have been calculated for two q-points not present in the q-mesh used for the real-space force constant calculations. 
The obtained phonon frequencies are compared in Fig.~\ref{fig_phqtest} with the frequencies obtained from the force constants calculated using the Fourier interpolation from the 7x7x3 q-grid . The difference between the “exact” and interpolated frequencies is unnoticeable confirming the convergence of the grid.

Convergence of the electron-phonon coupling: SCDFT calculations in the Superconducting-Toolkit require using the tetrahedron method for the matrix elements, which provides an additional measure of accuracy of the obtained electron-phonon coupling constant by comparing average $\lambda$ from two methods, tetrahedron (SCDFT) and smearing (QE calculations of the Eliashberg function use the smearing technique in the integrations).
Both calculations are independent and done on different k- and q- meshes. The SCDFT calculations were done on a coarser 14x14x6 k-grid for electron-phonon matrix elements because the use of shifted q-mesh requires a larger number of dynamical matrices to be calculated. In our case that was 32 dynamical matrices. Despite that the interpolated Fermi surface in fig.7 and fig.8 is very similar to the one in fig. 3 that was calculated on a denser 28x28x12 k-grid. 
The electron-phonon coupling parameter $\lambda$, obtained from tetrahedron calculations on a coarser k-grid and a shifted q-grid, is only 3\% larger than the one from the smearing method on a finer k-grid and an unshifted q-grid, so we can conclude that it is converged up to a few percent.

Convergence of $T_c$: To estimate the convergence of $T_c$ we use the two values of $\lambda$, obtained using the tetrahedron and smearing methods. Taking the average frequency $\omega_{ln}$=163 K from the equation (7), Coulomb pseudopotential $\mu^*=0.1$ and two values of average $\lambda$ we obtain $T_c(\lambda_1=0.868)=8.9 K$ and $T_c(\lambda_2=0.895)=9.4$ K. Based on those two cases the computational uncertainty of Tc is estimated to be about 6\%.

\newpage \pagebreak \cleardoublepage
\section*{The effect of hydrostatic pressure on the structure and the van Hove singularity}

\begin{figure}[htb]
    \centering
    \includegraphics[width=0.6\linewidth]{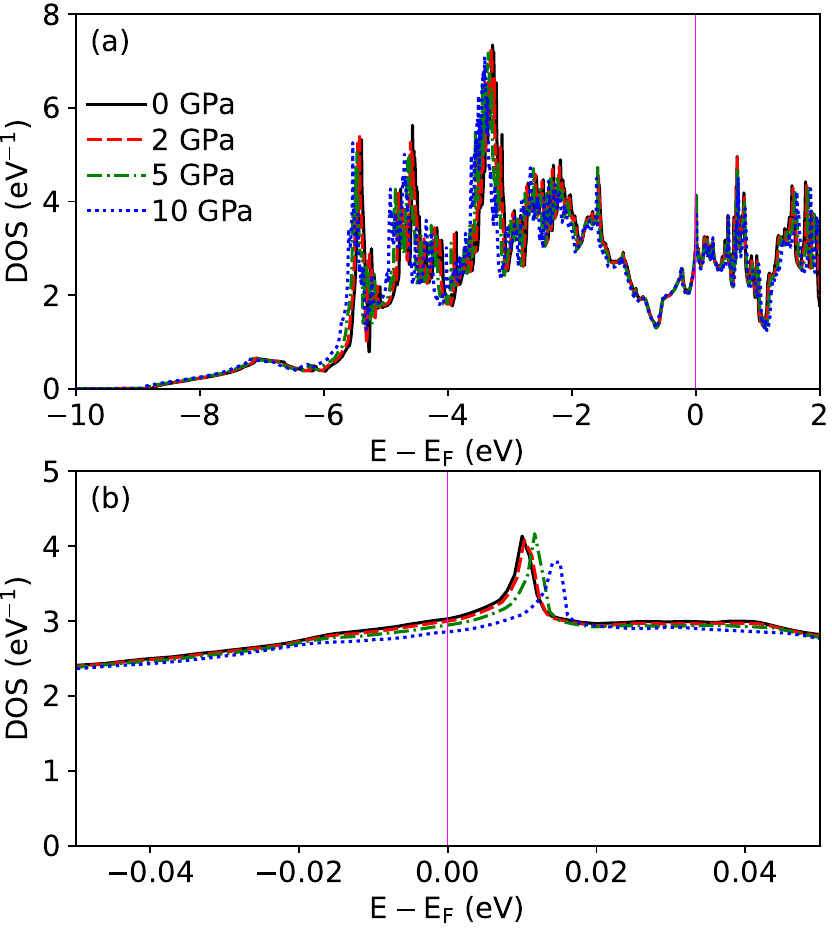}
    \caption{Electronic DOS of Re$_2$Zr under hydrostatic pressures 2, 5 and 10 GPa. Panel (b) magnifies a region near the van Hove singularity close to $E_F$, which forms also under pressure with a slightly modified lattice geometry.}
    \label{fig_press}
\end{figure}

\begin{table}[b]
\caption{Difference between relaxed atomic position at 10 GPa and ambient pressure in the cartesian coordinates and $a_B$. Relaxed lattice parameters at 10 GPa are $a$=9.874748 $a_B$ and $c/a$=1.639813. In the most compressed structure at 10 GPa, the $z$ coordinates change significantly more than $x$ and $y$ ones which is another argument for the importance of the geometrical frustration in the kagome lattice planes with Re2 atoms.}
    \centering
    \begin{ruledtabular}
    \begin{tabular}{cccc}
Re1 & 0.000000 & 0.000000 & 0.000000 \\
Re1 & 0.000000 & 0.000000 & -0.100438 \\
Re2 & 0.000000 & -0.035459 & -0.150656 \\
Re2 & -0.059850 & -0.068205 & -0.050219 \\
Re2 & -0.029142 & -0.085934 & -0.150656 \\
Re2 & -0.030708 & -0.017729 & -0.050219 \\
Re2 & 0.029142 & -0.085934 & -0.150656 \\
Re2 & -0.088992 & -0.017728 & -0.050219 \\
Zr & 0.000000 & -0.069109 & -0.086335 \\
Zr & -0.059850 & -0.034554 & -0.114541 \\
Zr & -0.059850 & -0.034554 & -0.186772 \\
Zr & 0.000000 & -0.069109 & -0.014103 \\
    \end{tabular}
    \end{ruledtabular}
    \label{tab_press}
\end{table}

\end{document}